\documentclass[twocolumn,showpacs,aps,prc,amsmath,amssymb,superscriptaddress,groupeaddress]{revtex4}

\usepackage{graphicx}
\usepackage{dcolumn}
\usepackage{bm}
\usepackage{epsfig}

\begin{document}

\preprint{APS/123-QED}

\title{$\beta$-decay half-lives and $\beta$-delayed neutron emission probabilities of nuclei in the region $A \lesssim 110$, relevant for the r-process}

\author{J.~Pereira}
\email{pereira@nscl.msu.edu}
\affiliation{National Superconducting Cyclotron Laboratory, Michigan State University, E.~Lansing, MI, USA}
\affiliation{Joint Institute for Nuclear Astrophysics, Michigan State University, E.~Lansing, MI, USA}
\author{S.~Hennrich}
\affiliation{Institut f\"{u}r Kernchemie, Universit\"{a}t Mainz, Mainz, Germany}
\affiliation{Joint Institute for Nuclear Astrophysics, Michigan State University, E.~Lansing, MI, USA}
\affiliation{Virtuelles Institut f\"{u}r Struktur der Kerne and Nuklearer Astrophysik, Mainz, Germany}
\author{A.~Aprahamian}
\affiliation{Institute of Structure and Nuclear Astrophysics, University of Notre Dame, South Bend, IN, USA}
\affiliation{Joint Institute for Nuclear Astrophysics, University of Notre Dame, South Bend, IN, USA}
\author{O.~Arndt}
\affiliation{Institut f\"{u}r Kernchemie, Universit\"{a}t Mainz, Mainz, Germany}
\affiliation{Virtuelles Institut f\"{u}r Struktur der Kerne and Nuklearer Astrophysik, Mainz, Germany}
\author{A.~Becerril}
\affiliation{National Superconducting Cyclotron Laboratory, Michigan State University, E.~Lansing, MI, USA}
\affiliation{Joint Institute for Nuclear Astrophysics, Michigan State University, E.~Lansing, MI, USA}
\affiliation{Department of Physics and Astronomy, Michigan State University, E.~Lansing, MI, USA}
\author{T.~Elliot}
\affiliation{National Superconducting Cyclotron Laboratory, Michigan State University, E.~Lansing, MI, USA}
\affiliation{Joint Institute for Nuclear Astrophysics, Michigan State University, E.~Lansing, MI, USA}
\affiliation{Department of Physics and Astronomy, Michigan State University, E.~Lansing, MI, USA}
\author{A.~Estrade}
\affiliation{National Superconducting Cyclotron Laboratory, Michigan State University, E.~Lansing, MI, USA}
\affiliation{Joint Institute for Nuclear Astrophysics, Michigan State University, E.~Lansing, MI, USA}
\affiliation{Department of Physics and Astronomy, Michigan State University, E.~Lansing, MI, USA}
\author{D.~Galaviz}
\altaffiliation{Present address: Centro de F\'isica Nuclear da Universidade de Lisboa, 1649-003, Lisboa, Portugal}
\affiliation{National Superconducting Cyclotron Laboratory, Michigan State University, E.~Lansing, MI, USA}
\affiliation{Joint Institute for Nuclear Astrophysics, Michigan State University, E.~Lansing, MI, USA}
\author{R.~Kessler}
\affiliation{Institut f\"{u}r Kernchemie, Universit\"{a}t Mainz, Mainz, Germany}
\affiliation{Joint Institute for Nuclear Astrophysics, Michigan State University, E.~Lansing, MI, USA}
\affiliation{Virtuelles Institut f\"{u}r Struktur der Kerne and Nuklearer Astrophysik, Mainz, Germany}
\author{K.-L.~Kratz}
\affiliation{Virtuelles Institut f\"{u}r Struktur der Kerne and Nuklearer Astrophysik, Mainz, Germany}
\affiliation{Max Planck Institut f\"{u}r Chemie, Otto-Hahn-Institut, Mainz, Germany}
\author{G.~Lorusso}
\affiliation{National Superconducting Cyclotron Laboratory, Michigan State University, E.~Lansing, MI, USA}
\affiliation{Joint Institute for Nuclear Astrophysics, Michigan State University, E.~Lansing, MI, USA}
\affiliation{Department of Physics and    Astronomy, Michigan State University, E.~Lansing, MI, USA}
\author{P.~F.~Mantica}
\affiliation{National Superconducting Cyclotron Laboratory, Michigan State University, E.~Lansing, MI, USA}
\affiliation{Department of Chemistry, Michigan State University, E.~Lansing, MI, USA}
\author{M.~Matos}
\affiliation{National Superconducting Cyclotron Laboratory, Michigan State University, E.~Lansing, MI, USA}
\affiliation{Joint Institute for Nuclear Astrophysics, Michigan State University, E.~Lansing, MI, USA}
\author{P.~M\"oller}
\affiliation{Theoretical Division, Los Alamos National Laboratory, NM, USA}
\author{F.~Montes}
\affiliation{National Superconducting Cyclotron Laboratory, Michigan State University, E.~Lansing, MI, USA}
\affiliation{Joint Institute for Nuclear Astrophysics, Michigan State University, E.~Lansing, MI, USA}
\author{B.~Pfeiffer}
\affiliation{Institut f\"{u}r Kernchemie, Universit\"{a}t Mainz, Mainz, Germany}
\affiliation{Virtuelles Institut f\"{u}r Struktur der Kerne and Nuklearer Astrophysik, Mainz, Germany}
\author{H.~Schatz}
\affiliation{National Superconducting Cyclotron Laboratory, Michigan State University, E.~Lansing, MI, USA}
\affiliation{Joint Institute for Nuclear Astrophysics, Michigan State University, E.~Lansing, MI, USA}
\affiliation{Department of Physics and Astronomy, Michigan State University, E.~Lansing, MI, USA}
\author{F.~Schertz}
\affiliation{Institut f\"{u}r Kernchemie, Universit\"{a}t Mainz, Mainz, Germany}
\affiliation{Joint Institute for Nuclear Astrophysics, Michigan State University, E.~Lansing, MI, USA}
\affiliation{Virtuelles Institut f\"{u}r Struktur der Kerne and Nuklearer Astrophysik, Mainz, Germany}
\author{L.~Schnorrenberger}
\affiliation{Joint Institute for Nuclear Astrophysics, Michigan State University, E.~Lansing, MI, USA}
\affiliation{Institut f\"{u}r Kernphysik, TU Darmstadt, Darmstadt, Germany}
\author{E.~Smith}
\affiliation{Joint Institute for Nuclear Astrophysics, Michigan State University, E.~Lansing, MI, USA}
\affiliation{Department of Physics, Ohio State University, Columbus, OH, USA}
\author{A.~Stolz}
\affiliation{National Superconducting Cyclotron Laboratory, Michigan State University, E.~Lansing, MI, USA}
\author{M.~Quinn}
\affiliation{Institute of Structure and Nuclear Astrophysics, University of Notre Dame, South Bend, IN, USA}
\affiliation{Joint Institute for Nuclear Astrophysics, University of Notre Dame, South Bend, IN, USA}
\author{W.~B.~Walters}
\affiliation{Department of Chemistry and Biochemistry, University of Maryland, College Park, MD, USA}
\author{A.~W\"{o}hr}
\affiliation{Institute of Structure and Nuclear Astrophysics, University of Notre Dame, South Bend, IN, USA}
\affiliation{Joint Institute for Nuclear Astrophysics, University of Notre Dame, South Bend, IN, USA}


\date{\today}

\begin{abstract}
Measurements of the $\beta$-decay properties of $A \lesssim 110$ r-process nuclei have been completed at the National Superconducting Cyclotron Laboratory, at Michigan State University. $\beta$-decay half-lives for $^{105}$Y, $^{106,107}$Zr and $^{111}$Mo, along with $\beta$-delayed neutron emission probabilities of $^{104}$Y, $^{109,110}$Mo and upper limits for $^{105}$Y, $^{103-107}$Zr and $^{108,111}$Mo have been measured for the first time. Studies on the basis of the quasi-random phase approximation are used to analyze the ground-state deformation of these nuclei.
\end{abstract}



\maketitle
\section{Introduction}~\label{sec:Introduction}
The rapid neutron-capture process (r-process)~\cite{B2FH,Cam57} remains as one of the most exciting and challenging questions in nuclear astrophysics. In particular, the theoretical quest to explain the production of r-process isotopes and the astrophysical scenario where this process occurs have not yet been satisfactorily solved (for a general review see, for instance, Ref.~\cite{Cow91,Tru02,Cow06}). R-process abundance distributions are typically deduced by subtracting the calculated s- and p-process contributions from the observed solar system abundances. Furthermore, elemental abundances originated in the early Galaxy can be directly observed in metal-poor, r-process-enriched stars (MPRES) (i.e., $[Fe/H]$$\lesssim$$-2$, $[Ba/Eu]$$\lesssim$$-0.7$, $[Eu/Fe]$$\gtrsim$$+1$). These combined observations reveal disparate behavior for light and heavy nuclei: MPRES abundance-patterns are nearly consistent from star to star and with the relative solar system r-process abundances for the heavier neutron-capture elements $A$$\gtrsim$130 (Ba and above), suggesting a rather robust main r-process operating over the history of the Galaxy. Such a consistent picture is not seen for light neutron-capture elements in the range 39$\leq$$Z$$\leq$50, as the solar system Eu-normalized elemental abundances in MPRES show a scattered pattern~\cite{Joh02,Sne03,Aok05,Bar05}. Anti-correlation trends between elemental abundances and Eu richness at different metalicities have been suggested to provide a hint for an additional source of isotopes below $A$$\simeq$130~\cite{Tra04,Mon07,Kra07,Far08,Kra08}. Measured abundances of $^{107}$Pd and $^{129}$I, for $A$$\textless$130, and $^{182}$Hf, for $A$$\textgreater$130, trapped in meteorites in the early solar system formation~\cite{Rey60,Lee95}, further reinforce the idea of different origins for isotopes lighter and heavier than $A$$=$130 (see e.g. Ref.~\cite{Ott08}).

Reliable nuclear physics properties for the extremely neutron-rich nuclei along the r-process path are needed to interpret the observational data in the framework of proposed astrophysical models. The r-process abundance region around $A$$\simeq$110, prior to the $A$$=$130 peak, is one region of intense interest, where the astrophysical models underestimate the abundances by an order of magnitude or more.
It has been shown that the underproduction of abundances can be largely corrected under the assumption of a reduction of the $N$$=$82 shell gap far from stability~\cite{Kra93,Che95,Pea96,Pfe97}. Whereas such quenching effect was suggested in elements near Sn---via analysis of the $\beta$ decay of $^{130}$Cd into $^{130}$In~\cite{Dil03}---its extension towards lighter isotones $^{129}_{47}$Ag$_{82}$, $^{128}_{46}$Pd$_{82}$ and below would have crucial consequences in the search for the r-process site.
In this sense, self-consistent mean-field model calculations predict that the $N$$=$82 shell quenching might be associated with the emergence of a harmonic-oscillator-like doubly semi-magic nucleus $^{110}_{40}$Zr$_{70}$, arising from the weakening of the energy potential surface due to neutron skins~\cite{Dob94,Dob96,Pfe96,Pfe03}. Thus, it is imperative to characterize the evolution of nuclear shapes in the region of $^{110}$Zr.

The goal of the experiment reported here was to use measured $\beta$-decay properties of nuclei in the neighborhood of $^{110}$Zr to investigate its possible spherical character arising from new semi-magic numbers~\cite{Dob94,Dob96}, or even a more exotic tetrahedral-symmetry type predicted by some authors~\cite{Dud02,Sch04}. In particular, the measured half-lives ($T_{1/2}$) and $\beta$-delayed neutron-emission probabilities ($P_{\rm n}$) can be used as first probes of the structure of the $\beta$-decay daughter nuclei in this mass region, where more detailed spectroscopy is prohibitive owing to the low production rates at present radioactive beam facilities. A similar approach has already been used in the past~\cite{Kra84,Sor93,Meh96,Wan99,Kra00,Han00,Woh02,Mon06}. Besides the nuclear-structure interest, our measurements also serve as important direct inputs in r-process model calculations. The $T_{1/2}$ values of r-process waiting-point nuclei determine the pre freeze-out isotopic abundances and the speed of the process towards heavier elements. The $P_{\rm n}$ values of r-process isobaric nuclei define the decay path towards stability during freeze-out, and provide a source of late-time neutrons.

In the present paper, the measurements of $T_{1/2}$ and $P_{\rm n}$ of $^{100-105}$Y, $^{103-107}$Zr, $^{106-109}$Nb, $^{108-111}$Mo and $^{109-113}$Tc are reported. The work is contextualized amid a series of $\beta$-decay r-process experimental campaigns, carried out at the National Superconducting Cyclotron Laboratory (NSCL) at Michigan State University (MSU). Details of the experiment setup and measurement techniques are provided in Sec.~\ref{sec:Experiment}, followed by a description of the data analysis in Sec.~\ref{sec:Analysis}. The results are further discussed in Sec.~\ref{sec:Discussion} on the basis of the quasi-random phase approximation (QRPA)~\cite{Kru84,Mol90,Mol97,Mol03} using nuclear shapes derived from the finite-range droplet model (FRDM)~\cite{Mol95} and the latest version of the finite-range liquid-drop model (FRLDM)~\cite{Mol07,Mol08}. The paper is closed with the presentation of the main conclusions and future plans motivated by the current measurements.

\section{Experiment}\label{sec:Experiment}

\subsection{Production and separation of nuclei}\label{sec:Production}
Neutron-rich Y, Zr, Nb, Mo and Tc isotopes were produced by fragmentation of a 120 MeV/u $^{136}$Xe beam in a 1242~mg/cm$^{2}$ Be target. The primary beam was produced at the NSCL Coupled Cyclotrons~\cite{CCF} at an average intensity of 1.5~pnA ($\sim$1$\times$10$^{10}$~s$^{-1}$). Forward-emitted fragmentation reaction products were separated in-flight with the A1900 fragment separator~\cite{Mor03}---operated in its achromatic mode---using the $B\rho$-$\Delta E$-$B\rho$ technique~\cite{Sch87}. Two plastic scintillators located at the intermediate (dispersive) focal plane and at the experimental area were used to measure the time-of-flight ($ToF$), related to the velocity of the transmitted nuclei. The first of these scintillators provided also the transversal positions $x_{d}$ of the transmitted nuclei at the dispersive plane. A Kapton wedge was mounted behind this detector to keep the achromatism of the A1900. Energy-losses experienced by nuclei passing through this degrader system of 22.51~mg/cm$^{2}$ (Kapton) and 22.22~mg/cm$^{2}$ (BC400) thickness, provided a further filter to select a narrower group of elements. A total of 29 neutron-rich isotopes ($^{100-105}$Y, $^{102-107}$Zr, $^{104-109}$Nb, $^{106-111}$Mo and $^{109-113}$Tc) defined the cocktail beam that was transmitted to the implantation station.

The trajectory followed by the nuclei at the A1900 dispersive focal plane depended on their magnetic rigidities $B\rho$, which, in turn, were related to the corresponding velocities and mass-over-charge ratios. An event-by-event separation of the transmitted nuclei according to their mass $A$ and proton $Z$ numbers was achieved by combining the measured $ToF$ and $x_{d}$ with the energy-loss $\Delta E$ in a silicon PIN detector located in the implantation station. The latter quantity had to be corrected from its velocity dependence (described by the Bethe-Bloch equation~\cite{Sig06}); the identification of nuclei transmitted through the A1900 is shown in Fig.~\ref{fig:PID}. In this figure, the variable $ToF^{*}$ corresponds to the $ToF$ corrected from its $x_{d}$ dependence, and the $\Delta E$ signal was measured with the most upstream PIN detector of the implantation station.

The maximum 5$\%$ $B\rho$ acceptance of the A1900 included the nuclei of interest, along with three primary-beam charge-states $^{136}$Xe$^{+51}$ ($B\rho$$=$3.8251~Tm), $^{136}$Xe$^{+50}$ ($B\rho$$=$3.9016~Tm) and $^{136}$Xe$^{+49}$ ($B\rho$$=$3.9812~Tm), with particle rates of 8.7$\times$10$^{6}$~s$^{-1}$, 3.5$\times$10$^{4}$~s$^{-1}$ and $80$~s$^{-1}$, respectively. As these contaminants could reach the intermediate image plane, resulting in the damage of the plastic scintillator, it was necessary to stop them with the standard A1900 slits and a 17.4~mm-wide tungsten finger located in the first image plane. This slit configuration blocked the most intense contaminants with a minor reduction of the $B\rho$ acceptance of the fragments of interest. Thus, the $^{136}$Xe$^{+50}$ charge-state was blocked by the finger at the central position of the transversal plane, while $^{136}$Xe$^{+51}$ stopped in one of the slits, closed at 39.88~mm from the optical axis (in the low-$B\rho$ side). The high-$B\rho$ slit was fully opened so that the fragment of interest (along with the low-intensity contaminant $^{136}$Xe$^{+49}$) were transmitted through the second half of the A1900. The resulting overall $B\rho$ acceptance was about 4$\%$.

\subsection{Implantation station}\label{sec:Impsetup}
Nuclei transmitted through the A1900 and beam transport system were implanted in the NSCL beta counting system (BCS)~\cite{Pri03} for subsequent analysis. The BCS consisted of a stack of four silicon PIN detectors (PIN1--4) of total thickness 2569~$\mu$m, used to measure the energy-loss of the exotic species, followed by a 979~$\mu$m-thick 40$\times$40-pixel double-sided silicon strip detector (DSSD) wherein implanted nuclei were measured along with their subsequent position-correlated $\beta$ decays. Located 9~mm downstream of the DSSD, a 988~$\mu$m-thick 16-strip single sided Si detector (SSSD) and a 10~mm-thick Ge crystal, separated 2~mm from each other, were used to veto particles whose energy-loss in the DSSD was similar to that left by $\beta$ decays. The signals from each DSSD and SSSD strip were processed by preamplifiers with high- and low-gain outputs, each defined over a scale-range equivalent to 100~MeV and 3~GeV, respectively. Energy thresholds for the low-gain signals were set to values around 300~MeV.
High-gain signals from every DSSD and SSSD strips were energy-matched in the beginning of the experiment using a $^{228}$Th $\alpha$-source as reference, whereas independent threshold were adjusted using a $^{90}$Sr $\beta$-source. Two of the DSSD strips (the front-side 31$^{\rm st}$ and the back-side $12^{\rm th}$) were damaged and had consequently to be disabled during the experiment, raising their high/low-gain thresholds to maximum values. Energy thresholds for the PIN detector were set to 2000~MeV for the first detector and around 300~MeV for the rest of PIN detectors. A dedicated un-wedged setting of the A1900, transmitting a large amount of light fragments, was used in the beginning of the experiment to energy-calibrate the PIN detectors and low-gain signals from DSSD. The measured energy losses were compared with calculations performed with the LISE program~\cite{Baz02}, using the Ziegler energy-loss formulation~\cite{Zie85}.

The BCS was surrounded by the neutron emission ratio observer (NERO) detector~\cite{Lor06,Lor08,Hos04}, which was used to measure $\beta$-delayed neutrons in coincidence with the $\beta$-decay precursor. This detector consisted of sixty (16 $^{3}$He and 44 B$_{3}$F) proportional gas-counter tubes, embedded in a 60$\times$60$\times$80~cm$^{3}$ polyethylene moderator matrix. The detectors were set parallel to the beam direction and arranged in three concentric rings around a 22.4~cm-diameter vacuum beam line that accommodated the BCS. $\beta$-delayed neutrons were thermalized in the polyethylene moderator to maximize the neuron capture cross section in the $^{3}$He and B$_{3}$F gas counters. Gains and thresholds of the neutron counters were adjusted using a $^{252}$Cf post-fission neutron source.

The master-event trigger was provided by the PIN1 detector or by a coincidence between the DSSD front and back high-gain outputs. The master trigger opened a 200~$\mu$s-time window~\cite{Mon06,Hos04}, during which signals from the NERO counters were recorded by an 80~ms-range multi-hit TDC. The 200~$\mu$s interval was chosen on the basis of an average moderation time of about 150~$\mu$s, measured for neutrons emitted from a $^{252}$Cf source. Energy signals of moderated neutrons were also collected in an ADC. The closure of the time window was followed by the readout of the BCS and NERO electronics.

\subsection{Identification of nuclei}\label{sec:PIDsetup}
The particle identification (PID) was performed with a dedicated setup installed upstream of the BCS/NERO end station. Several nuclei having $\mu$s isomers with known $\gamma$-decay transitions were selected in a specific A1900 setting, and implanted in a 4~mm-thick aluminum foil, surrounded by three $\gamma$-ray detectors from the MSU segmented germanium array (SeGA)~\cite{Mue01}. These detectors were energy and efficiency calibrated using sources of $^{57}$Co, $^{60}$Co and $^{152}$Eu. The $\gamma$-peak efficiency was about 6$\%$ at 1~MeV. A 50$\times$50~mm$^{2}$, 503~$\mu$m-thick silicon PIN detector (PIN0) upstream of the aluminium foil was used to measure $\Delta E$, which, combined with $ToF$ and $x_{d}$, allowed for an identification of the various nuclei transmitted in the setting according to their $Z$ and $A$ numbers. The measurement of these three signals, gated in specific known isomeric $\gamma$-decay lines, provided a filter to identify the corresponding $\mu$s isomers.

Four settings of the A1900 were used to identify the nuclei of interest in a stepwise fashion based on the observation of the $\mu$s isomers. In each of these settings, the first half of the A1900 was tuned to $B\rho$=3.9016~Tm---defined by the rigidities of the nuclei of interest---while the second half was tuned to $B\rho$=3.7881~Tm, $B\rho$=3.7929~Tm, $B\rho$=3.7976~Tm and $B\rho$=3.8024~Tm. The four values were chosen to have fragments transmitted in two consecutive settings, and far enough to cover the entire range of rigidities of interest. In the first setting $B\rho$=3.7881~Tm, several $\gamma$ lines were seen for the $\mu$s-isomeric states of $^{121}$Pd (135~keV), $^{123}$Ag (714~keV), $^{124}$Ag (156~keV), $^{125}$Ag (684~keV) and $^{125}$Cd (720~keV and 743~keV)~\cite{Tom06,Tom07}. From these references, it was possible to identify the more exotic nuclei present in the subsequent settings. The particle identification was further confirmed by detecting the 135~keV $\gamma$ line from the isomer $^{121}$Pd transmitted in the settings $B\rho$=3.7929~Tm and $B\rho$=3.7976~Tm. The fourth setting $B\rho$=3.8024~Tm was chosen to optimize the transmission of the nuclei of interest to the final end station.

After the PID was completed, the Al catcher and PIN0 detector were retracted. The cocktail beam was then transmitted to the final BCS/NERO end station, where the $\Delta E$ necessary for the PID spectrum was provided by the PIN1 detector.

\begin{figure}[h!]
\begin{center}
\includegraphics[width=8cm]{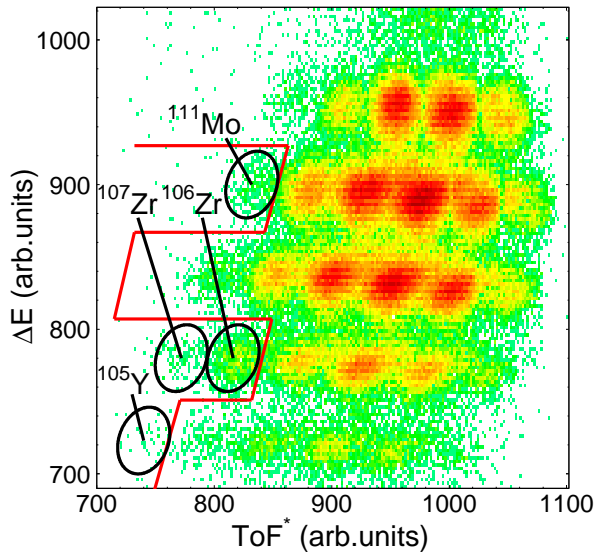} 
\caption{(Color online) Particle-identification (PID) spectra collected during 112 hours of beam time, showing the different nuclei characterizing the cocktail beam transmitted to the end station. The solid line marks the r-process region of interest (left side). The r-process waiting-point nuclei analyzed in the present experiment are indicated by the ellipses.}
\label{fig:PID}
\end{center}
\end{figure}

\begin{figure*}[t!]
\begin{center}
\includegraphics[width=3cm]{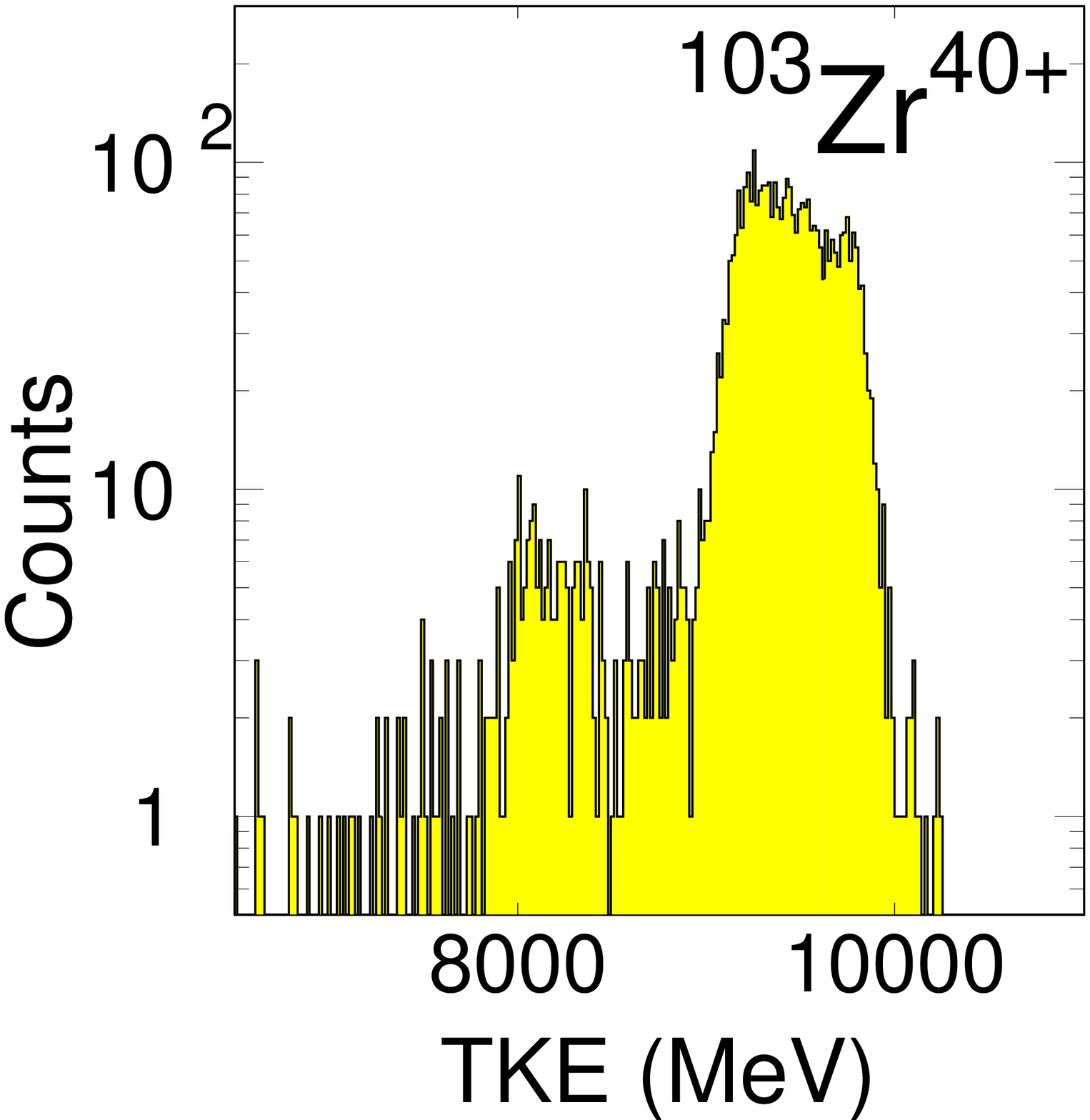}
\includegraphics[width=3cm]{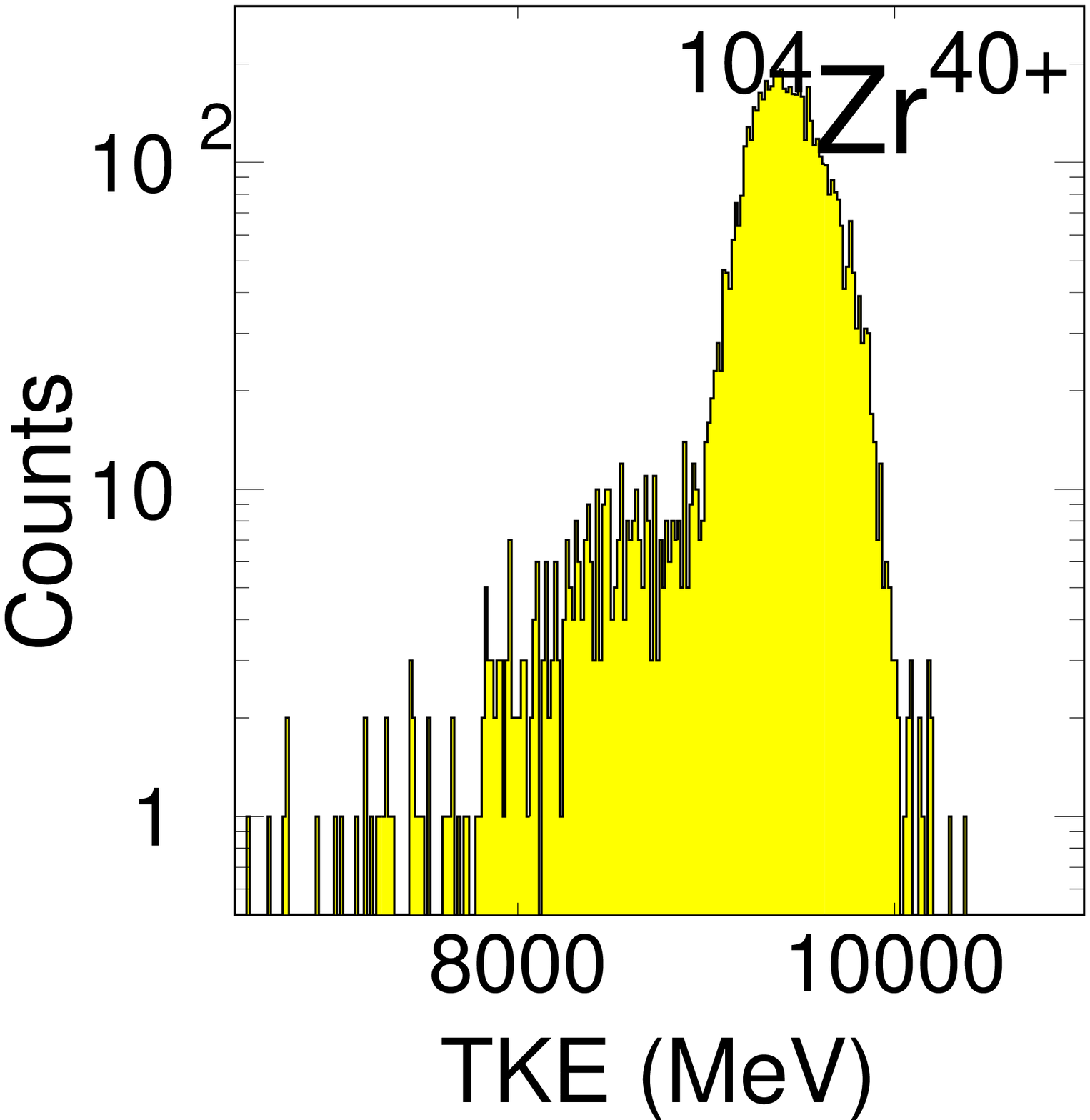}
\includegraphics[width=3cm]{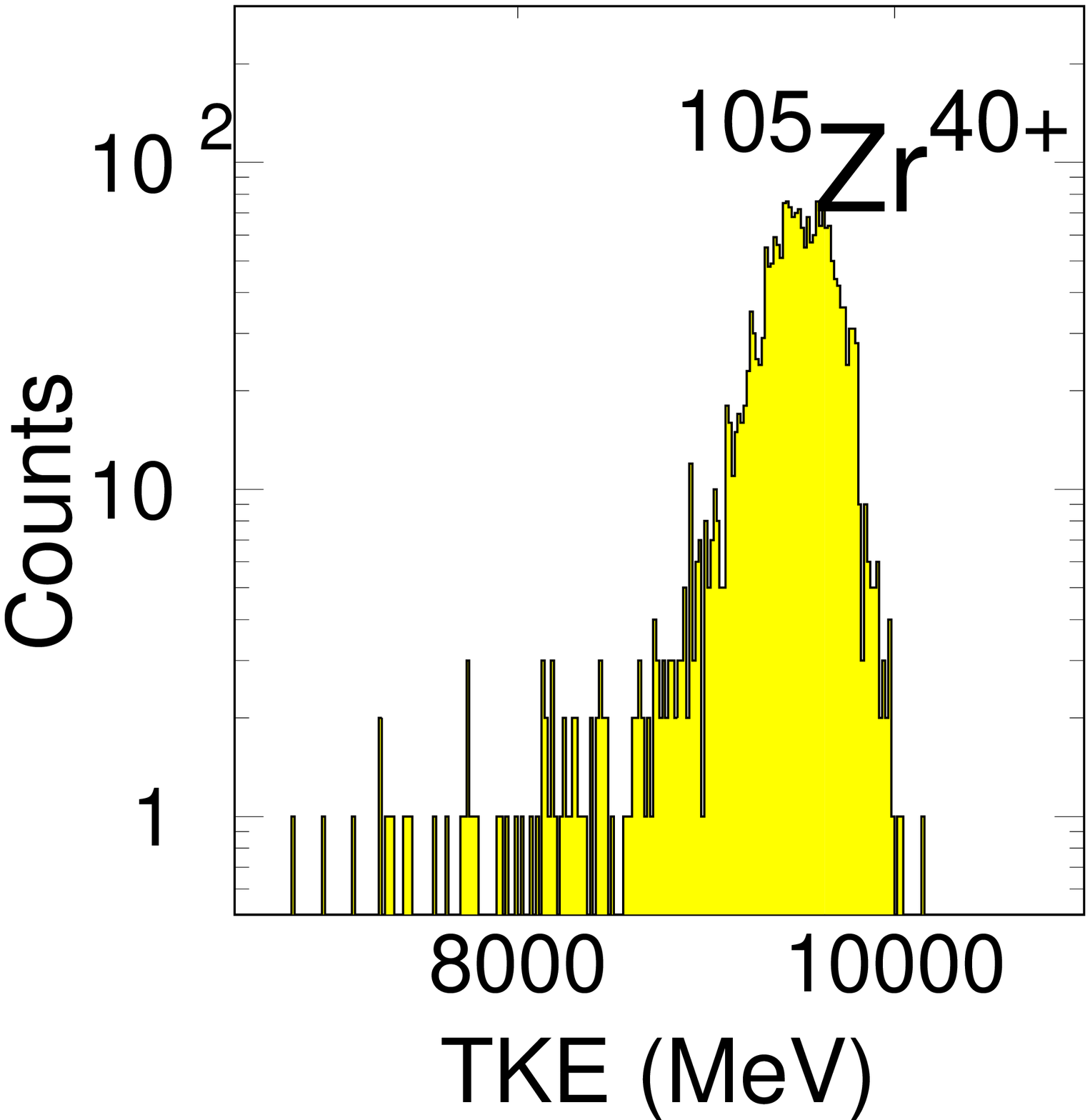}
\includegraphics[width=3cm]{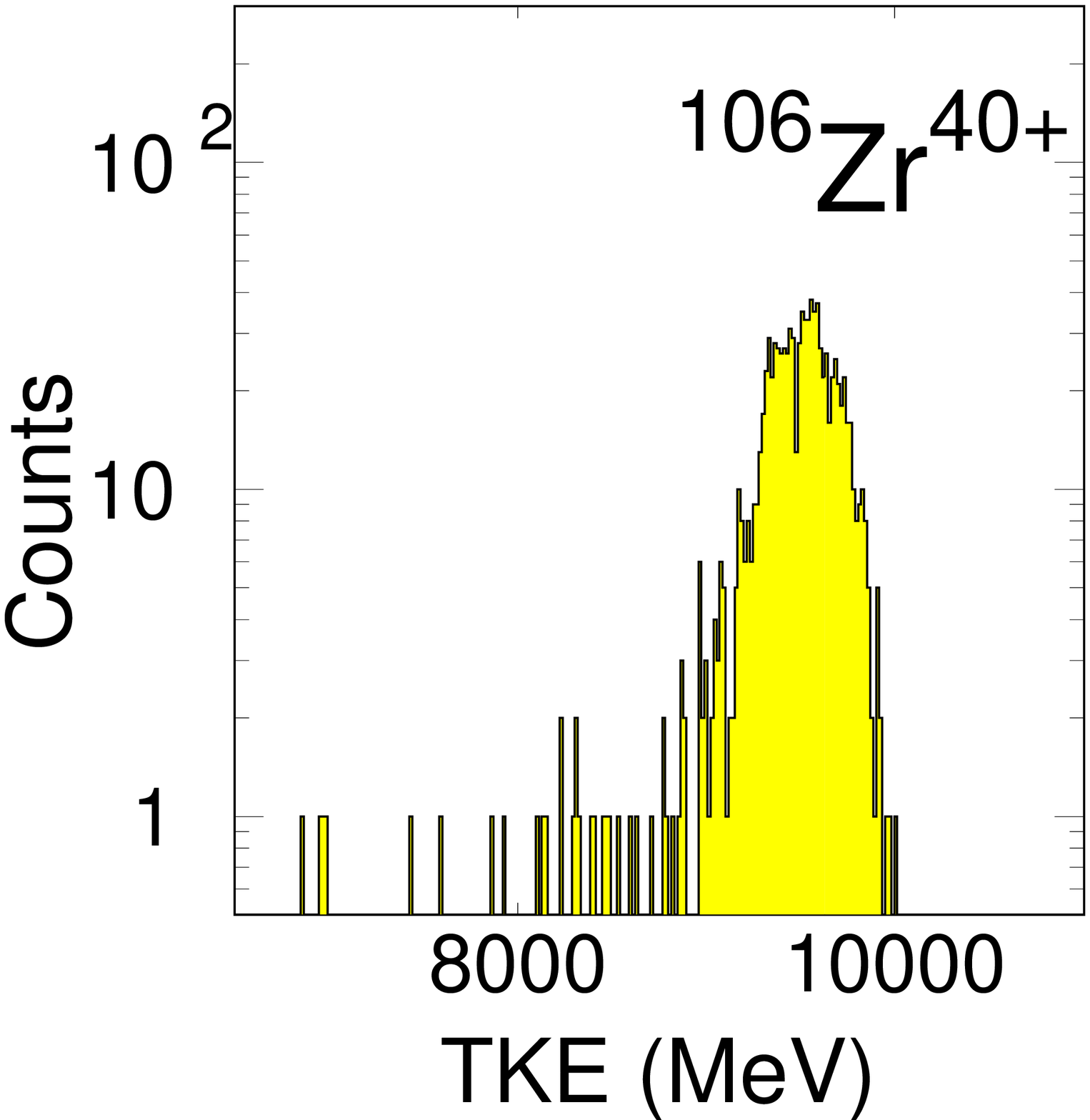}
\includegraphics[width=3cm]{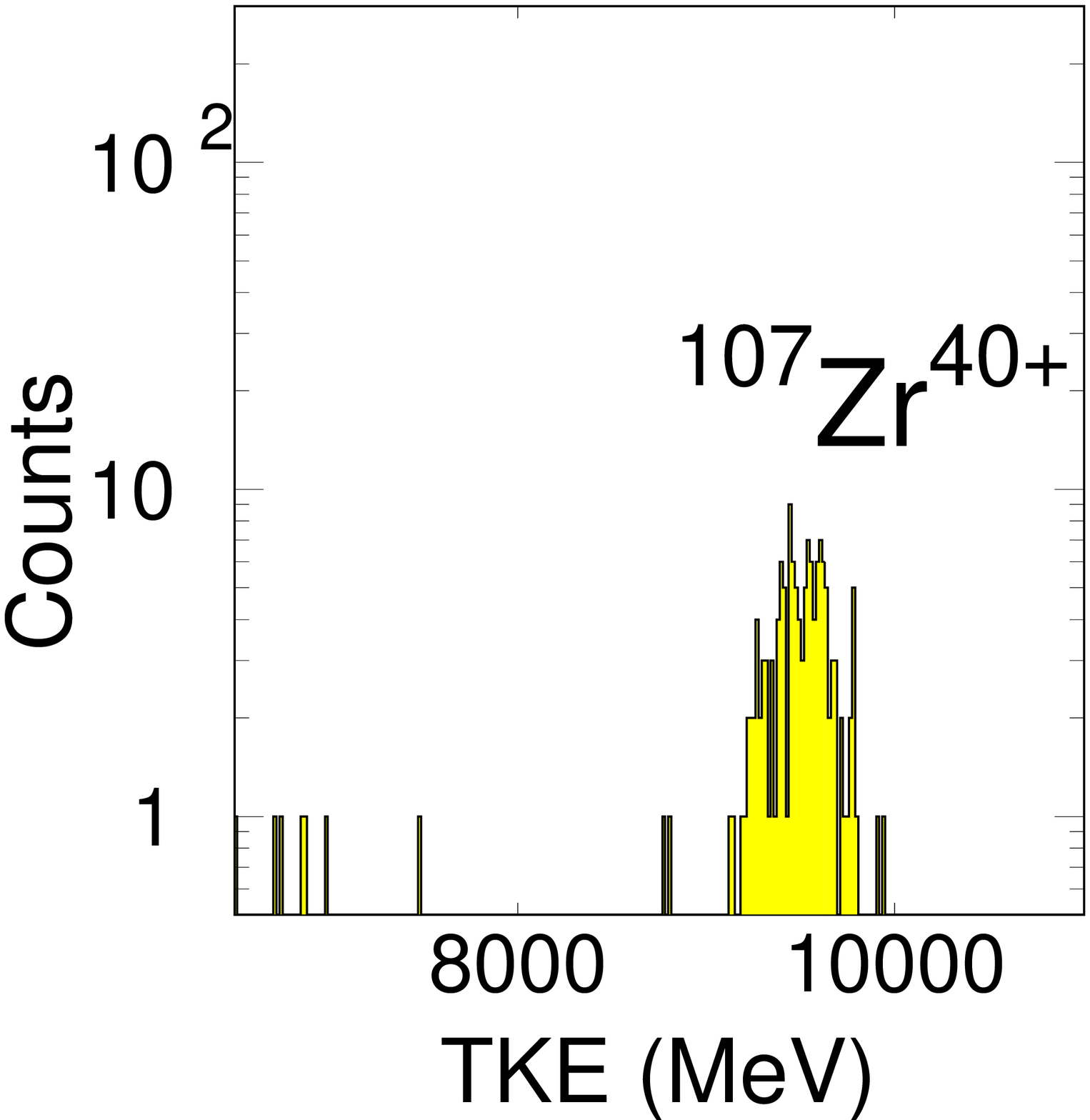} \\
\includegraphics[width=3cm]{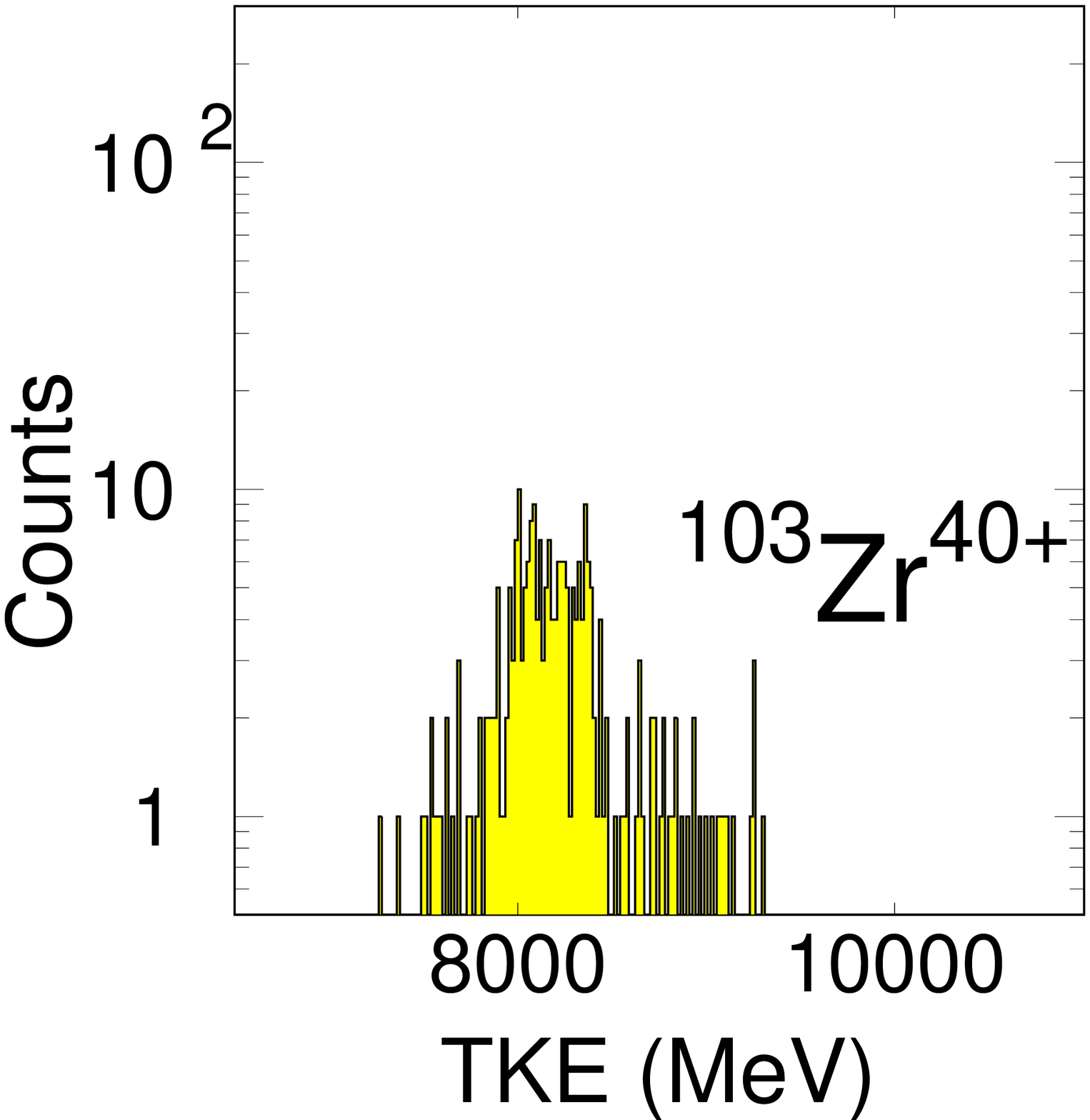}
\includegraphics[width=3cm]{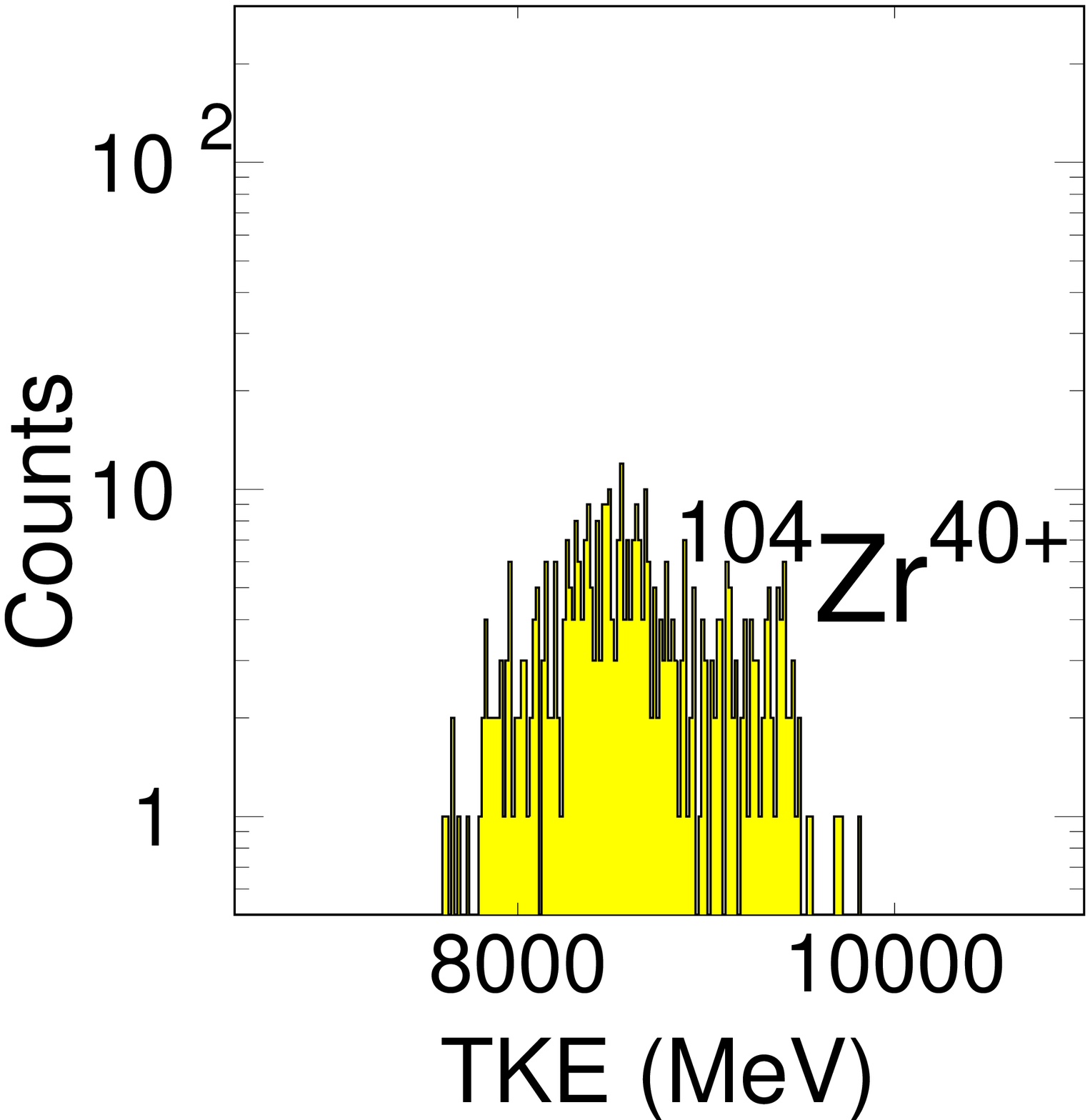}
\includegraphics[width=3cm]{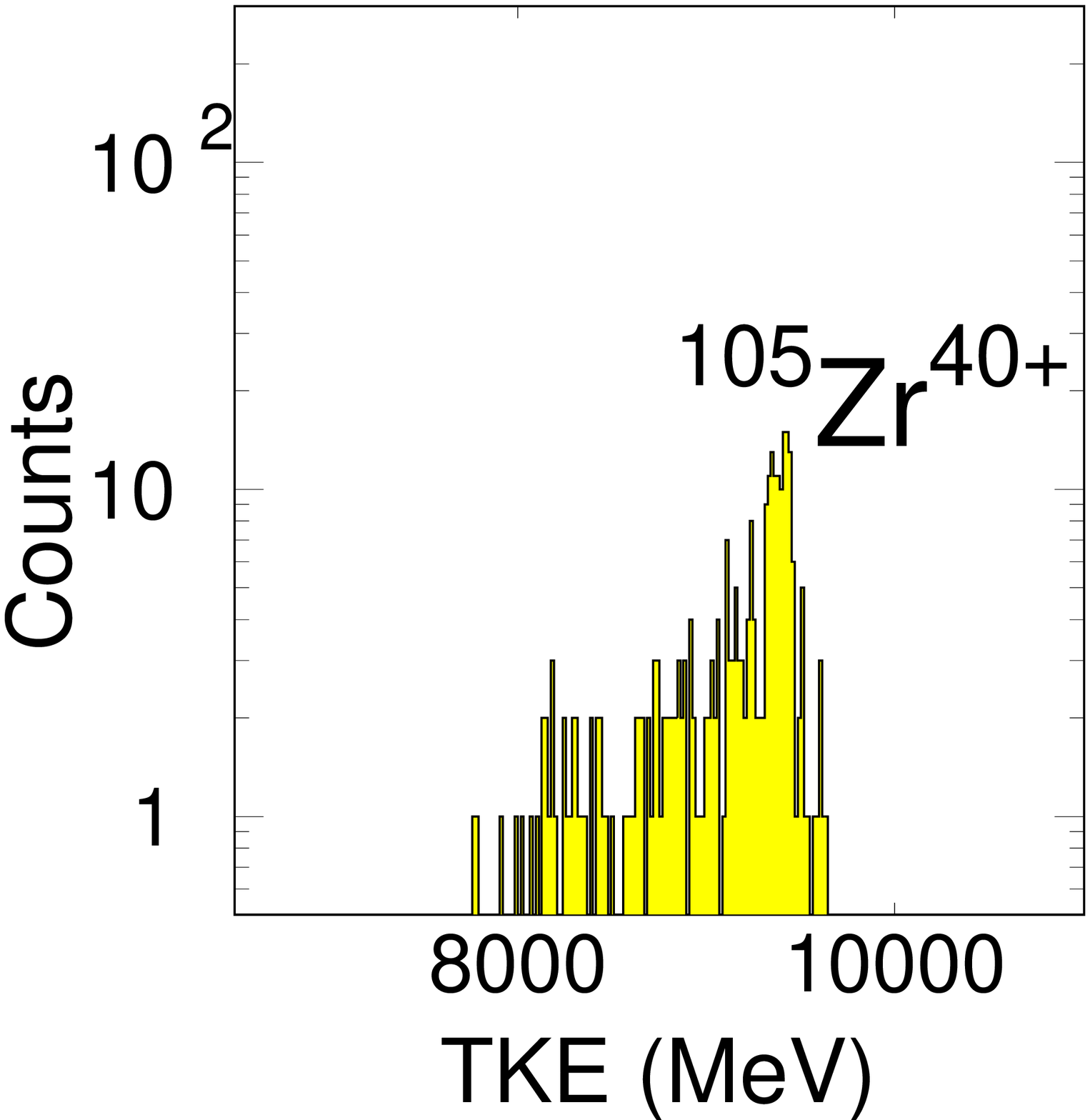}
\includegraphics[width=3cm]{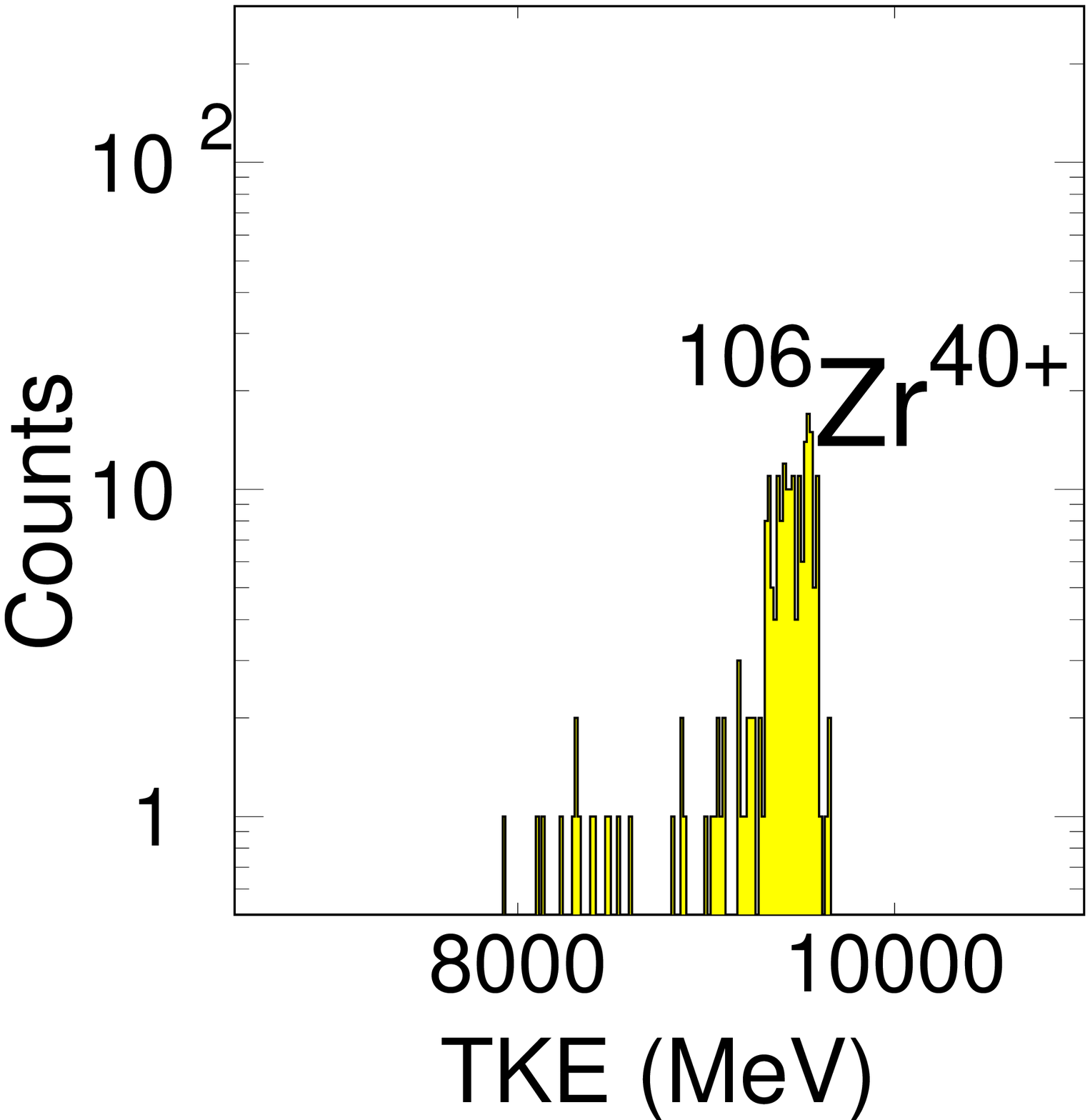}
\includegraphics[width=3cm]{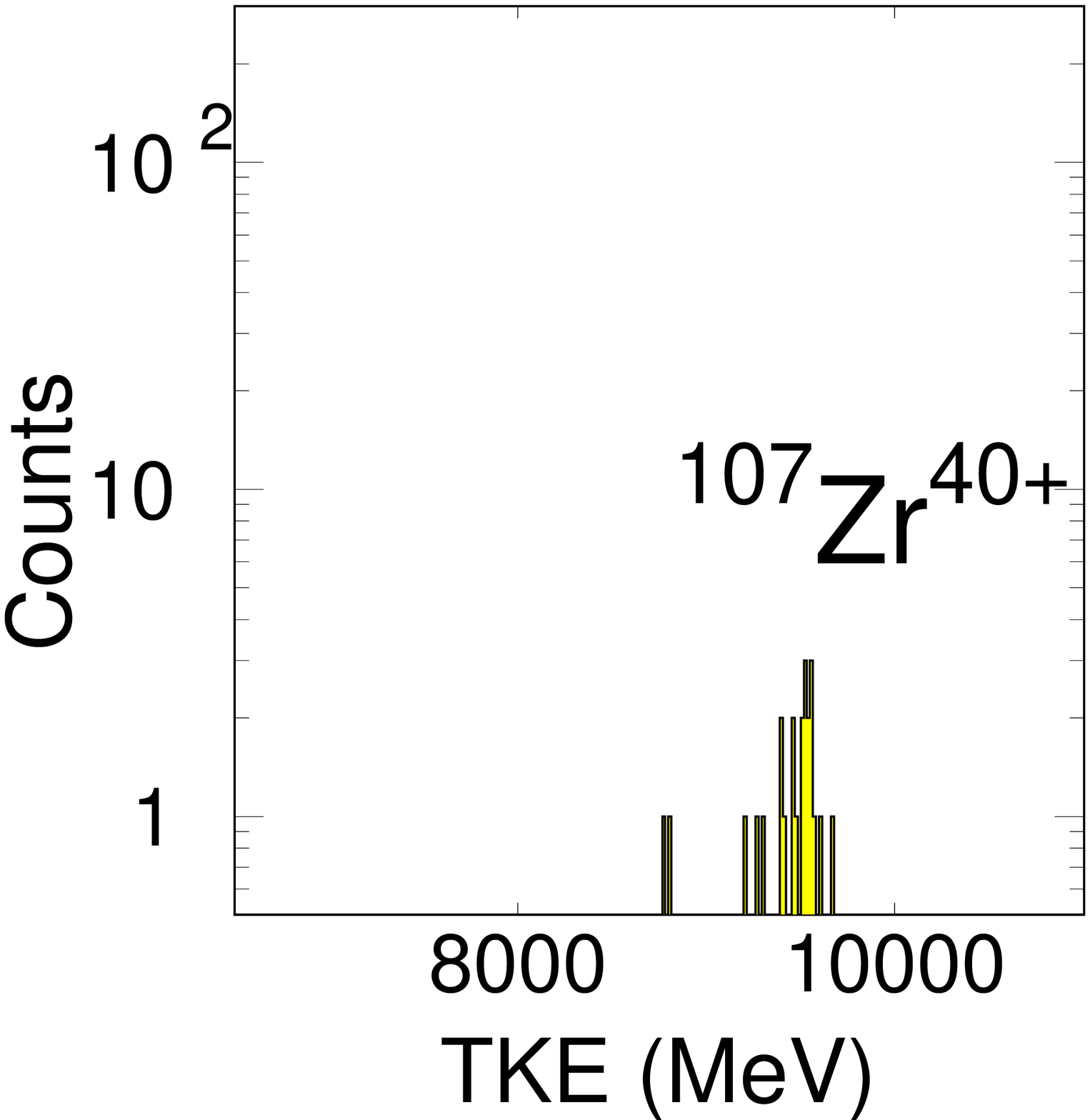} \\
\includegraphics[width=3cm]{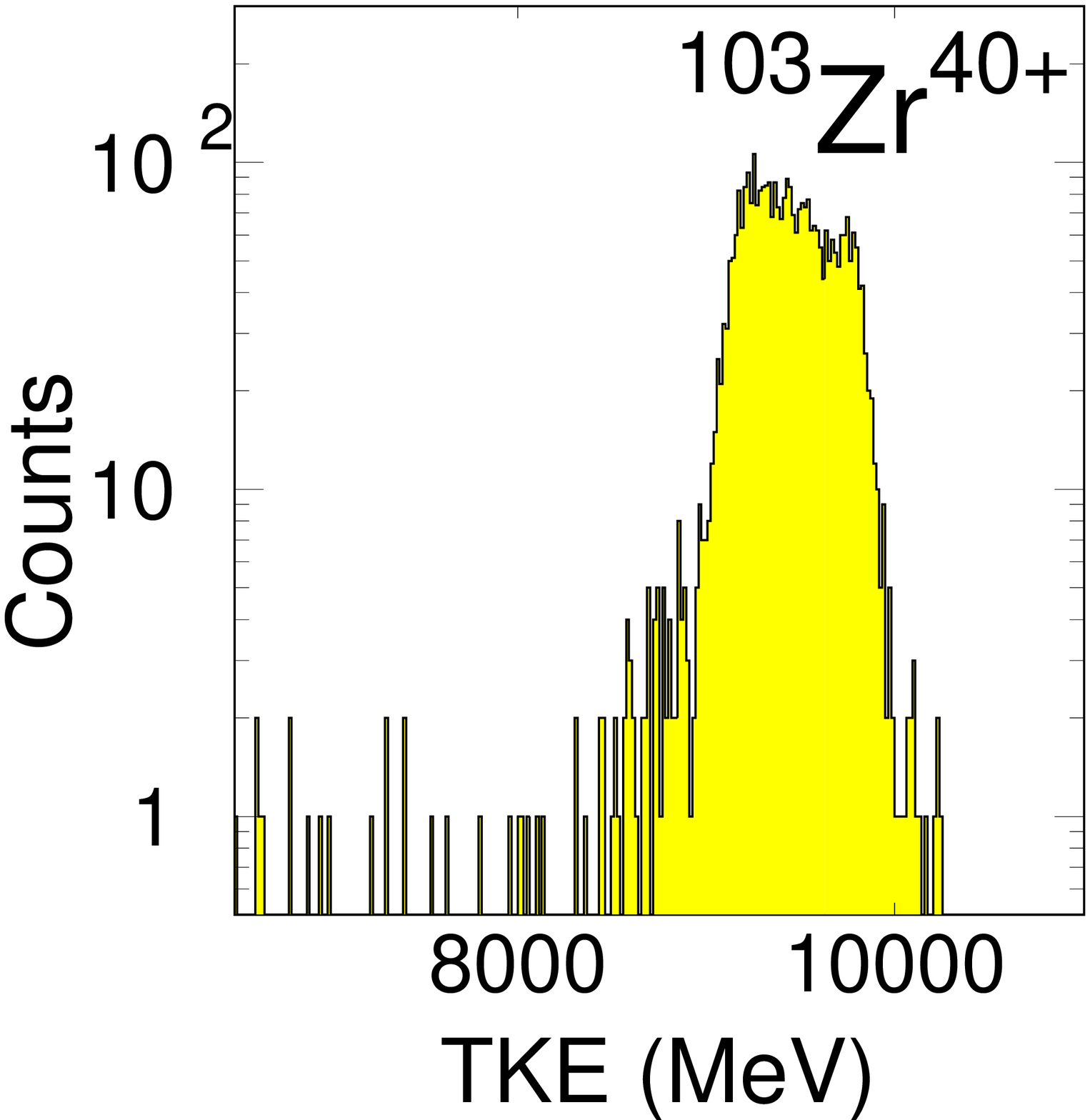}
\includegraphics[width=3cm]{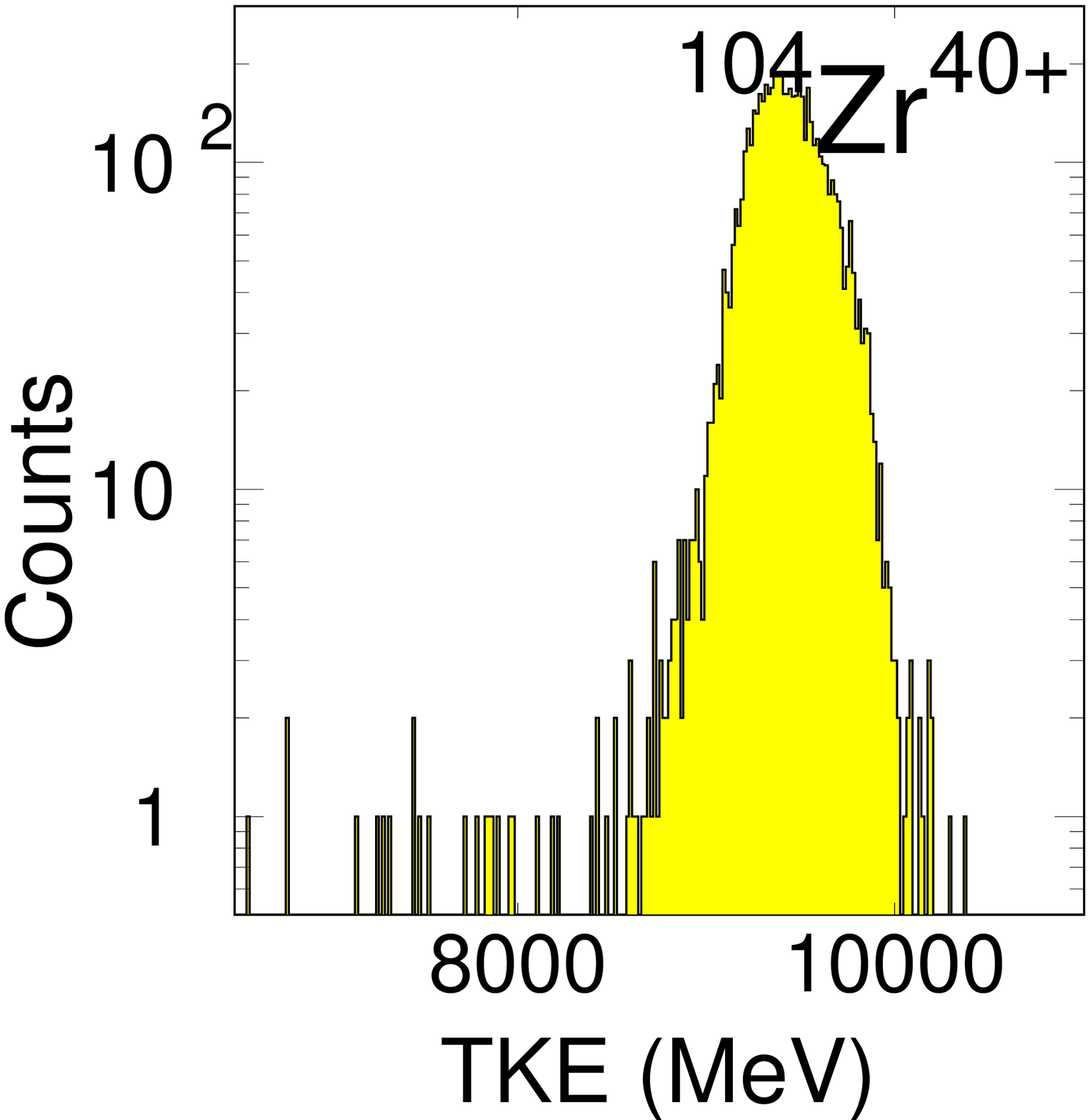}
\includegraphics[width=3cm]{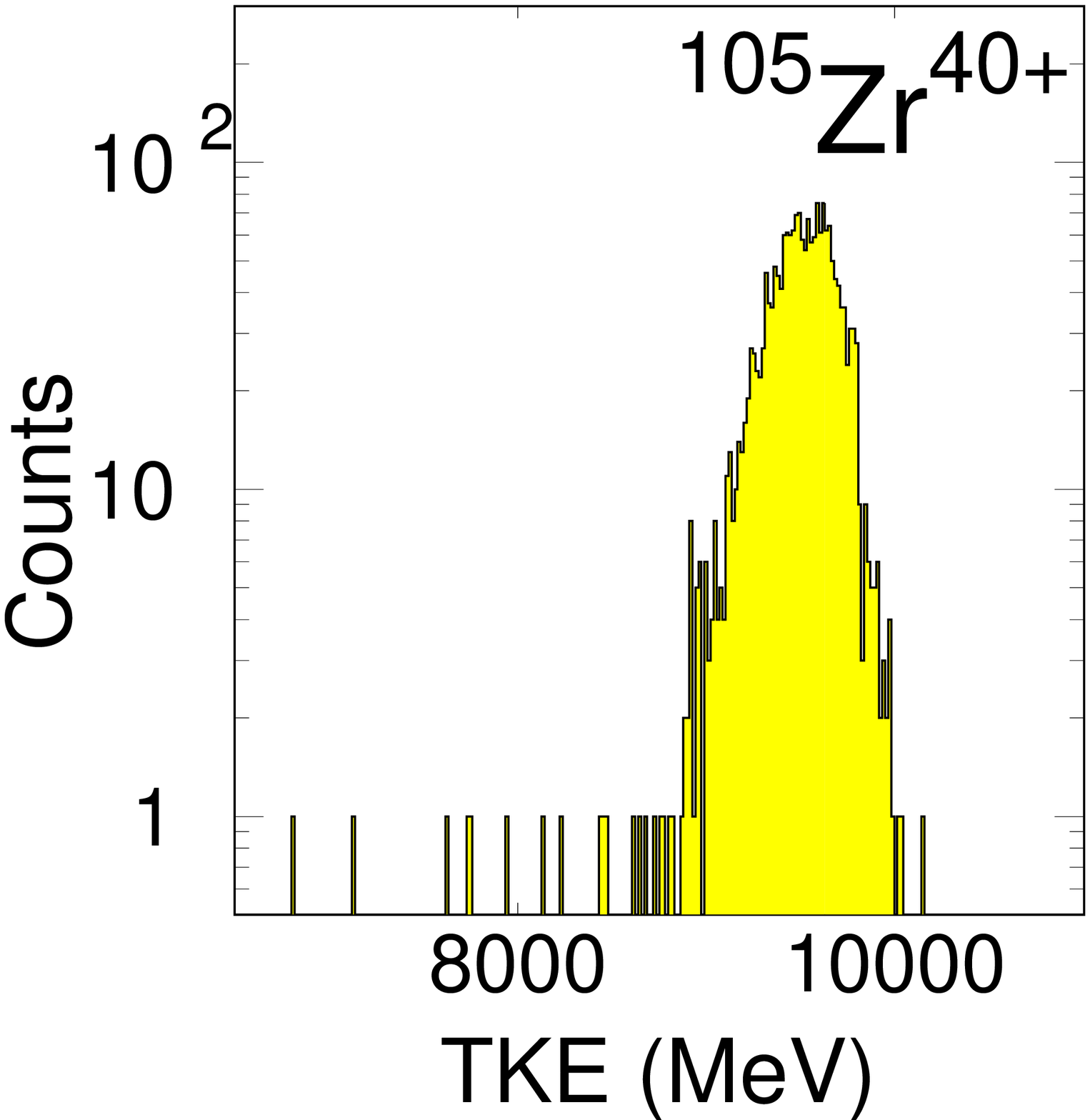}
\includegraphics[width=3cm]{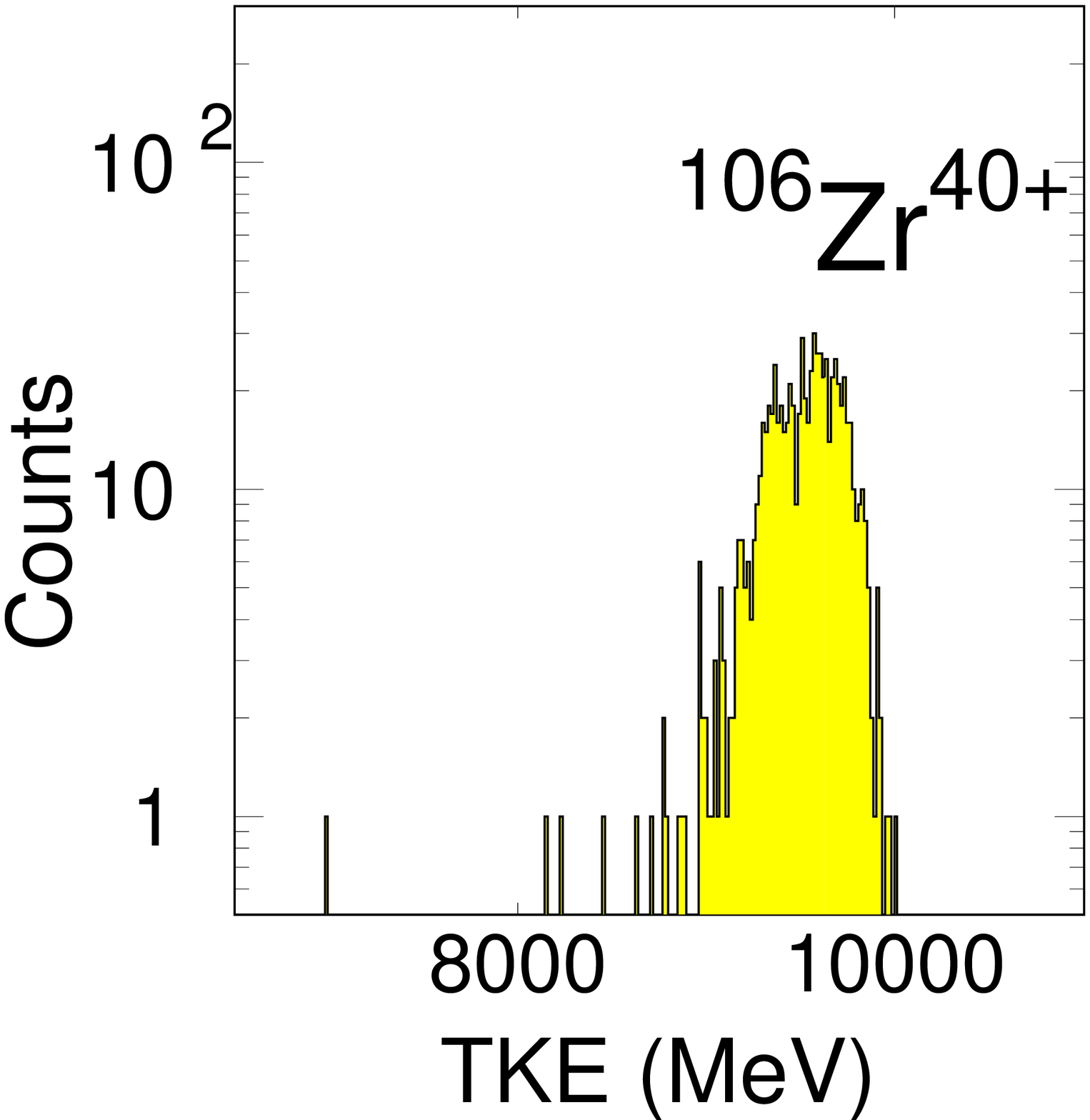}
\includegraphics[width=3cm]{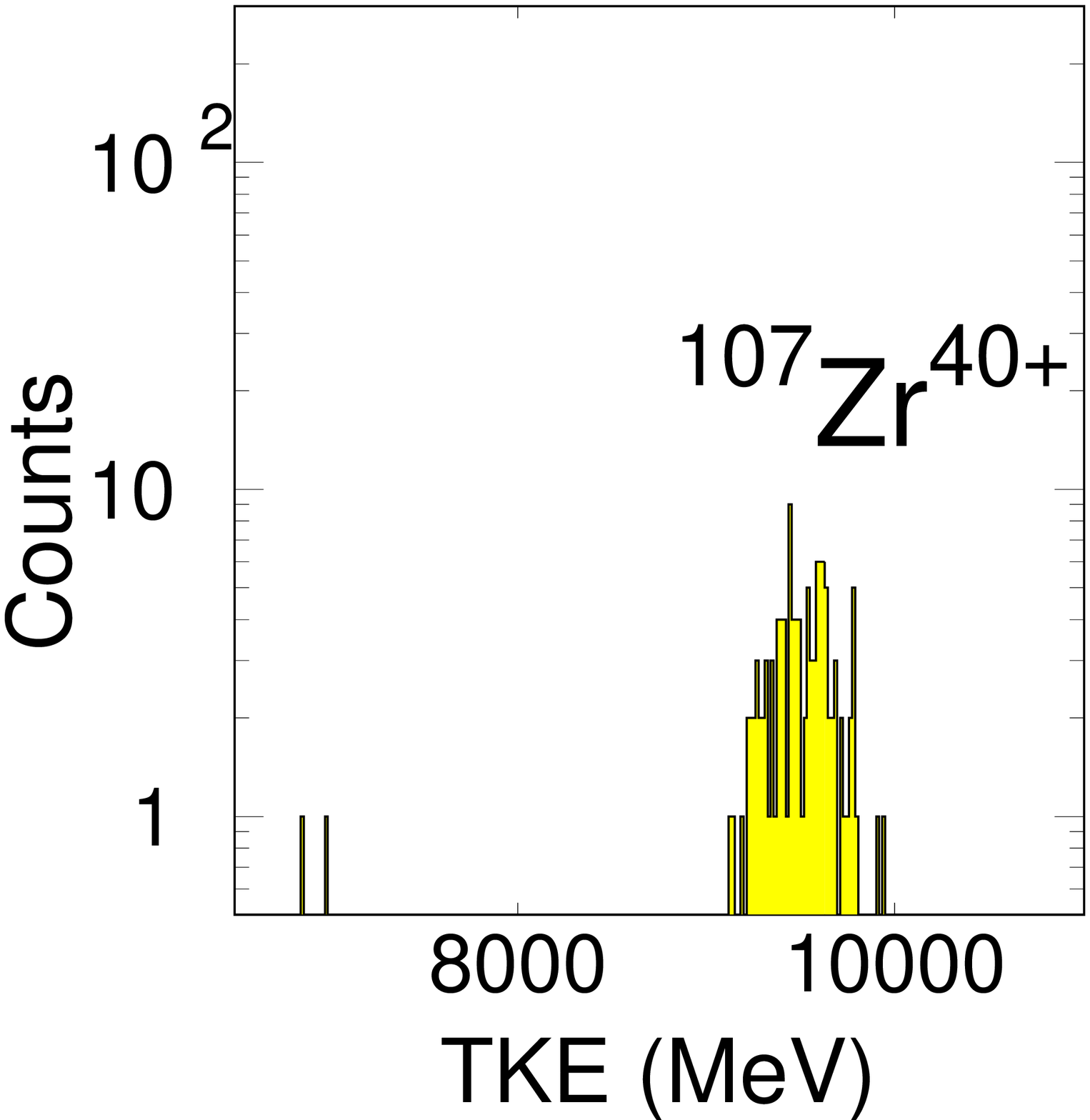}
\caption{Total kinetic energy ($TKE$) of Zr isotopes identified in the PID spectrum. The different spectra rows correspond to: all detected events in PIN1 (upper row); nuclei that were implanted in PIN4 (central row); and nuclei that were implanted in DSSD (lower row). The fully-stripped ions are labeled on top of the corresponding high-$TKE$ peaks.}
\label{fig:TKE}
\end{center}
\end{figure*}

The PID spectrum shown in Fig.~\ref{fig:PID} includes the fully-stripped ions of the nuclei of interest, overlapped with a small fraction of charge-states contaminants from lighter isotopes. Given the high $B\rho$ selected in the A1900 setting, only the hydrogen-like ions with mass numbers $A-2$ and $A-3$ reached the experimental area. These contaminants were disentangled from the fully-stripped nuclei by measuring the total kinetic energy ($TKE$) of the transmitted ions compiled from signals of the PIN detectors and DSSD. The $TKE$ spectra of different Zr isotopes is shown in Fig.~\ref{fig:TKE}: the upper row of panels corresponds to the nuclei that were detected with the first PIN detector of the BCS. The double-peak structure in the $TKE$ spectra arises from the fully-stripped species (high-$TKE$ peak) and the corresponding hydrogen-like contaminants (low-$TKE$ peak). The central and lower row of panels show the same spectra with the additional requirement of being implanted in either the most downstream PIN detector in the stack (i.e., PIN4) (central row) or in the DSSD (lower row). Nearly only the fully-stripped nuclei reached the latter, whereas the hydrogen-like components were mainly implanted in the last PIN. A small fraction of the fully-stripped components did not reach the DSSD due to a slightly overestimated Si thickness. Furthermore, no low-gain signals from the SSSD were observed, demonstrating that no nuclei reached this detector.


\section{Data analysis}\label{sec:Analysis}

\subsection{$\beta$-decay half-lives}\label{sec:halflives}
Specific conditions in the different detectors of the BCS were required to distinguish implantations, decays and light-particle events: a signal registered in each of the four PIN detectors, in coincidence with the low-gain output from at least one strip on each side of the DSSD, and in anti-coincidence with the SSSD, was identified as an implantation event. Decay events were defined as high-gain output signals from at least one strip on each side of the DSSD, in anti-coincidence with signals from PIN1. Software thresholds were set separately for each DSSD strip in order to cut off noise. Decay-like events accompanied by a preamplifier overflow signal from the Ge crystal downstream of the DSSD were identified as light particles and consequently rejected. According to LISE-based calculations, these light particles were mainly tritium nuclei and, to a lesser degree, $^{8}$Li nuclei, with energy-loss signals in the DSSD high-gain output comparable to the decays of interest.

\begin{figure}[h!]
\begin{center}
\includegraphics[width=7cm]{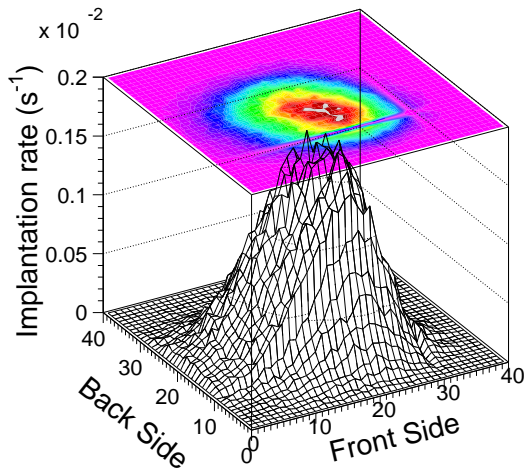}
\caption{(Color online) Implantation-event distribution measured over 112 hours of beam on target. The damaged 12$^{\rm th}$ back strip and 31$^{\rm st}$ front strip in the DSSD can be observed as respective gaps on the top-plane projection.}
\label{fig:impdssd}
\end{center}
\end{figure}

For each implantation event, the strip location on each side of the DSSD---defining the implanted pixel---was determined from the average of the strips weighted by their respective energy signal amplitude. The resulting average pixel was recorded along with the implantation time taken from a continuously counting 50~MHz clock. The last beam-line quadrupoles in front of the BCS were adjusted to illuminate a wide area of the DSSD cross-section; the resulting distribution of implantation events is shown in Fig.~\ref{fig:impdssd}. Subsequent decay events occurring in the same or neighboring pixels (defining a cluster of nine pixels) within a give correlation-time window ($t_{c}$) were associated with the previous implantation, and their times and pixels recorded. The value of $t_{c}$ was chosen to be around 10 times the expected $T_{1/2}$. Whenever a decay-like event was correlated with more than one implantation, all the events within the sequence (i.e., decay and implantations) were rejected. Such scenario would be possible if different implantations occur in the same cluster within the correlation-time window. Given the low maximum implantation rate per pixel of about 1.8$\times$10$^{-3}$~$s^{-1}$ and the $T_{1/2}$ values (typically below 1~s), the probability of multiple implantation events correlated with a decay was negligible.

\subsubsection{$\beta$-decay background}\label{sec:bkgrate}
The implantation-decay correlation criterion did not prevent the occurrence of spurious correlations arising from sources other than the actual decays of interest. Possible background sources included: light particles that did not lead to overflows in the Ge detector (either because they missed the detector or because they deposited only a fraction of their energy); real decays from longer-lived implanted nuclei and from nuclei implanted in neighboring pixels; and electronic noise signals above the thresholds. A detailed study of this decay-like background was necessary to extract $T_{1/2}$ values.

Background rates were determined separately for each DSSD cluster and 1-hour data-collection run, counting the number of decay-like events that were not correlated with any implantation. During this background measurement, each time an implantation was detected in a given pixel, the corresponding 9-pixel cluster was blocked to any subsequent decay event during a time interval chosen to be longer than $t_{c}$. Decay-like events detected in that cluster after the closure of the post-implantation blocking time (i.e., those that were not correlated with the previous implantation) were recorded as background events. Background rates were calculated for each 1-hour run as the ratio of the number of uncorrelated decay events in each cluster to the unblocked time in that cluster. The resulting rates were position-dependent and nearly constant over different runs. A critical factor in this analysis was the length of the post-implantation blocking time which had to be chosen so as to minimize the probability of recording real correlated decays. A blocking-time window of 40~s was found to fulfill this requirement. Finally, the DSSD cluster-averaged $\beta$-decay background was about 0.01~s$^{-1}$, nearly constant throughout the experiment.


The total number of $\beta$-decay background events $B_{\beta}$ for each isotope was calculated from the measured background rates in the runs and pixels where the isotopes were implanted, multiplied by $t_{c}$. This number was divided by the total unblocked time, calculated as the product of the number of implantations and $t_{c}$, giving the background rate for that nucleus. The statistical error for the background rate of each nucleus, derived from the number of background events recorded in each DSSD cluster and run, was about 5$\%$.




\subsubsection{$\beta$-decay half-lives from decay-curve fits}\label{sec:decaycurve}
\begin{figure}[h!]
\begin{center}
\includegraphics[width=6.5cm]{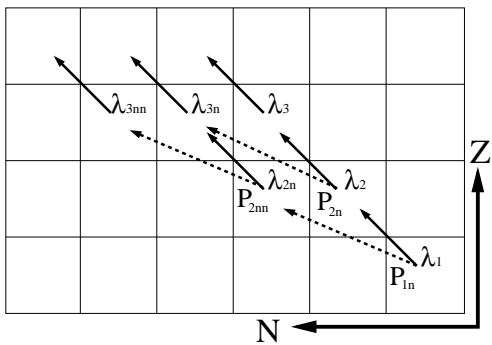}
\caption{Schematic representation of the possible decay-paths followed by a decaying mother nucleus (1), including first (2) and second (3) descendant-nuclei generations (see text for details).}
\label{fig:decscheme}
\end{center}
\end{figure}

The time differences between implantation and correlated decay events were accumulated for each nucleus in separated histograms and fitted by least-squares to a multi-parameter function derived from the Batemann equations~\cite{Cet06}. Due to the limited detection efficiency of the DSSD, some of the recorded $\beta$-decay events may come from descendant nuclei following a missed decay of the nuclei of interest. Given the low probability of missing consecutive decays, the correlation times used in the analysis, and the values of the half-lives of the descendant nuclei, up to three generations were included in the fit functions, along with contributions from background events. The paths that define the possible decay sequences following the decay of a mother nucleus are schematically illustrated in Fig.~\ref{fig:decscheme}.

\begin{figure}[t!]
\begin{center}
\includegraphics[width=6cm]{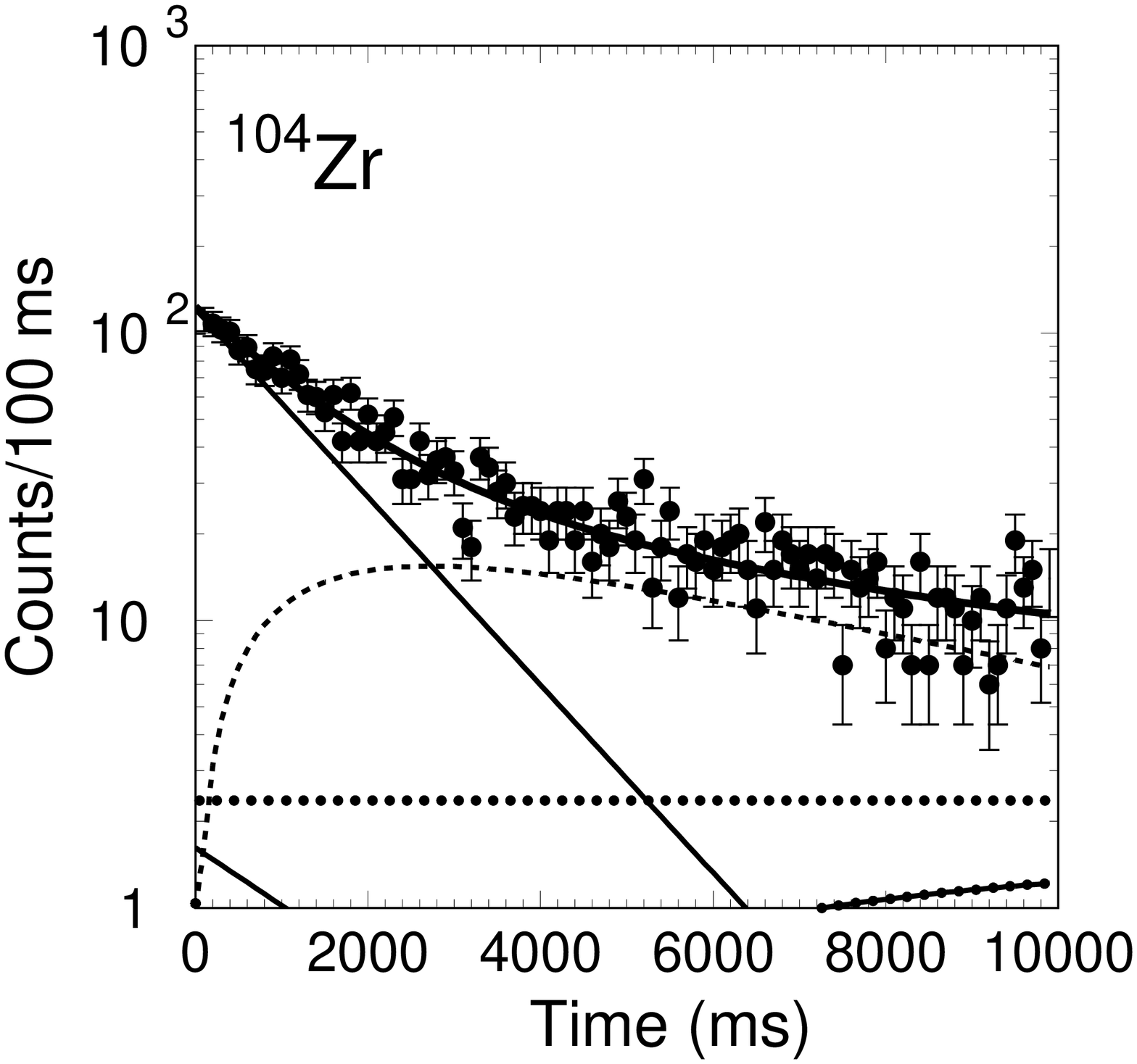} \\
\includegraphics[width=6cm]{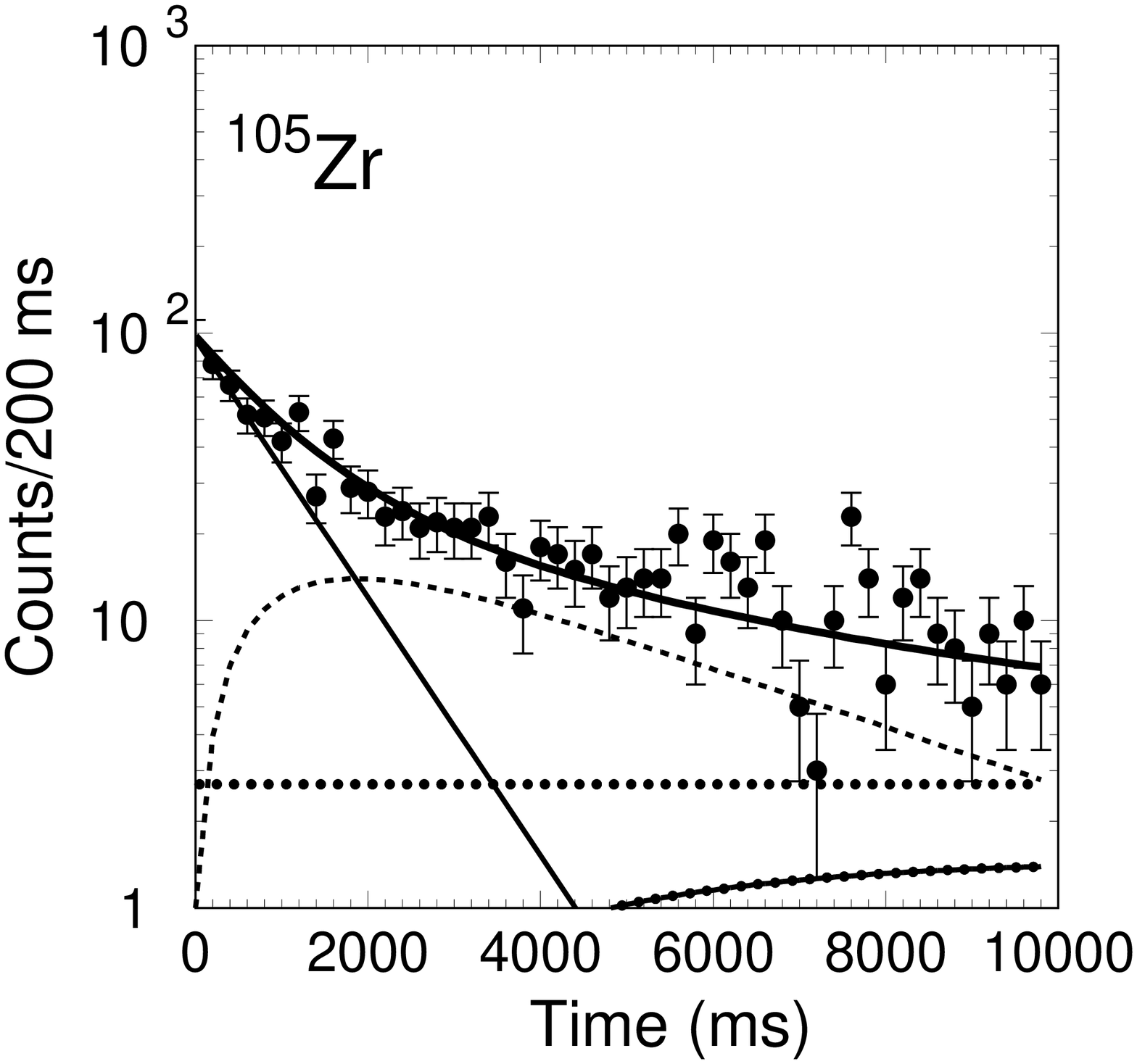}
\caption{Decay curve data (full circles) for $^{104,105}$Zr$_{64,65}$ isotopes. Included are fit functions for mother (solid thin line), daughter (dashed line), granddaughter (dot-dashed line), background (dotted line). The sum of these functions is represented by the solid thick line.}
\label{fig:decaycurves}
\end{center}
\end{figure}

The fit equation included a total of eleven parameters, eight of them fixed to constant values, namely: the decay constants of the daughter ($\lambda_{2}$) and granddaughter ($\lambda_{3}$); the neutron-emission probability of the mother ($P_{1\rm{n}}$), daughter ($P_{2\rm{n}}$), and neutron-emitted daughter ($P_{2\rm{nn}}$); and the decay constants of the single neutron-emitted daughter ($\lambda_{2\rm{n}}$) and granddaughter ($\lambda_{3\rm{n}}$), and the double neutron-emitted granddaughter ($\lambda_{3\rm{nn}}$). These fit constants were taken from the literature or, in the case of some unknown $P_{1\rm{n}}$, calculated using the FRDM$+$QRPA model~\cite{Mol90,Mol97,Mol03,Mol95}. The remaining three parameters were treated as free variables to be determined from the fit algorithm; two of them were the decay constant of the mother nucleus ($\lambda_{1}$) and the initial number of mother decaying nuclei ($N_{0}$). The third parameter, namely the background constant, was treated as a constrained \textquotedblleft free$\textquotedblright$ variable, defined within $\pm$10$\%$ of the calculated value (see Sec.~\ref{sec:bkgrate}). Finally, after determining the $P_{\rm n}$ as described in Sec.~\ref{sec:Pnvalues}, the decay curves were re-fitted replacing $P_{1\rm{n}}$ by the newly measured values. The decay curves of some selected Zr isotopes are presented in Fig.~\ref{fig:decaycurves} with the different contributions to the total fit curve.

The least-squares method used to fit the decay-curve histograms requires that the number of events per bin size $\Delta t$ is described by Gaussian statistics. More formally, the individual probability for one event to be recorded in a given bin must be $\ll$1, and the total number of events per bin $N(\Delta t)$ must be large, typically $\sim$20. Taking the time scale of the histograms as $t_{c}$, the latter condition can be expressed as $N(\Delta t)=(\Delta t/t_{c}) N\sim 20$, where $N$ is the total number of decay events in the histogram. Thus, since $\Delta t/t_{c}$ $\ll$1, $N$ must be $\gg$200 for the least-squares method to be valid. Table~\ref{tab:halflives-results} shows the half-lives of those nuclei that fulfilled the Gaussian-statistics requirement, along with their corresponding $N$. For cases with lower statistics, an alternative analysis based on the Maximum Likelihood method was used (see Sec.~\ref{sec:mlh}).
\begin{figure*}[t!]
\begin{center}
\includegraphics[width=5cm]{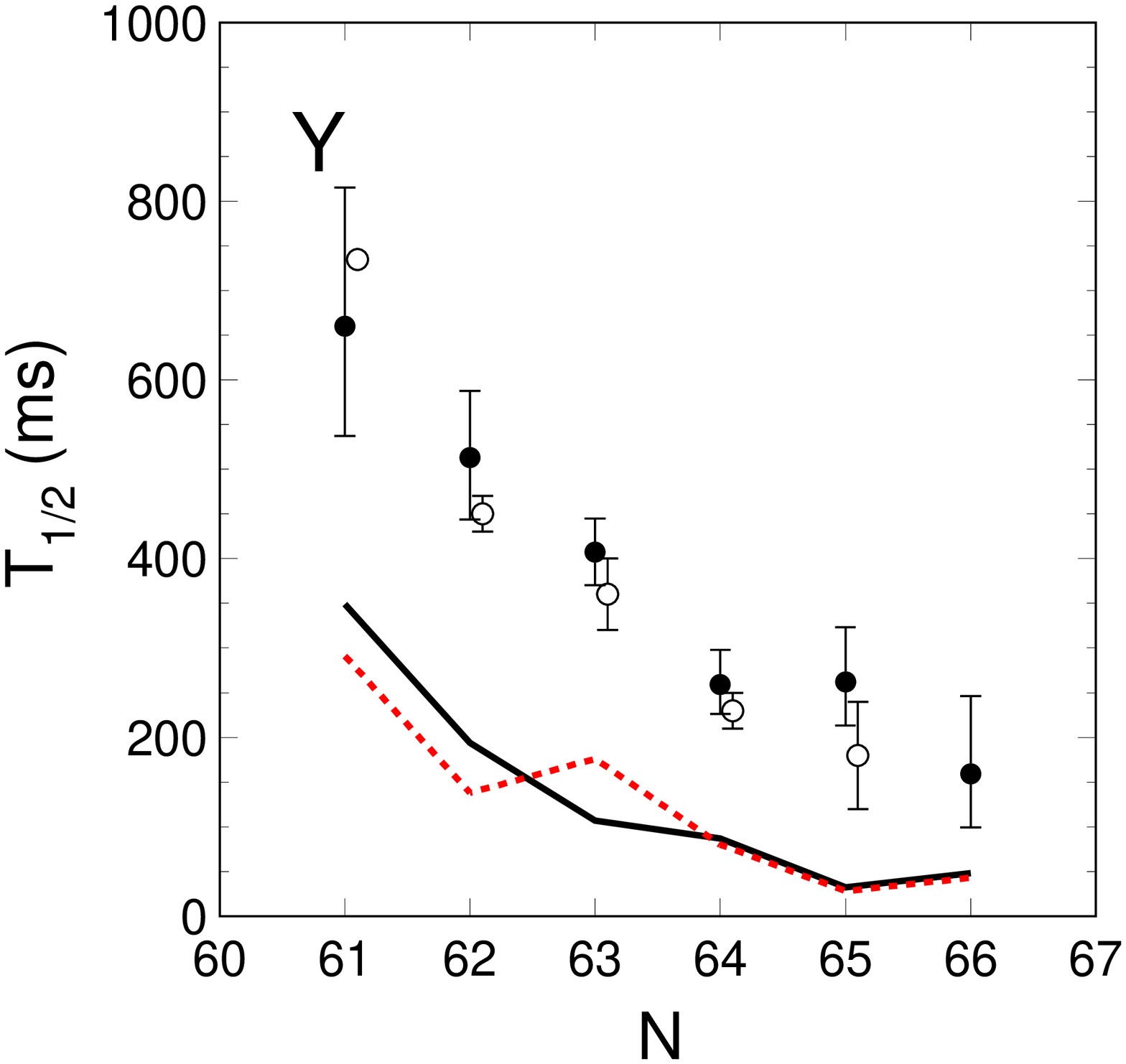}
\includegraphics[width=5cm]{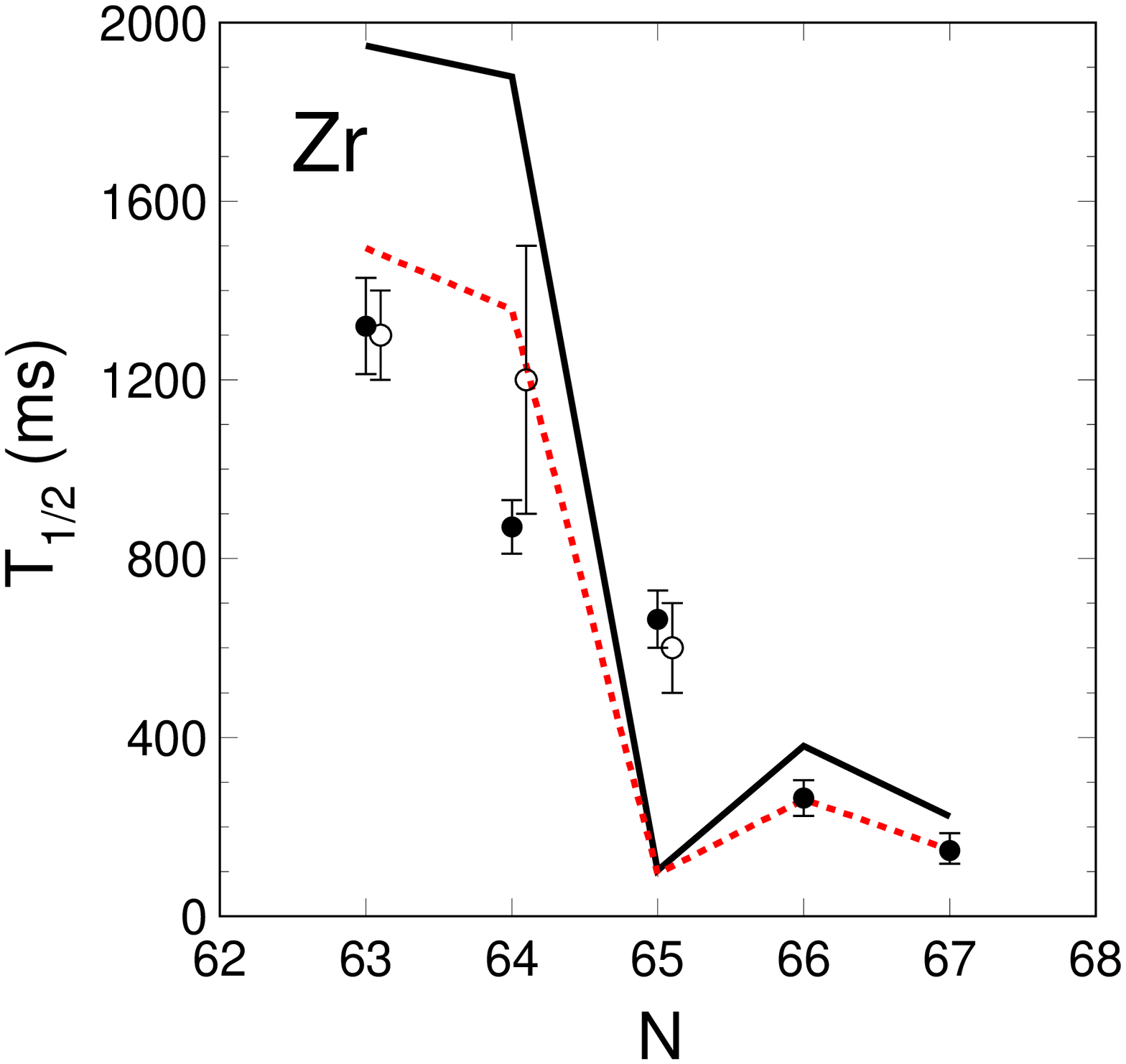}
\includegraphics[width=5cm]{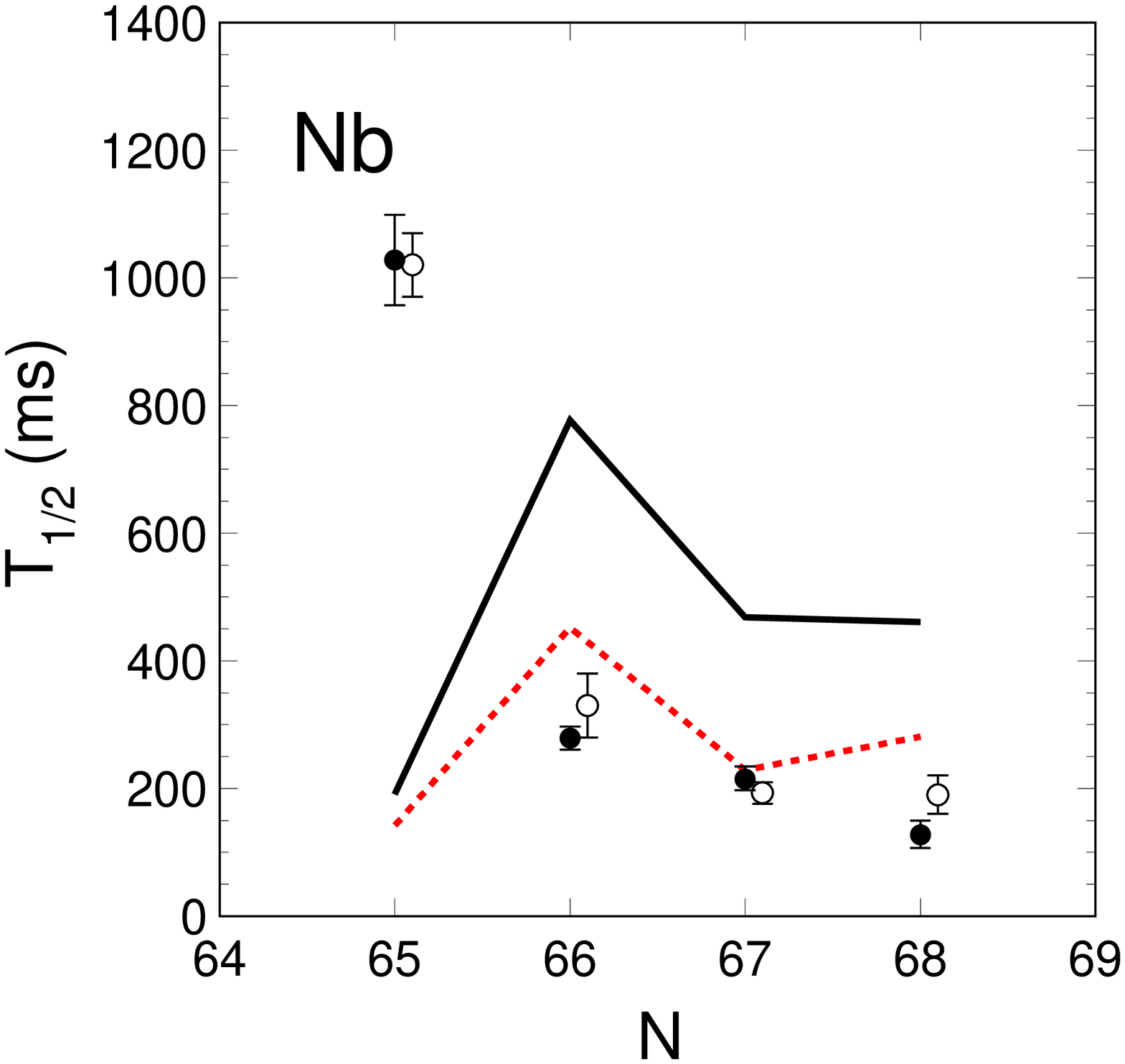} \\
\includegraphics[width=5cm]{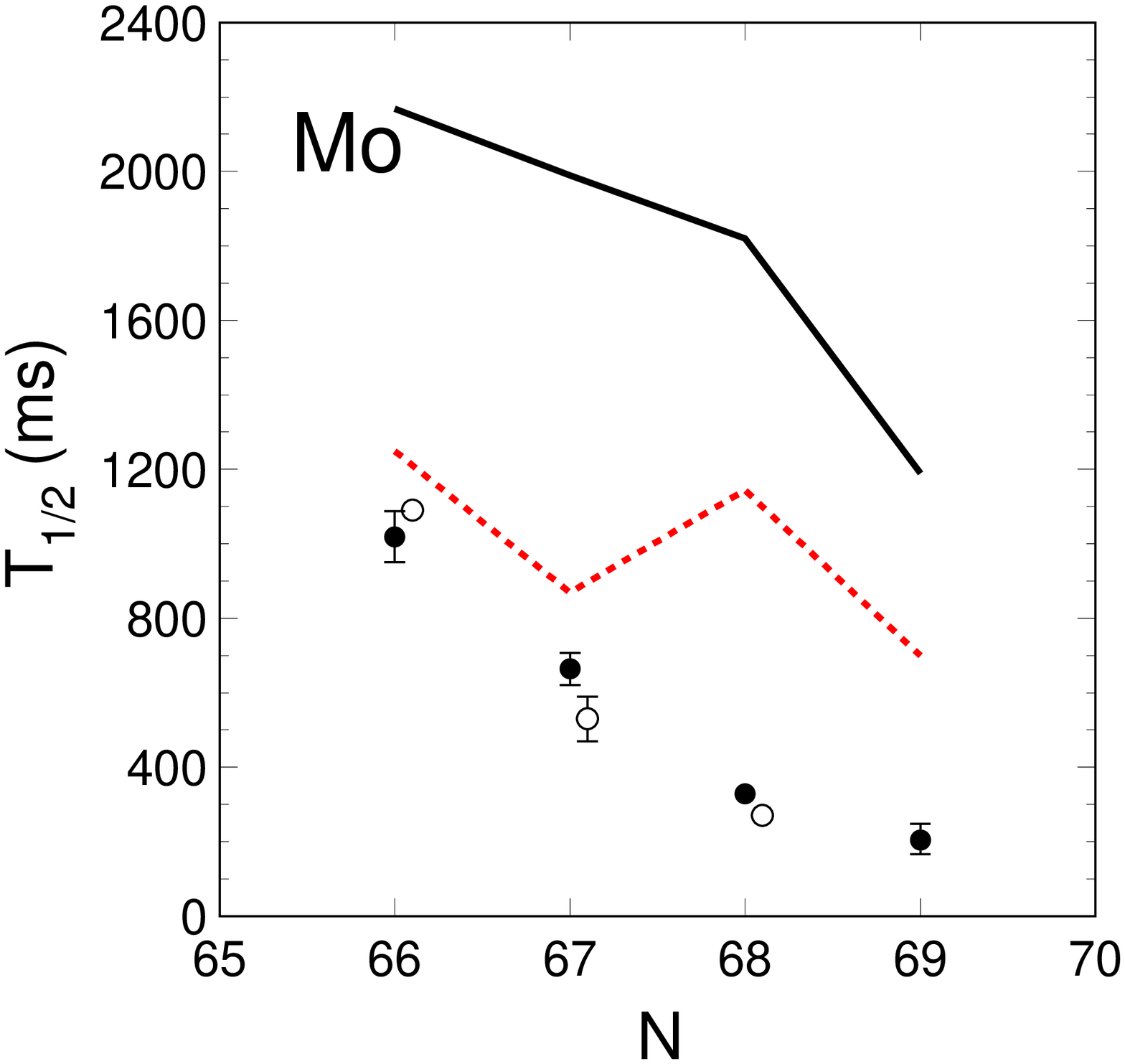}
\includegraphics[width=5cm]{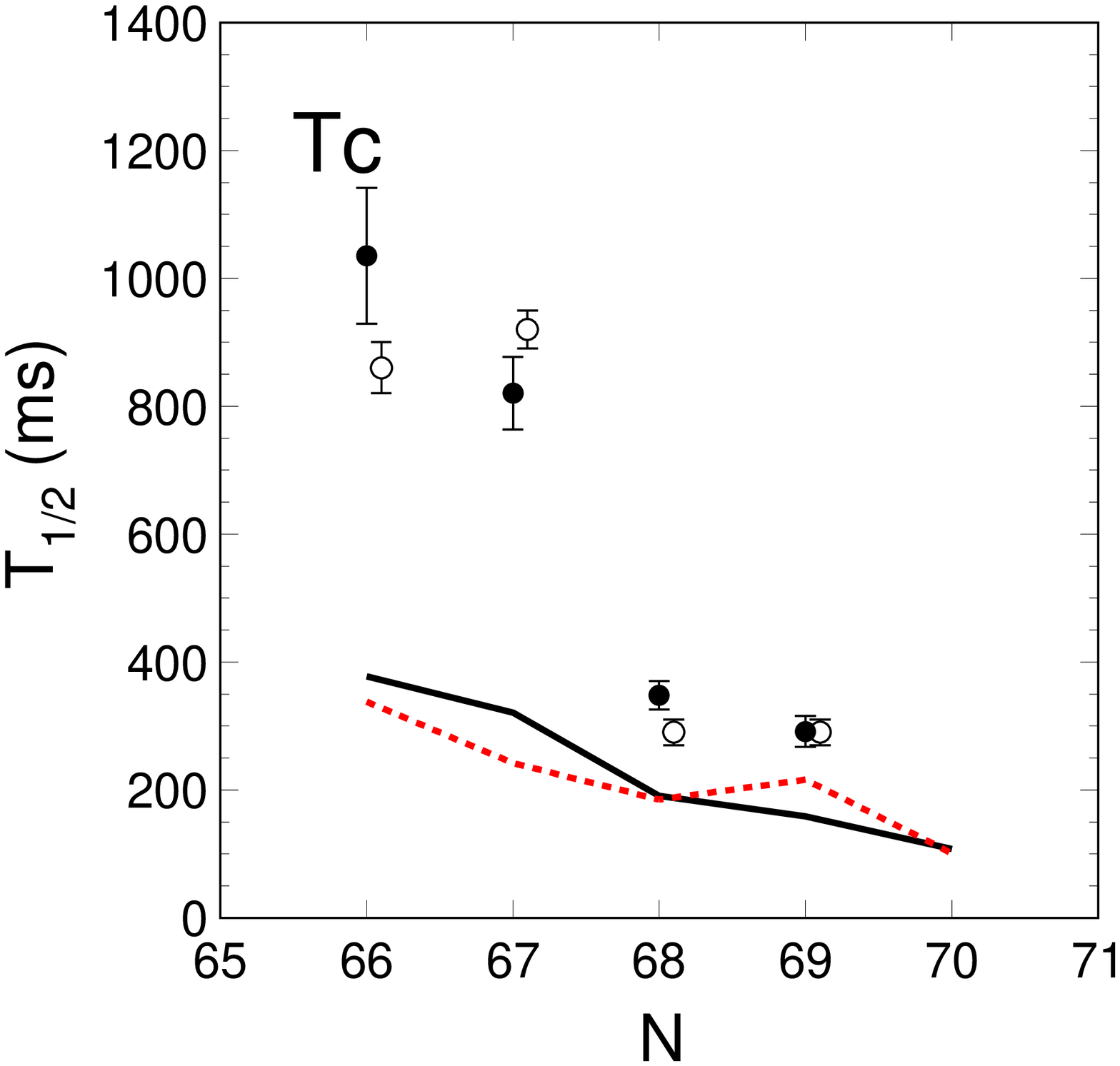}

\caption{$\beta$-decay half-lives obtained from the Maximum Likelihood method for Y, Zr, Nb, Mo and Tc isotopes (filled circles), compared with results from previous experiments~\cite{Kha77,Woh86,ENSDF,Shi83a,Hil91,Meh96,Wan99,Sch80,Shi83b,Ays91,Jok95,Ays92,Wan04,Ays90} (open circles). For the sake of clarity, the latter were shifted to the right by 0.1 units.
The data are compared with two versions of the QRPA model of M\"oller \emph{et al.}: the version described in Ref.~\cite{Mol03} (solid line), and the interim version QRPA06 described in Sec.~\ref{sec:QRPA} (dashed line). (See text for details).}
\label{fig:halflives-results}
\end{center}
\end{figure*}

Different sources of systematic error were included in the decay-curve analysis: uncertainties in the input parameters (half-lives and neutron-emission probabilities of the descendant nuclei) were accounted for by comparing the fit results obtained with these input values scanned over their respective error intervals. The resulting errors depended on the half-lives of the mother and, to a lesser degree, descendant nuclei. Uncertainties were typically below 5$\%$. In addition, comparisons of fit half-lives using background rates varied over their corresponding uncertainty showed differences below 1$\%$. Absolute systematic and statistical errors are shown in Table~\ref{tab:halflives-results}.

\begin{table*}[t]
\caption{Total number of implantations, number of events per histogram $N$ and experimental $\beta$-decay half-lives obtained from least-squares fit (Least squares) and the Maximum Likelihood method (MLH) with systematic and statistical errors. The results are compared with available data from previous experiments (Literature), and with the versions QRPA03 and QRPA06 of M\"oller's QRPA model. (See text for more details).}
\begin{ruledtabular}
\begin{tabular}{cccccccc}
 &  &  &  &  &  &  &\\
 Isotope  &    Implantations   &  $N$& \multicolumn{5}{c}{Half-life (ms)} \\
 &  &  &  &  &  &  &\\
          \cline{4-8}
 &  &  &  &  &  &  &\\
 &  &  & Least squares & MLH & Literature & QRPA03~\cite{Mol03} & QRPA06 \\ \hline
 &  &  &  &  &  &  &\\\vspace{1mm}
 $^{100}$Y  & 188   & 107   &              & $660(25)^{+150}_{-120}$& 940(32)~\cite{Kha77}, 735(7)~\cite{Woh86} & 349  & 291     \\\vspace{1mm}
 $^{101}$Y  & 746   & 453   &              & $510(30)^{+70}_{-60}$  & 450(20)~\cite{ENSDF}   & 194  & 138   \\\vspace{1mm}
 $^{102}$Y  & 1202  & 976   &              & $410(20)(30)$          & 300(10)~\cite{Hil91}, 360(40)~\cite{Shi83a} & 107  & 176 \\\vspace{1mm}
 $^{103}$Y  & 596   & 538   &              & $260(10)^{+40}_{-30}$  & 230(20)~\cite{Meh96}   & 87   & 80      \\\vspace{1mm}
 $^{104}$Y  & 128   & 116   &              & $260(10)^{+60}_{-50}$  & 180(60)~\cite{Wan99}   & 32   & 28     \\\vspace{1mm}
 $^{105}$Y  & 27    & 21    &              & $160(15)^{+85}_{-60}$  &                        & 48   & 43     \\\vspace{1mm}
 $^{103}$Zr & 2762  & 1842  & 1380(60)(40) & $1320(90)(60)$         & 1300(100)~\cite{Sch80} & 1948 & 1495 \\\vspace{1mm}
 $^{104}$Zr & 4743  & 3158  & 920(20)(20)  & $870(50)(30)$          & 1200(300)~\cite{Sch80} & 1879 & 1358   \\\vspace{1mm}
 $^{105}$Zr & 1707  & 1118  & 670(20)(20)  & $660(45)^{+50}_{-45}$  & 600(100)~\cite{Meh96}  & 102  & 95    \\\vspace{1mm}
 $^{106}$Zr & 643   & 570   &              & $260(20)^{+35}_{-30}$  &                        & 381  & 261    \\\vspace{1mm}
 $^{107}$Zr & 90    & 91    &              & $150(5)^{+40}_{-30}$   &                        & 223  & 149    \\\vspace{1mm}
 $^{106}$Nb & 10445 & 8182  & 1240(15)(15) & $1030(65)(30)$         & 1020(50)~\cite{Shi83b} & 191  & 142    \\\vspace{1mm} $^{107}$Nb & 6672  & 5384  & 290(10)(5)   & $280(15)(10)$          & 330(50)~\cite{Ays91}   & 777  & 452   \\\vspace{1mm}
 $^{108}$Nb & 1479  & 1731  & 210(2)(5)    & $220(10)(15)$          & 193(17)~\cite{ENSDF}   & 468  & 229    \\\vspace{1mm}
 $^{109}$Nb & 268   & 340   &              & $130(5)(20)$           & 190(30)~\cite{Meh96}   & 461  & 281    \\\vspace{1mm}
 $^{108}$Mo & 17925 & 11732 & 1110(5)(10)  & $1020(65)(20)$         & 1090(20)~\cite{Jok95}  & 2168 & 1249   \\\vspace{1mm}
 $^{109}$Mo & 9212  & 7013  & 700(10)(10)  & $660(40)(20)$          & 530(60)~\cite{Ays92}   & 1989 & 869   \\\vspace{1mm}
 $^{110}$Mo & 2221  & 2453  & 340(5)(10)   & $330(20)(20)$          & 270(10)~\cite{Wan04}   & 1820 & 1144   \\\vspace{1mm}
 $^{111}$Mo & 167   & 210   &              & $200(10)^{+40}_{-35}$  &                        & 1189 & 699   \\\vspace{1mm}
 $^{109}$Tc & 2922  & 1623  & 1140(10)(30) & $1040(95)(50)$         & 860(40)~\cite{ENSDF}   & 378  & 338    \\\vspace{1mm}
 $^{110}$Tc & 9549  & 6256  & 910(10)(10)  & $820(50)(25)$          & 920(30)~\cite{Ays90}   & 321  & 242    \\\vspace{1mm}
 $^{111}$Tc & 5433  & 4626  & 350(10)(5)   & $350(15)(15)$          & 290(20)~\cite{Meh96}   & 191  & 185 \\\vspace{1mm}
 $^{112}$Tc & 1198  & 1206  & 290(5)(10)   & $290(10)(20)$          & 280(30)~\cite{Ays90}   & 159  & 216\\\vspace{1mm}
 $^{113}$Tc & 84    & 80    &              & $160(5)^{+50}_{-40}$   & 170(20)~\cite{Wan99}   & 108  & 101  \\\vspace{1mm}
\end{tabular}
\end{ruledtabular}
\label{tab:halflives-results}
\end{table*}

\subsubsection{$\beta$-decay half-lives from Maximum Likelihood Method}\label{sec:mlh}
The Maximum Likelihood analysis (MLH) is well suited for determining decay half-lives in cases with low statistics~\cite{Mon06,Hos05,Sch84,Ber90,Sch95,Sum97,Ian04}. The method, described in appendix~\ref{sec:AppendixA}, defines decay sequences for up to three generations following an implantation. The probability of observing a given decay sequence was calculated by summing up the probabilities for all possible scenarios leading to the detection of the decay-event members. The scenarios were evaluated by considering the occurrence of up to three $\beta$ decay events, including $\beta$-delayed neutron branching, the contributions from background events, and the \textquotedblleft missing$\textquotedblright$ decays due to the limited detection efficiency. A joint probability density, the likelihood function $\mathcal{L}$, was calculated by multiplying the probabilities for all the measured decay sequences. The resulting $\mathcal{L}$ was a function of the measured decay times $t_{i}$ of the different members of the decay-sequence, their decay constants and neutron-emission probabilities, the correlation time $t_{c}$, the background rate of the corresponding DSSD cluster and run where the decay sequence was detected, and the $\beta$-decay detection efficiency $\epsilon_{\beta}$.
The half-lives of the nuclei of interest were determined from the maximization of $\mathcal{L}$, using the decay constant of the mother nucleus $\lambda_{1}$ as free parameter.

\label{fig:LHF}

All of the descendant decay parameters necessary to define $\mathcal{L}$ were taken from previous measurements or---in the case few $P_{\rm n}$ values---calculated from theory. Similar to the least-squares fit method described in Sec.~\ref{sec:decaycurve}, the half-lives were re-calculated with the newly measured $P_{\rm n}$ values, once known. The $\beta$-decay detection efficiency $\epsilon_{\beta}$ was determined as the ratio of the number of detected $\beta$-decays $N_{\beta}$ attributed to a given nucleus to the number of implantations of that nucleus. The former was given by $N_{\beta}=N_{0}/(\lambda_{1}  \Delta t)$, where $N_{0}$ and $\lambda_{1}$ were obtained from the decay-curves fits, for the cases where the least-squares method was valid. No systematic trend for $\epsilon_{\beta}$ was observed within a given isotopic chain, so a weighted average
efficiency per DSSD cluster of (31$\pm$4)$\%$ was used. Finally, the background rate was determined for each DSSD cluster and run, as described in Sec.~\ref{sec:bkgrate}.

The sources of systematic error included contributions from uncertainties in the experimental descendant-nuclei $T_{1/2}$ and $P_{\rm n}$, background and $\epsilon_{\beta}$. The systematic error of $T_{1/2}$ was calculated for each nucleus as described in Sec.~\ref{sec:decaycurve}, yielding typical values below 10$\%$. The statistical error was directly calculated from the MLH analysis using the prescription described by W.~Br\"uchle~\cite{Bru03}. Since the $\mathcal{L}$ distributions were typically asymmetric, the shortest possible interval containing the maximum of the $\mathcal{L}$-distribution and 68$\%$ (i.e., 1-$\sigma$) of the total integrated density probability was used~\cite{Bru03,Mon06}. The calculated systematic and statistical uncertainties are listed in Table~\ref{tab:halflives-results}. The total error shown in Fig.~\ref{fig:halflives-results} was obtained summing up the contributions from systematic and statistical uncertainties according to the method described in Ref.~\cite{Bar04}. The $T_{1/2}$ obtained from the MLH are in agreement with the decay-curve fits for the cases where the least-squares fit method was valid.

\subsection{$\beta$-delayed neutron emission probabilities}~\label{sec:Pnvalues}
The $\beta$ decay of a neutron-rich nucleus can populate levels in the daughter nucleus above the neutron separation energy $S_{\rm n}$, thus opening the $\beta$-delayed neutron-emission channel. The probability of observing a neutron associated with the $\beta$ decay of a nucleus is given by the neutron-emission probability or $P_{\rm n}$ value (called $P_{1\rm n}$ in Sec.~\ref{sec:decaycurve}). $\beta$-delayed neutrons were detected in coincidence with $\beta$ decays using the NERO detector in conjunction with the BCS. $P_{\rm n}$ values were determined for each nucleus according to:
\begin{equation}
P_{\rm n}=\frac{N_{\beta \rm n}-B_{\rm n}-N_{\beta \beta \rm n}}{\epsilon_{\rm n}  N_{\beta}},
\label{eq:Pn}
\end{equation}
where $N_{\beta \rm n}$ is the number of detected neutrons in coincidence with $\beta$ decays correlated with previous implantations; $B_{\rm n}$ is the number of background $\beta$-neutron coincidences; $\epsilon_{\rm n}$ is the neutron detection efficiency; and $N_{\beta}$ is the number of $\beta$-decaying mother nuclei. $N_{\beta \beta \rm n}$ is the number of detected $\beta$-delayed neutrons from descendant nuclei; it should be consequently subtracted from the total number of detected neutrons in order to determine the actual number of neutrons associated with the nucleus of interest. For the nuclear species discussed in this paper, $\beta$-neutron coincidences associated with descendant nuclei other than the $\beta$-decay daughter were negligible. Using the Batemann equations~\cite{Cet06}, it is possible to write explicitly the value of $N_{\beta \beta \rm n}$:
\begin{equation}
N_{\beta \beta \rm n} = (1-P_{\rm n})  C,
\label{eq:Nbbn}
\end{equation}
where $C$ is a constant given by:
\begin{equation}
C = \frac{\lambda_{2} P_{\rm{nn}} N_{\beta} \epsilon_{\rm n}}{\lambda_{2}-\lambda_{1}} \left[
1-e^{{-\lambda_{1}t_{c}}}-\frac{\lambda_{1}}{\lambda_{2}} \left( 1 - e^{{-\lambda_{2} t_{c}}} \right)
\right].
\label{eq:C}
\end{equation}
In this equation, $P_{\rm{nn}}$ is the neutron-emission probability of the daughter nucleus (called $P_{2\rm n}$ in Sec.~\ref{sec:decaycurve}), and $\lambda_{2}$ and $\lambda_{1}$ are the decay constants of the daughter and mother nuclei, the latter being extracted from the analysis discussed in the previous sections. Inserting Eq.~\ref{eq:Nbbn} and Eq.~\ref{eq:C} into Eq.~\ref{eq:Pn}, and rearranging terms:
\begin{equation}
P_{\rm n}=
\frac{N_{\beta \rm n}-B_{\rm n}-C}{N_{\beta}  \epsilon_{\rm n}-C}.
\label{eq:Pnlong}
\end{equation}
The value of $N_{\beta}$ for a given nucleus was calculated as the product of the total number of implantations by the average $\epsilon_{\beta}$. The number of neutrons detected by NERO in coincidence with $\beta$ decays were recorded in a multi-hit TDC. $P_{\rm n}$ values were first determined for the less exotic nuclei, taking $\lambda_{1}$ from Table~\ref{tab:halflives-results}, and $\lambda_{2}$ and $P_{\rm{nn}}$ from Ref.~\cite{ENSDF} (as their corresponding daughters were not included in the present experiment). The newly calculated values of $P_{\rm n}$ were then included in Eq.~\ref{eq:Pnlong} as $P_{\rm{nn}}$, to calculate $P_{\rm n}$ for the next exotic nuclei.

\subsubsection{Neutron detection efficiency}~\label{sec:NEROEff}
\begin{figure}[t!]
\begin{center}
\includegraphics[width=6cm]{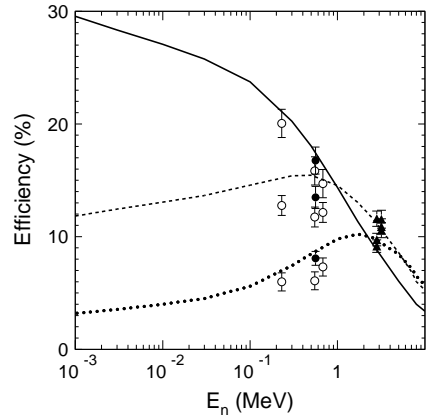}
\caption{
MCNP-calculated efficiencies as a function of neutron energy $E_{\rm n}$ for the innermost ring (solid line), intermediate ring (dashed line) and external ring (dotted line), compared to the measured values from the reactions  $^{11}$B($\alpha$,$n$) (dots), $^{13}$C($\alpha$,$n$) (triangles) and $^{51}$V($p$,$n$) (empty circles), and from a neutron post-fission $^{252}$Cf source (square). See Ref.~\cite{Lor06,Lor08} for more details.}
\label{fig:NEROeff}
\end{center}
\end{figure}

The design of the NERO detector was optimized to achieve a large and energy-independent efficiency, at least in the typical range of energies of the measured $\beta$-delayed neutrons. The efficiency response of NERO was determined at the Nuclear Structure Laboratory, at the University of Notre Dame, by detecting neutrons produced at different energies $E_{\rm n}$ from resonant and non-resonant reactions, and from a $^{252}$Cf source, as described in Ref.~\cite{Hos04}. In that analysis, eight different values of $E_{\rm n}$ ranging from about 0.2~MeV to 5~MeV were covered. The experimental results were extrapolated to a wider energy range, using the MCNP code~\cite{MCNP}. The NERO efficiency is nearly constant for $E_{\rm n}$ below 0.5~MeV, and gradually decreases beyond this value, as discussed in Refs.~\cite{Lor06,Lor08}. Further analysis of the detector rings showed that, for energies below 1~MeV, where NERO is most efficient, the total efficiency was mainly governed by the innermost detector ring followed by the intermediate and external rings. This result suggested that, at those energies, the most efficient thermalization of neutrons takes place during the first interactions with the polyethylene moderator. Conversely, the three rings converge to nearly the same efficiency at energies above 1~MeV, where the total efficiency drops significantly (see Fig.~\ref{fig:NEROeff}).

In the present experiment, the energies of the $\beta$-delayed neutrons $E_{\beta \rm n}$ ranged from zero to $Q_{\beta}-S_{\rm n}$, where $Q_{\beta}$ is the $\beta$-decay $Q$ value of the mother nucleus and $S_{\rm n}$ is the neutron separation energy of the daughter nucleus. The distribution of $E_{\beta \rm n}$ between these two values follows $\sim S_{\beta}(E) f(Q_{\beta}-E)$, where $S_{\beta}(E)$ is the $\beta$-decay strength function for a decay into the daughter's level at energy $E=S_{\rm n}+E_{\beta \rm n}$, and
\begin{equation}
f(Q_{\beta}-E) \sim (Q_{\beta}-S_{\rm n}-E_{\beta \rm n})^{5}.
\label{eq:fFermi}
\end{equation}
The strong energy dependence of $f$ largely favors $S_{\beta}$ to excited levels of the daughter nucleus near its $S_{\rm n}$. Moreover, as discussed in Refs.~\cite{Kra79a, Kra79b, Kra82}, high-resolution spectroscopic studies of $\beta$-delayed neutron-emitter nuclei produced by fission showed that $E_{\beta \rm n}$ was always much lower than $Q_{\beta}-S_{\rm n}$ (e.g. 199~keV for $^{87}$Br, 450~keV for $^{98}$Rb, and 579~keV for $^{137}$I). This trend was also observed by the same authors in the total spectra of $^{235}$U and $^{239}$Pu, with average $E_{\beta \rm n}$ of 575~keV and 525~keV, respectively, and with little neutron intensities at
$E_{\beta \rm n}$$\gtrsim$800~keV~\cite{Kra79a,Eng88}. The reason for these \textquotedblleft compressed$\textquotedblright$ $E_{\beta \rm n}$ spectra is the strong, often preferred population of the lowest excited states in the final nuclei~\cite{Kra82}. Since the region investigated in the present work includes strongly deformed nuclei, the respective expected low-laying excited states are rather low. Thus, it was safe to assume the $E_{\beta \rm n}$ values of the nuclei of interest to be typically below 500~keV. For these low energies, a constant value of (37$\pm$5)$\%$ for the NERO efficiency was assumed.




\subsubsection{Neutron background}
\begin{figure*}[t!]
\begin{center}
\includegraphics[width=5cm]{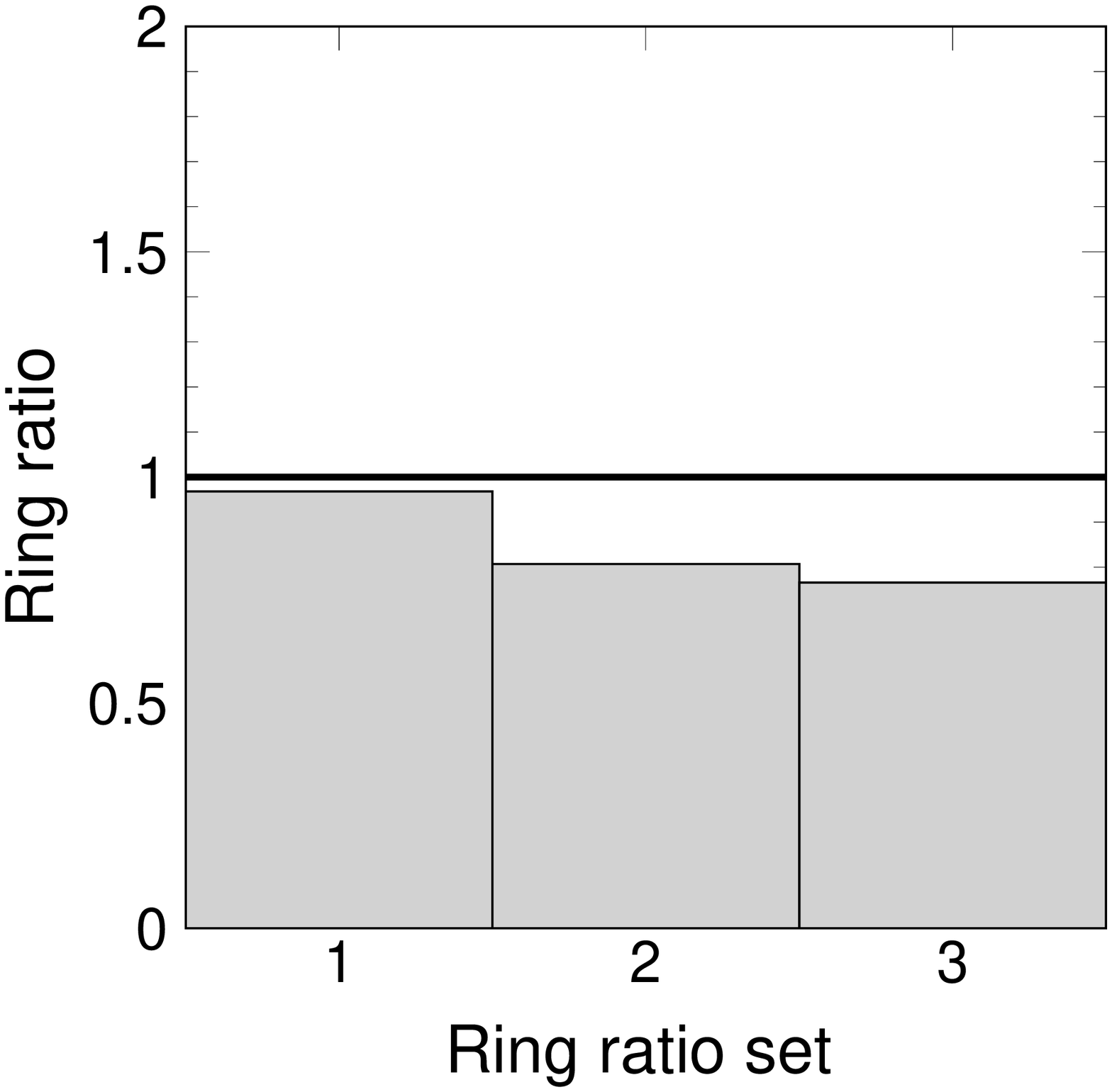}
\includegraphics[width=5cm]{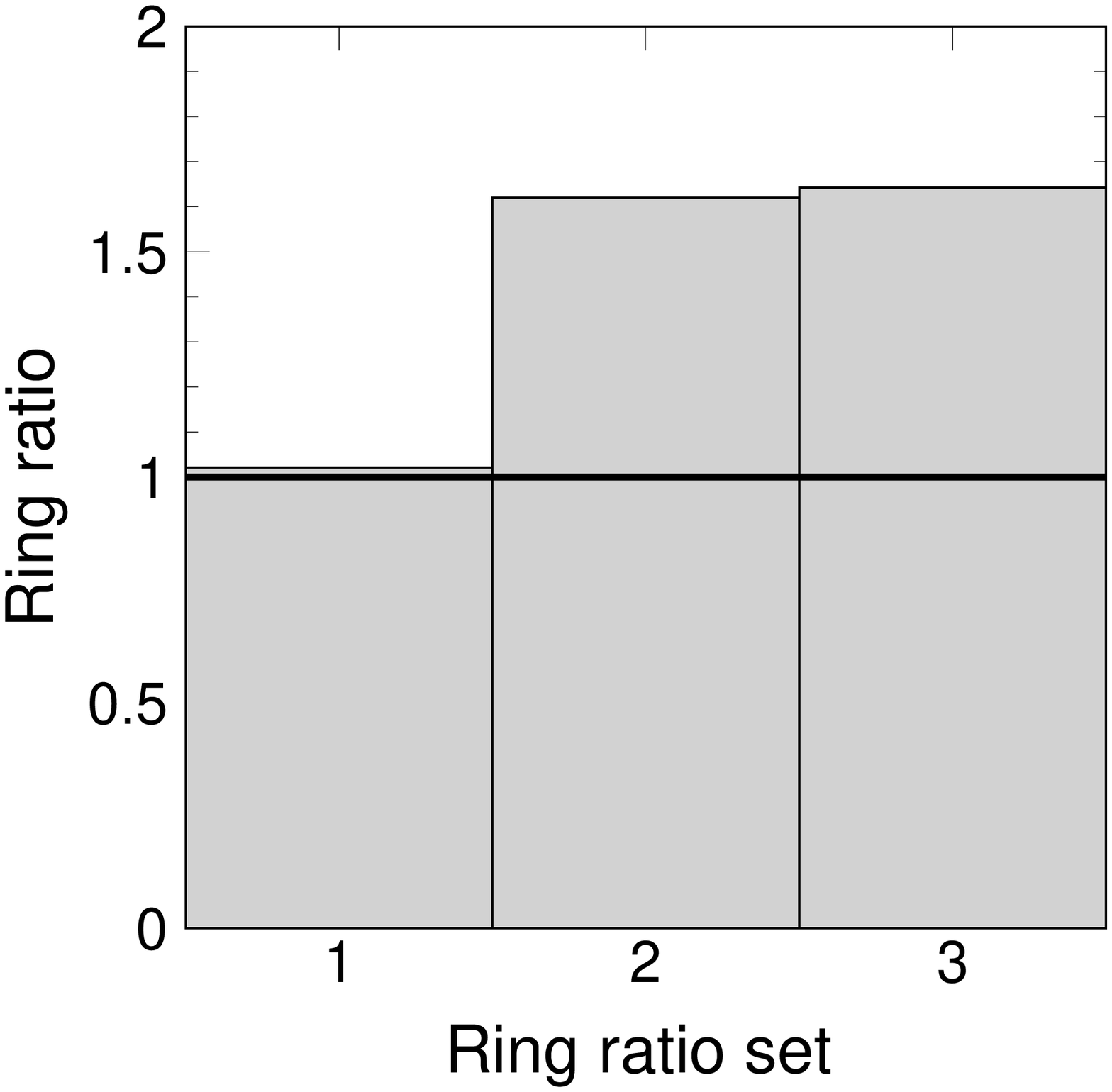}
\includegraphics[width=5cm]{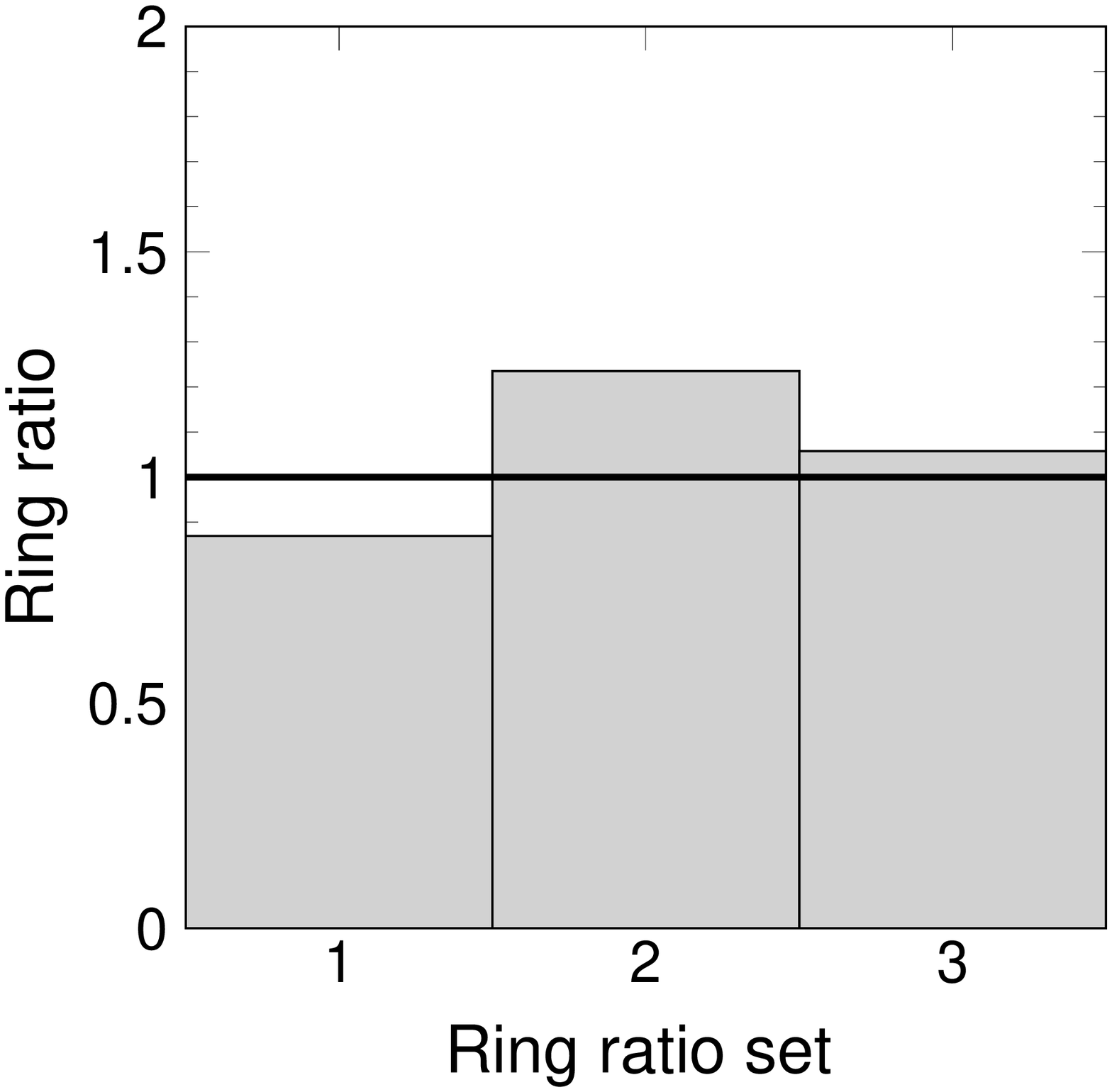}
\caption{Ratio of neutrons detected with different NERO rings for three different runs: production (left), background with beam off (center), and background with beam on (right). Histogram bin numbers 1, 2 and 3 correspond to ring ratios $R_{2}/R_{1}$,  $R_{3}/R_{2}$, and $R_{3}/R_{1}$, where $R_{1-3}$ are the innermost, intermediate and external rings (see text for details).}
\label{fig:NEROrings}
\end{center}
\end{figure*}

Free neutron background rates were independently recorded throughout the experiment using NERO in self-triggering mode. Four of these measurements were taken without beam, and two with beam on target. The background rates doubled from 4~s$^{-1}$ to 8~s$^{-1}$, when fragments were sent into the experimental setup, revealing the existence of two different background sources. The energy spectra recorded during the background measurements proved that the background origin could be attributed to actual neutrons. One of the neutron-background sources was intrinsic to the detector and its environment, while the other had a beam-linked origin. Analysis of the ring-counting ratios for background and production runs supported this idea (see Fig.~\ref{fig:NEROrings}). Measurements of $\beta$-delayed neutrons emitted from the implanted nuclei showed that the NERO counting rates were higher for the innermost ring (i.e., the closest to the DSSD) and systematically decreased for the next external rings. This results is compatible with MCNP simulations summarized in Fig.~\ref{fig:NEROeff}. Background runs with beam off showed the opposite trend, with high rates in the most external ring, which gradually decreased for the next internal ones. Such a result suggests that these runs were mainly affected by an external background source, most probably related to cosmic rays. Background runs with beam on target showed an intermediate situation that could be explained as arising from a combination of external and internal sources.

The value of $B_{\rm n}$ in Eq.~\ref{eq:Pn} included contributions from neutron-$\beta$-background events (i.e., neutrons in coincidence with $\beta$-decay background events) $B_{\rm n}(B_{\beta})$, and from random coincidences between free NERO background events and real $\beta$ decays $B_{\rm n}({\beta})$. The value of $B_{\rm n}(B_{\beta})$ for each nucleus was calculated as the product of the neutron-$\beta$-background rate measured on each implanted DSSD-cluster, and the neutron-detection time following the corresponding implantations of that nucleus. Owing to the very low total number of neutron and $\beta$-decay background coincidences measured per DSSD-cluster, the neutron-$\beta$-background rate on each cluster was determined by scaling the $\beta$-decay background rate calculated in Sec.~\ref{sec:bkgrate}. The corresponding scaling factor, calculated as the DSSD cluster-averaged ratio of neutron-$\beta$-background coincidences to $\beta$-decay background events, was about 0.08 and nearly constant throughout the experiment. Besides this background source, $B_{\rm n}({\beta})$ was approximately calculated as the product of the number of mother $\beta$-decays $N_{\beta}$, and the probability for at least one free neutron background with a rate 8~s$^{-1}$ to be detected in random coincidence with a $\beta$ decay. This latter approximation is not valid for coincidences of $\beta$-decays with  free background neutrons that were produced by fragmentation reactions induced by the same implanted mother nuclei. A calculated probability for this scenario, however, demonstrated that the occurrence of such a type of coincidences was negligible. Table~\ref{tab:Pn-results} shows the value of $N_{\beta}$, $N_{n}$, $B_{\rm n}(B_{\beta})$ and $B_{\rm n}(\beta)$.


\subsubsection{Error analysis}
The error analysis of $P_{\rm n}$ was derived from Eq.~\ref{eq:Pnlong}. In general, the main source came from uncertainties in the number of detected $\beta$-delayed neutrons $N_{\beta \rm n}$ and background events, with typical values about 20$\%$ for each. The former had statistical origin, whereas the latter was calculated from the $\beta$-background uncertainties, described in Sec.~\ref{sec:bkgrate}, and the error in the determination of the 0.08 scaling factor described in the previous section. An additional contribution of 15$\%$ to the total error came from uncertainties in the number of mother decays $N_{\beta}$, which were calculated by propagating the uncertainties in $\epsilon_{\beta}$, according to Sec.~\ref{sec:mlh}. Finally, an average 13.5$\%$ relative error in $\epsilon_{\rm n}$ was calculated as described in Ref.~\cite{Lor06, Lor08}.

In the case of $^{109}$Mo and $^{110}$Mo---where contributions from the daughter nuclei to the total number of $\beta$-delayed neutrons was significant---the systematic error was governed by uncertainties in the value of $C$ in Eq.~\ref{eq:Pnlong}. The latter was derived from the error propagation of all the variables in Eq.~\ref{eq:C}, being $P_{\rm{nn}}$ the main contribution. Relative uncertainties of about 46$\%$ and 35$\%$ were obtained for $^{109}$Mo and $^{110}$Mo, respectively.

\begin{figure*}[t!]
\begin{center}
\includegraphics[width=5cm]{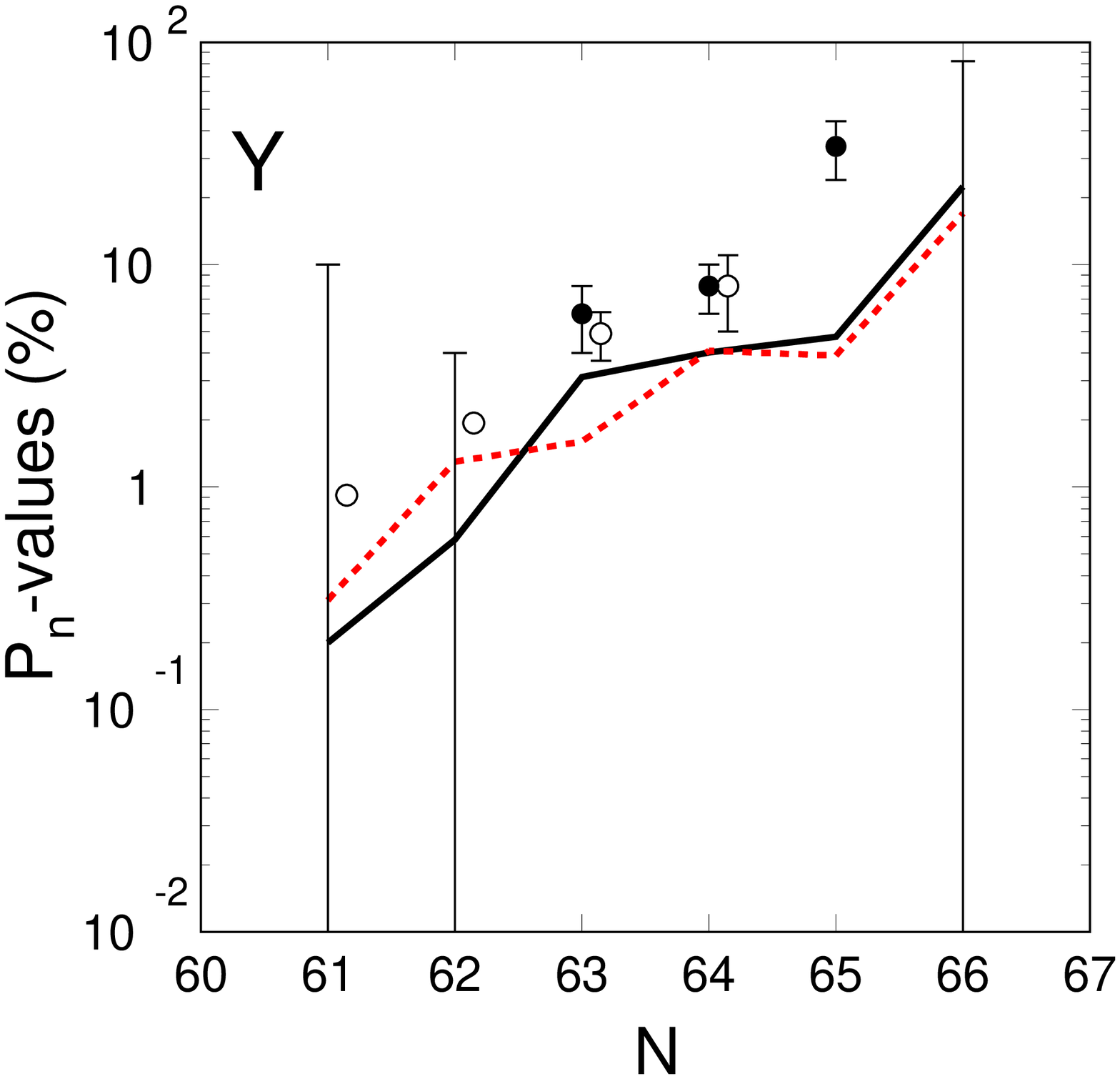}
\includegraphics[width=5cm]{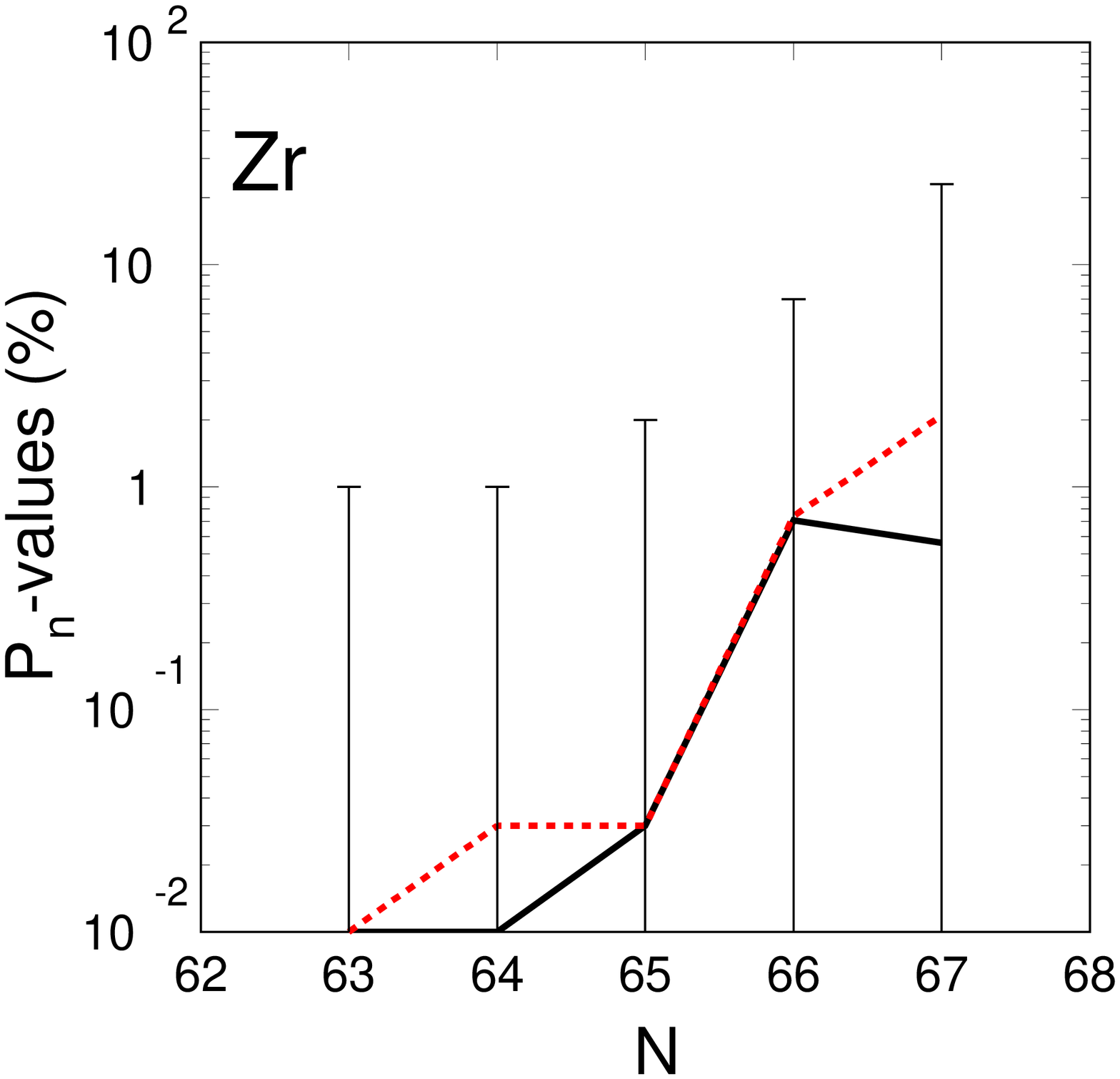}
\includegraphics[width=5cm]{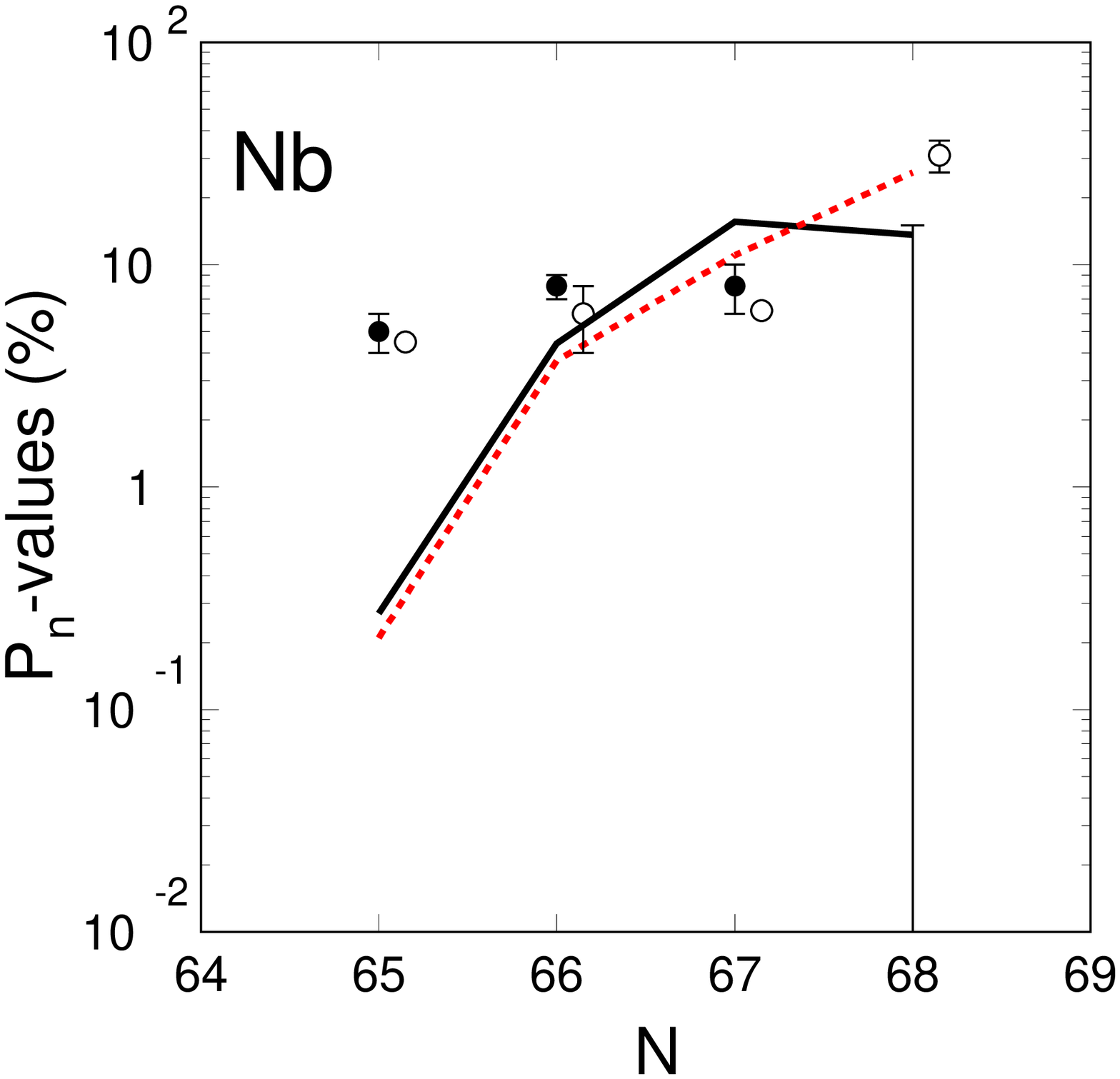} \\
\includegraphics[width=5cm]{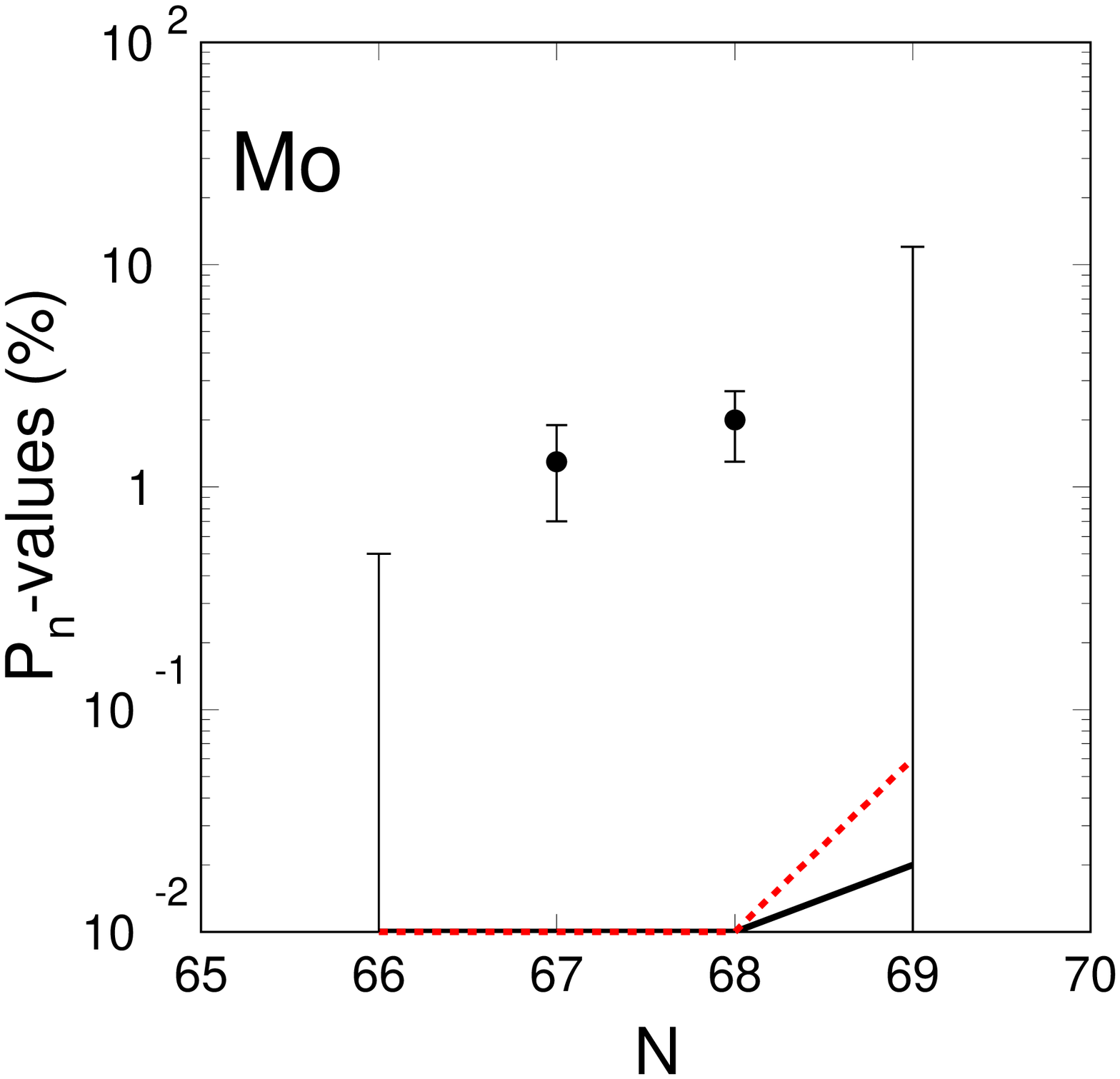}
\includegraphics[width=5cm]{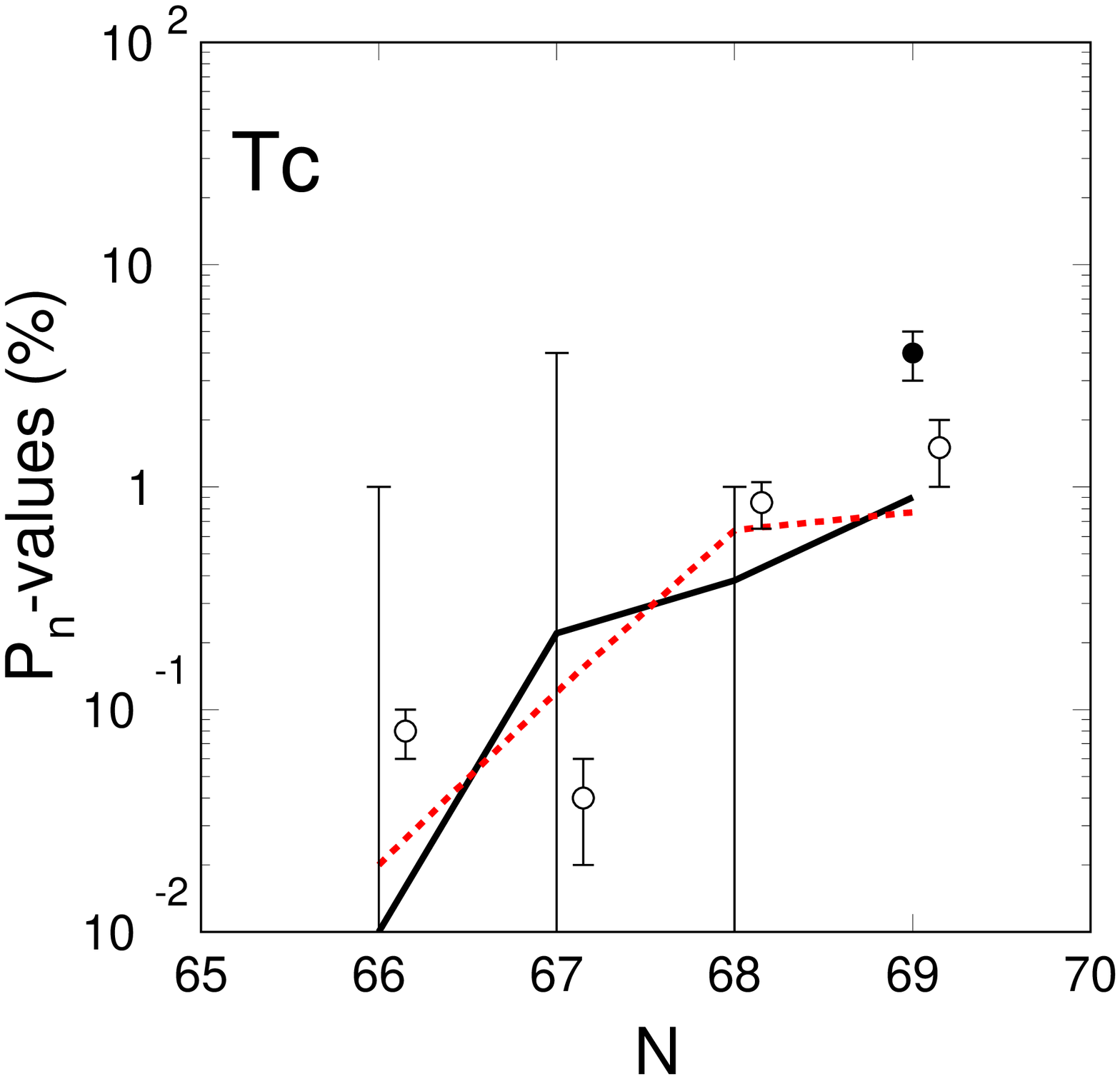}
\caption{Measured $P_{\rm n}$-values (full circles) and established upper limits (segments) for Y, Zr, Nb, Mo and Tc isotopes, compared with results from previous experiments~\cite{ENSDF,Meh96,Wan99} (empty circles). For the sake of clarity, the latter were shifted to the right by 0.15 units. The data are compared with two versions of the QRPA model of M\"oller \emph{et al.}: the version described in Ref.~\cite{Mol03} (solid line), and the interim version QRPA06 described in Sec.~\ref{sec:QRPA} (dashed line). (See text for details).}
\label{fig:Pn-results}
\end{center}
\end{figure*}

The $P_{\rm n}$ values and their errors obtained in the present experiment are listed in table~\ref{tab:Pn-results} and systematically presented for each isotopic chain in Fig.~\ref{fig:Pn-results}. $P_{\rm n}$ values were only deduced for nuclei with a statistically significant number of detected neutrons, i.e., with a number of detected neutrons above the number of background neutrons plus uncertainties within a 1-$\sigma$ confidence level. Otherwise, only upper limits of $P_{\rm n}$ were deduced using the method described in Ref.~\cite{Cow98}---for a Poisson distribution of detected neutrons---with the extension proposed by Hebbeker to include systematic uncertainties in the input quantities~\cite{Heb01} (see vertical lines in Fig.~\ref{fig:Pn-results}). The upper limits were calculated for a confidence level of 68$\%$.

\begin{table*}[t]
\caption{$P_{\rm n}$ values obtained in the present experiment. The results are compared with available data from previous experiments (Literature), and with the versions QRPA03 and QRPA06 of M\"oller's QRPA model. (See text for details).}
\begin{ruledtabular}
\begin{tabular}{ccccccccc}
&  &  &  &  &  &  &  &\\
Isotope   &  $N_{\beta}$ &  $N_{n}$  & $B_{\rm n}(B_{\beta})$ & $B_{\rm n}(\beta)$ & \multicolumn{4}{c}{$P_{\rm n}$ $(\%)$} \\
 &  &  &  &  &  &  &  &\\
          \cline{6-9}
 &  &  &  &  &  &  &  &\\
 &  &  &  &  &  Present exp. & Literature & QRPA03~\cite{Mol03} & QRPA06  \\ \hline
 &  &  &  &  &  &  &  &\\\vspace{1mm}
 $^{100}$Y  &   58   &  1  &  0.4  & 0.1  &  $\leq$ 10      & 0.92(8)~\cite{ENSDF}       &  0.2  &  0.3    \\\vspace{1mm}
 $^{101}$Y  &  231   &  3  &  0.6  & 0.4  &   $\leq$ 4      & 1.94(18)~\cite{ENSDF}      &  0.6  &  1.3    \\\vspace{1mm}
 $^{102}$Y  &  373   & 10  &  1.1  & 0.6  &          6(2)   & 4.9(12)~\cite{ENSDF}       &  3.4  &  1.6    \\\vspace{1mm}
 $^{103}$Y  &  185   &  6  &  0.5  & 0.3  &          8(2)   & 8(3)~\cite{Meh96}          &  4.0  &  4.1    \\\vspace{1mm}
 $^{104}$Y  &   40   &  5  &  0.1  & 0.1  &         34(10)  &                            &  4.8  &  3.9    \\\vspace{1mm}
 $^{105}$Y  &    8   &  1  &  0.03 & 0.01 &  $\leq$ 82      &                            & 22.4  & 17.0    \\\vspace{1mm}
 $^{103}$Zr &  856   &  5  &  2.4  & 1.4  &   $\leq$ 1      &                            &  0.0  &  0.0    \\\vspace{1mm}
 $^{104}$Zr & 1470   & 10  &  4.3  & 2.4  &   $\leq$ 1      &                            &  0.0  &  0.0    \\\vspace{1mm}
 $^{105}$Zr &  529   &  4  &  1.5  & 0.8  &   $\leq$ 2      &                            &  0.0  &  0.0    \\\vspace{1mm}
 $^{106}$Zr &  199   &  4  &  0.5  & 0.3  &   $\leq$ 7      &                            &  0.7  &  0.7    \\\vspace{1mm}
 $^{107}$Zr &   28   &  1  &  0.04 & 0.04 &  $\leq$ 23      &                            &  0.6  &  2.1    \\\vspace{1mm}
 $^{106}$Nb & 3238   & 70  &  9.1  & 5.2  &          5(1)   & 4.5(3)~\cite{Meh96}        &  0.3  &  0.2    \\\vspace{1mm}
 $^{107}$Nb & 2068   & 68  &  5.7  & 3.3  &          8(1)   & 6(2)~\cite{Meh96}          &  4.4  &  3.7    \\\vspace{1mm}
 $^{108}$Nb &  458   & 15  &  1.3  & 0.7  &          8(2)   & 6.2(5)~\cite{Meh96}        & 15.6  & 11.0    \\\vspace{1mm}
 $^{109}$Nb &   83   &  3  &  0.2  & 0.1  &  $\leq$ 15      & 31(5)~\cite{Meh96}         & 13.6  & 26.0    \\\vspace{1mm}
 $^{108}$Mo & 5557   & 35  & 24.2  & 8.9  &   $\leq$ 0.5    &                            &  0.0  &  0.0    \\\vspace{1mm}
 $^{109}$Mo & 2856   & 27  &  8.1  & 4.6  &          1.3(6) &                            &  0.0  &  0.0    \\\vspace{1mm}
 $^{110}$Mo &  689   &  8  &  1.9  & 1.1  &          2.0(7) &                            &  0.0  &  0.0    \\\vspace{1mm}
 $^{111}$Mo &   52   &  1  &  0.1  & 0.1  &  $\leq$ 12      &                            &  0.0  &  0.1    \\\vspace{1mm}
 $^{109}$Tc &  906   &  6  &  2.6  & 1.4  &   $\leq$ 1      & 0.08(2)~\cite{Meh96}       &  0.0  &  0.0    \\\vspace{1mm}
 $^{110}$Tc & 2960   & 14  &  8.5  & 4.7  &   $\leq$ 4      & 0.04(2)~\cite{Meh96}       &  0.2  &  0.1    \\\vspace{1mm}
 $^{111}$Tc & 1684   & 12  &  4.7  & 2.7  &   $\leq$ 1      & 0.85(20)~\cite{Meh96}     &  0.4  &  0.6    \\\vspace{1mm}
 $^{112}$Tc &  371   &  6  &  0.5  & 0.6  &          4(1)   & 1.5(5)~\cite{Wan99}        &  0.9  &  0.8    \\\vspace{1mm}
\end{tabular}
\end{ruledtabular}
\label{tab:Pn-results}
\end{table*}

\section{Results and discussion}~\label{sec:Discussion}
The newly measured $\beta$-decay half-lives included in Table~\ref{tab:halflives-results} and Fig.~\ref{fig:halflives-results} (as full circles) follow a systematic decreasing trend with neutron-richness, agreeing, for most of the cases, with previous measured values (empty circles). In some particular cases, half-lives from $\beta$-decay isomers were found in the literature. In particular, Khan \emph{et al.}~\cite{Kha77} reported two different half-lives for $^{100}$Y$_{61}$, presumably from low- and high-spin $\beta$-decaying isomers. The $T_{1/2}$ measured in the present experiment for this nucleus is compared in Fig.~\ref{fig:halflives-results} with the value found by Wohn \emph{et al.}~\cite{Woh86}, presently assumed to correspond to the ground state~\cite{ENSDF}. Similarly, in the case of $^{102}$Y$_{63}$, two half-lives were separately reported for the low-spin~\cite{Hil91} and high-spin~\cite{Shi83a} isomers. Interestingly enough, only the latter case is compatible with the value measured in the present experiment, thus indicating a favored production of this nucleus in a high-spin configuration.

Our results include new half-lives for the $N$$=$66 mid-shell isotopes $^{105}$Y$_{66}$ and $^{106}$Zr$_{66}$, as well as the more exotic $^{107}$Zr$_{67}$ and $^{111}$Mo$_{69}$. New $P_{\rm n}$ values were also deduced for $^{104}$Y$_{65}$ and $^{109,110}$Mo$_{67,68}$, and new upper limits for $^{105}$Y$_{66}$, $^{103-107}$Zr$_{63-67}$ and $^{108,111}$Mo$_{66,69}$. In the case of $^{104}$Y$_{65}$, the evolution of $P_{\rm n}$ with neutron number shows a pronounced increase compared with the smooth trend observed for lighter isotopes. Conversely, the sharp increase of $P_{\rm n}$ observed by Mehren and collaborators~\cite{Meh96} from $P_{\rm n}$$=$(6.2$\pm$0.5)$\%$, for $^{108}$Nb$_{67}$, to $P_{\rm n}$$=$(31$\pm$5)$\%$ for, $^{109}$Nb$_{68}$, is not supported by our measured upper limit $P_{\rm n}$$\leq$15$\%$ for $^{109}$Nb$_{68}$.

The small $P_{\rm n}$ values for $^{109,110}$Mo$_{67,68}$, which could not be observed in previous experiments, were detected here as a result of a lower neutron background rate of 0.001~$s^{-1}$. Additionally, the selectivity achieved in the present experiment, resulting from the combined in-flight separation technique and the event-by-event implantation-decay-neutron correlations, made it possible to rule out any potential neutron contaminant from neutron emitters in the cocktail beam. On the contrary, Wang \emph{et al.}~\cite{Wan99} pointed out the possible presence of neutron emitting contaminants from neighboring isobars in IGISOL-type experiments, which can only be detected by continuous monitoring of $\gamma$-lines from the separated beam. These authors use that argument as a possible explanation for the weak components of long-lived contaminants in the $A$$=$104 time-spectrum measured by Mehren \emph{et al.}~\cite{Meh96}.

\subsection*{QRPA results}~\label{sec:QRPA}
The experimental data shown in Fig.~\ref{fig:halflives-results}, for $T_{1/2}$, and in Fig.~\ref{fig:Pn-results}, for $P_{\rm n}$, are compared to two theoretical calculations. The solid lines are the results taken from Ref.~\cite{Mol03} (QRPA03), in which the allowed Gamow-Teller transition rates are calculated in a microscopic QRPA approach and in which the first-forbidden transition rates are
obtained from the statistical gross theory~\cite{Tak72,Tak73}. The second calculation (henceforth referred to as QRPA06), represented by the dashed lines, gives results from an identical model, but with the theoretical $Q_{\beta}$ values which enter in the phase-space integrals obtained from the improved finite-range liquid-drop model (FRLDM) of Ref.~\cite{Mol07} corresponding to the last line in Table~I of Ref.~\cite{Mol07}. In this interim global mass model, triaxial deformation of the nuclear ground state is taken into account, but there are also some other improvements. The agreement with the 2003 mass evaluation is 0.6038 MeV. As noted elsewhere~\cite{Mol08}, there are substantial effects from axial asymmetry on the ground-state masses in precisely the region of nuclei studied in this work. When experimental masses are available for both parent and daughter $Q_{\beta}$ and $S_{\rm n}$ are calculated from experimental data, otherwise from theory. In QRPA03, the 1995 mass evaluation from Ref.~\cite{Aud95} was used, in the QRPA06-interim calculation, the 2003 evaluation~\cite{Aud03} was used. Finally, both models calculate $S_{\beta}$ assuming the same deformation for the mother and daughter nuclei; an approximation that was discussed in detail by Krumlinde and M\"oller for some selected cases~\cite{Kru84}. For complete details about the model see Refs.~\cite{Kru84,Mol90,Mol97,Mol03}. Examples of how different types of  nuclear structure effects manifest themselves in the calculated $T_{1/2}$ and $P_{\rm n}$ are discussed in detail in, for example, Refs.~\cite{Kra84,Sor93,Meh96,Wan99,Kra00,Han00,Woh02,Mon06}).

\subsubsection*{General trends}
As shown in Figs.~\ref{fig:halflives-results} and~\ref{fig:Pn-results}, QRPA06 shows generally better agreement with the measured $T_{1/2}$ and $P_{\rm n}$ than the older QRPA03. The generally poor results for the half-lives of the less exotic isotopes are consistent with the fact that uncertainties in parameters such as $Q_{\beta}$
have a very strong impact for decays with small energy releases. This general behavior was already observed and discussed by Pfeiffer \emph{et al.} for different nuclei (see Figs.~8 and~9, and Tables~V and~VI of Ref.~\cite{Pfe03}). Beyond these general observations, the level of agreement between measured data and calculations shows no clear general systematic behavior. For instance, the half-lives predicted by both models are too short for all Y isotopes, and too long for all Mo isotopes. A similar trend is seen within the same isotopic chain such as Mo, where the half-life of $^{108}$Mo$_{66}$ is well reproduced, while the more exotic $^{109}$Mo$_{67}$, $^{110}$Mo$_{68}$ and $^{111}$Mo$_{69}$ are significantly overestimated. Such \textquotedblleft fluctuating$\textquotedblright$ behavior stems from the wide variety of nuclear shapes in this shape-transition region.

Both QRPA03 and QRPA06 predict half-lives for all $N$$=$65 isotones that are too short relative to the observed data, as was already pointed out by Wang \emph{et al.} in their analysis of $^{104}$Y$_{65}$~\cite{Wan99}. According to these authors, the coupling of the proton orbital $\pi [422]5/2^{+}$ to the neutron valence orbital $\nu [413]5/2^{+}$---which is in near proximity to $\nu [532]5/2^{-}$ at quadrupole deformation $\epsilon_{2}$$\simeq 0.3$---would give rise to the allowed $\beta$-decay transition from $^{104}$Y$_{65}$ $0^{+}$ into the $^{104}$Zr$_{64}$ $0^{+}$ ground state with a very short half-life. This interpretation explains also the disagreement between our measured and calculated $P_{\rm n}$ for $N$$=$65 isotones. In this case, the too low $P_{\rm n}$ values predicted by QRPA reflect an overestimated $\beta$-decay feeding into levels below the neutron separation energy $S_{\rm n}$.

\subsubsection*{Analysis of nuclear deformations}
\begin{figure*}[t!]
\begin{center}
\includegraphics[width=4.0cm]{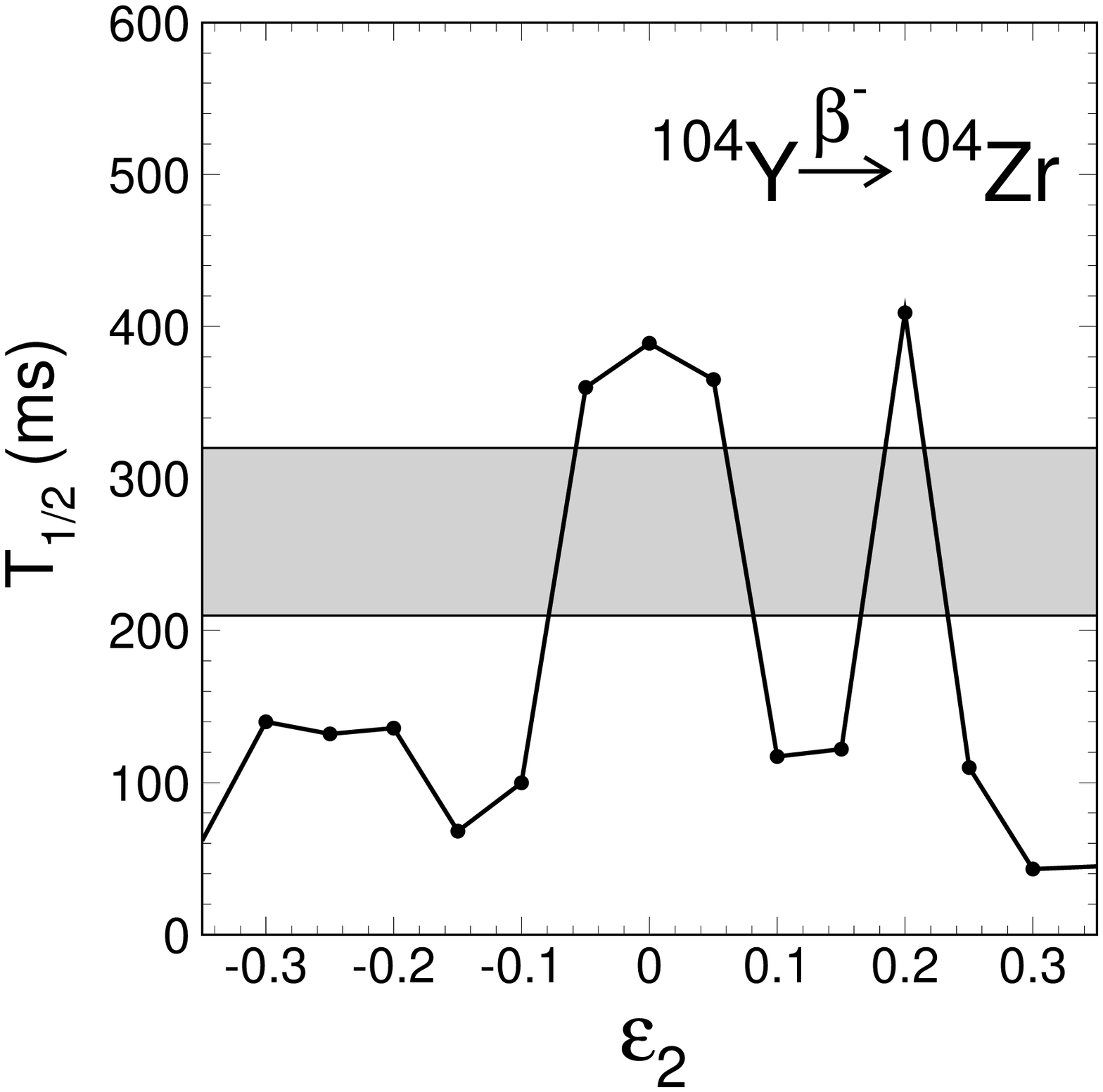}
\includegraphics[width=4.0cm]{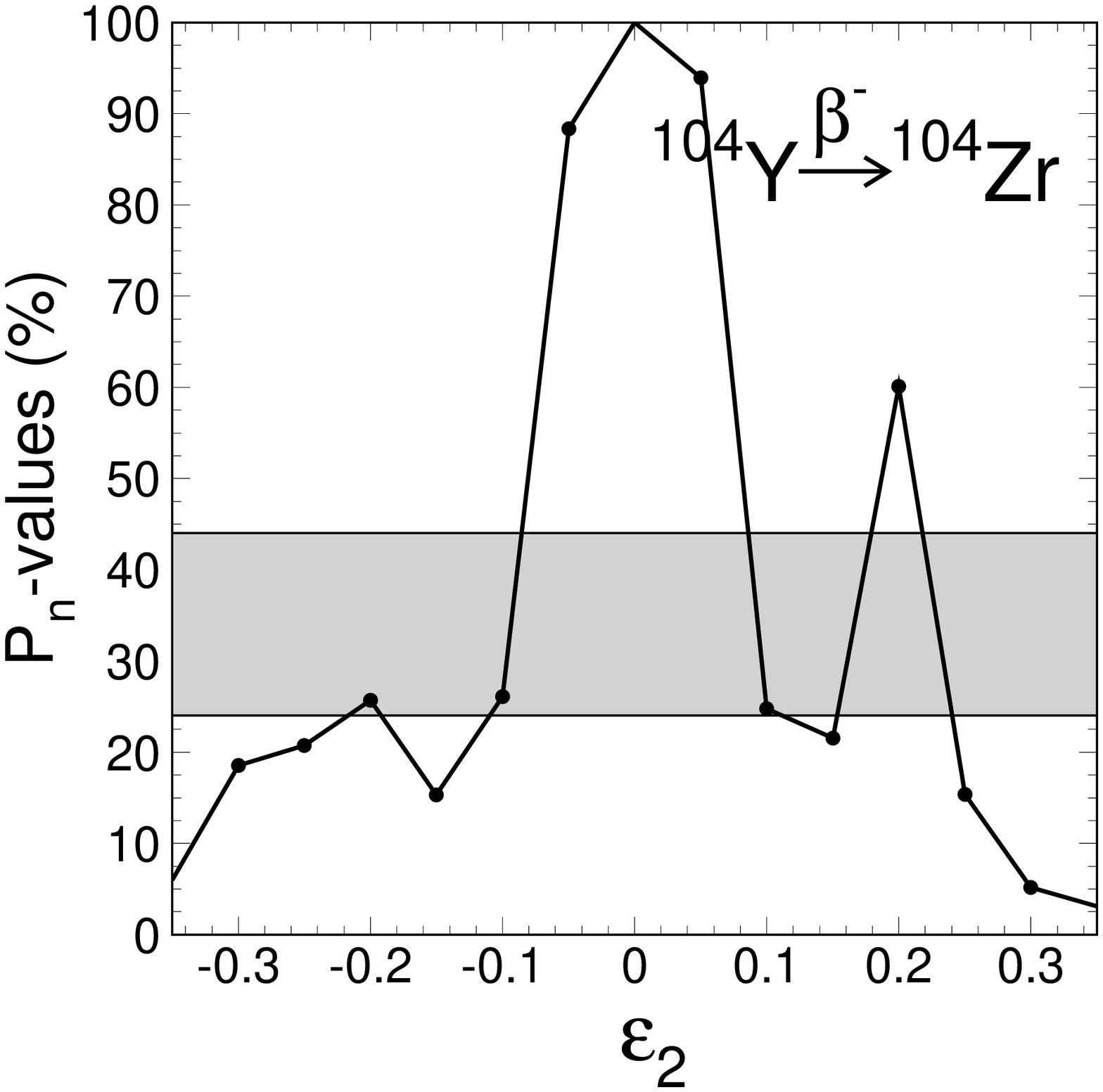}
\includegraphics[width=4.0cm]{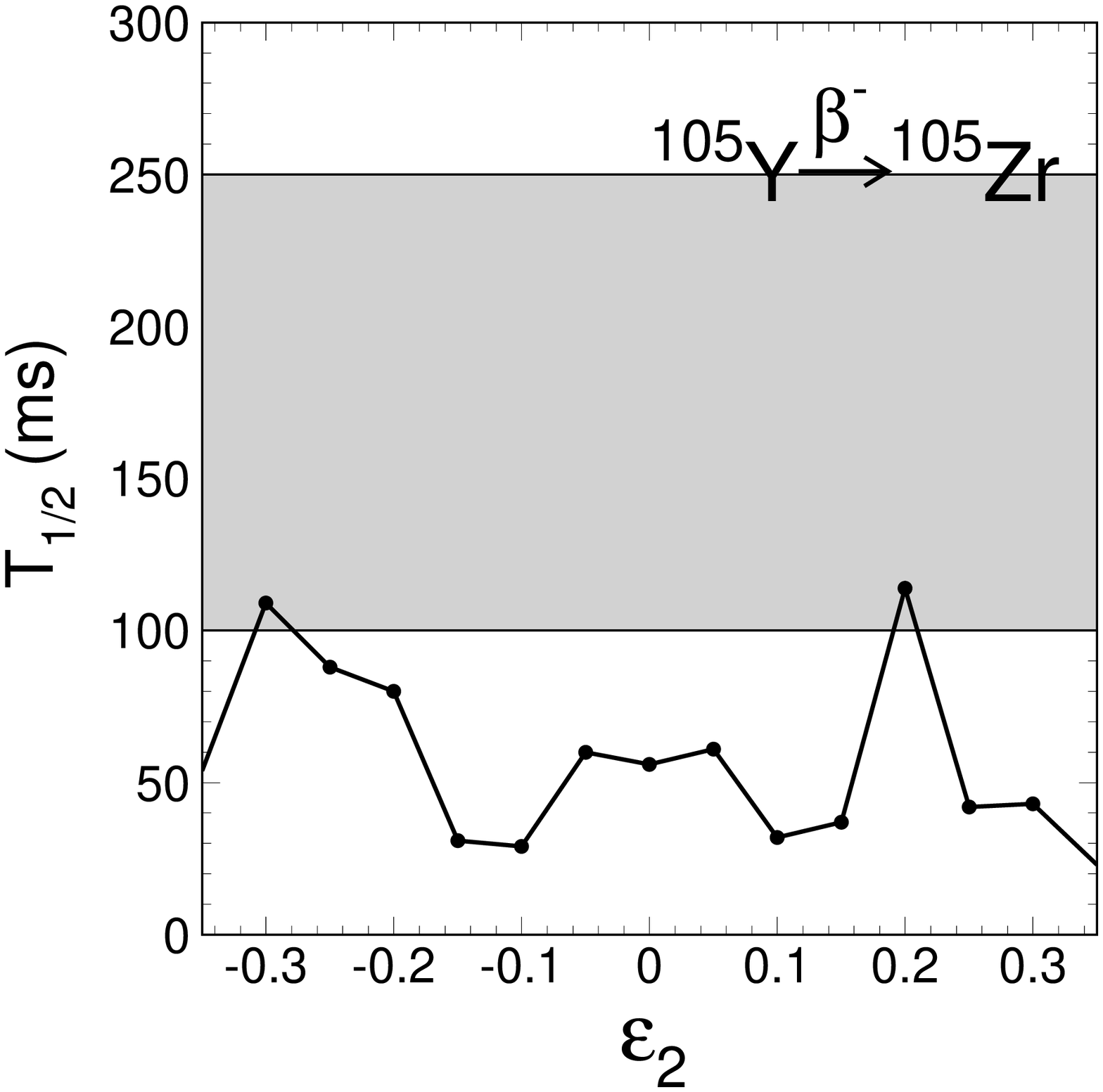}
\includegraphics[width=4.0cm]{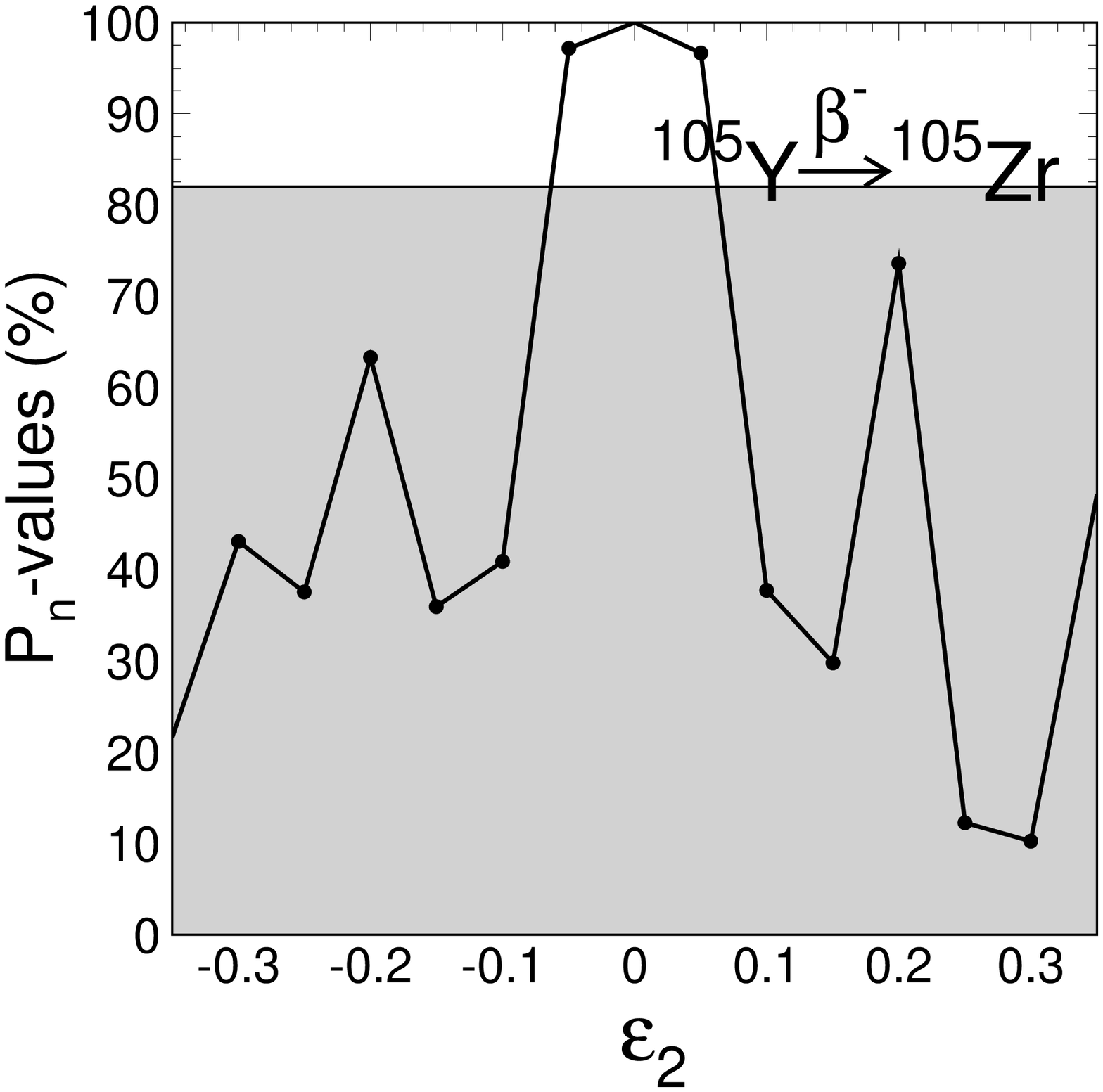} \\
\includegraphics[width=4.0cm]{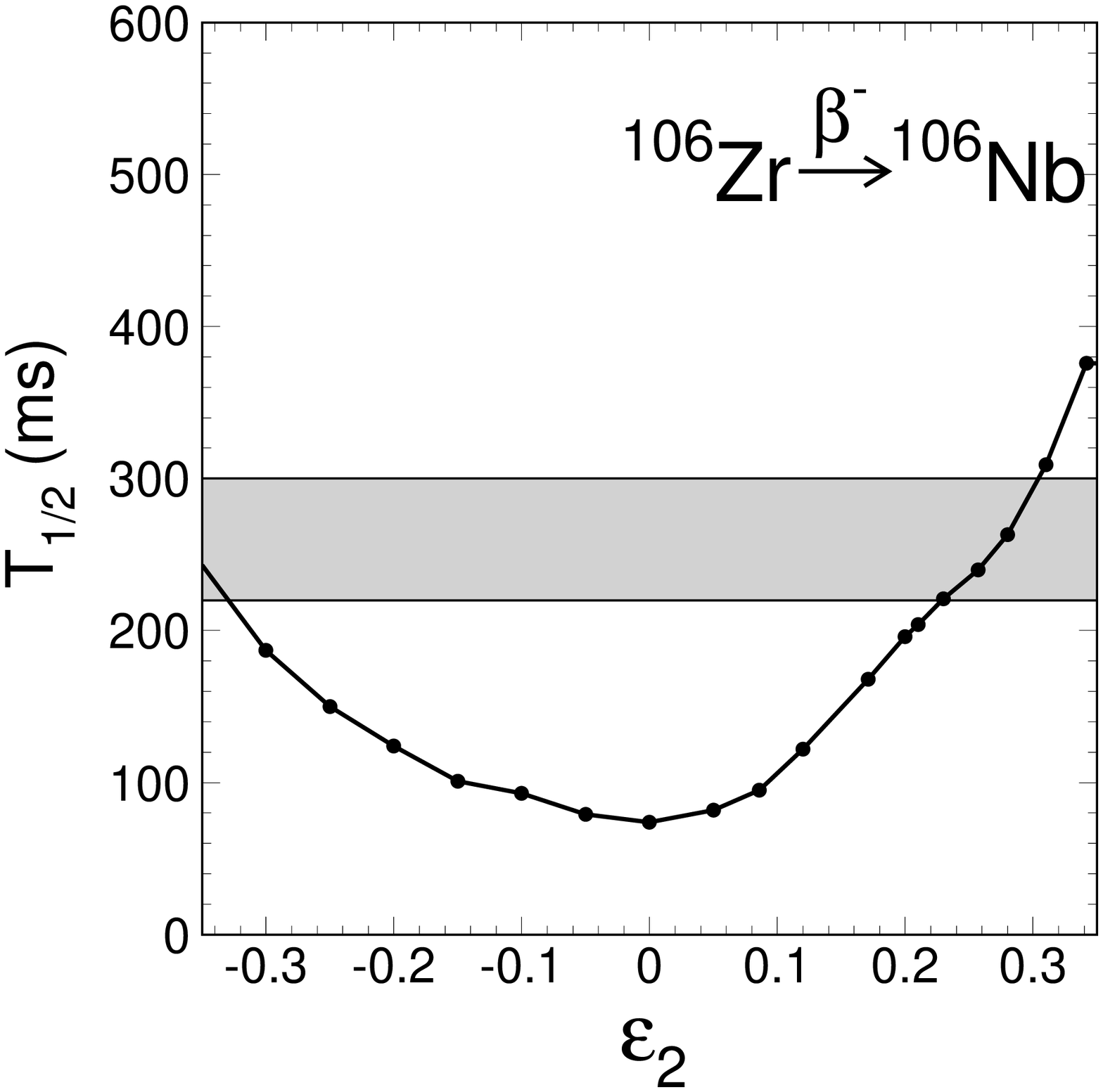}
\includegraphics[width=4.0cm]{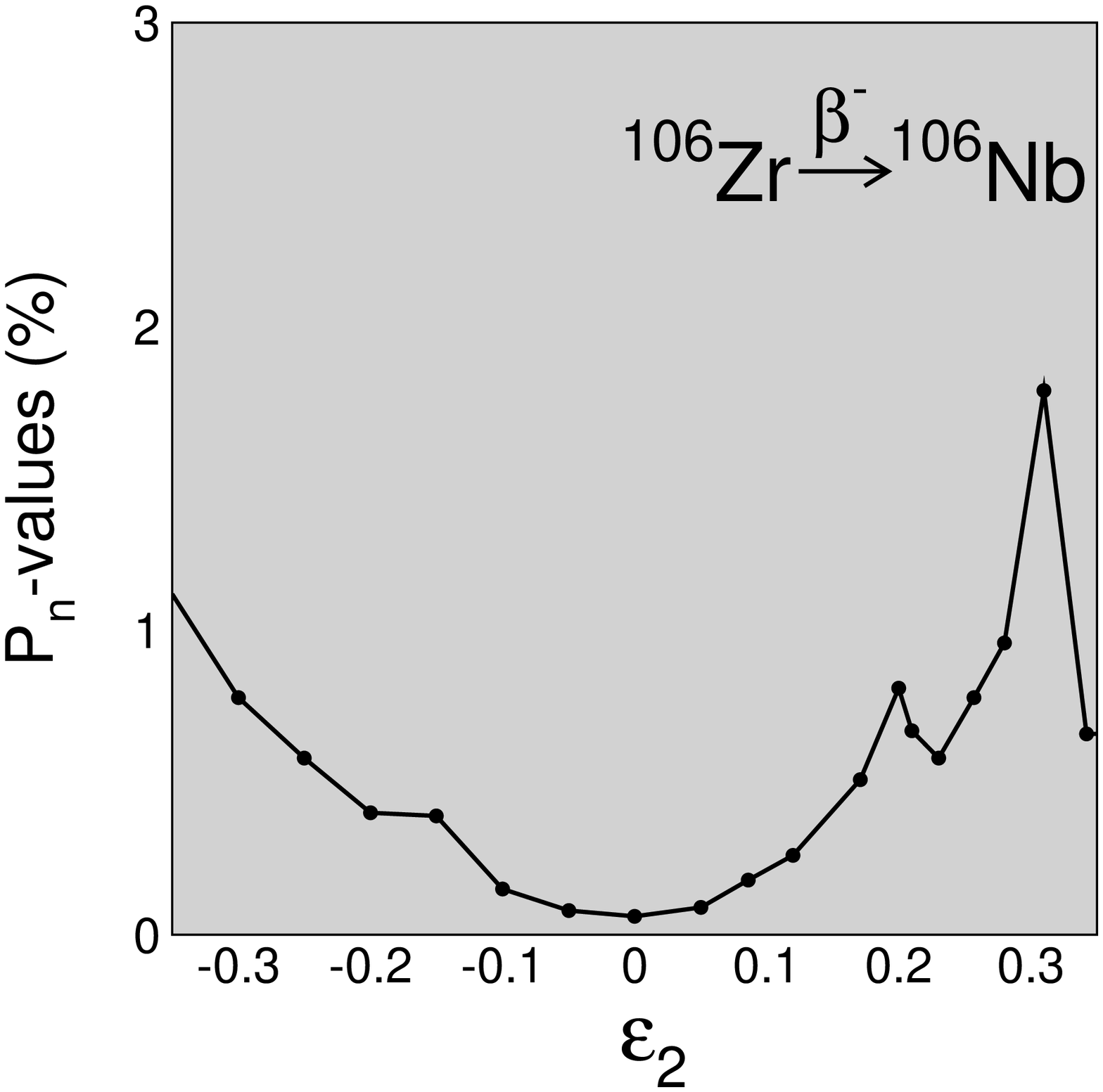}
\includegraphics[width=4.0cm]{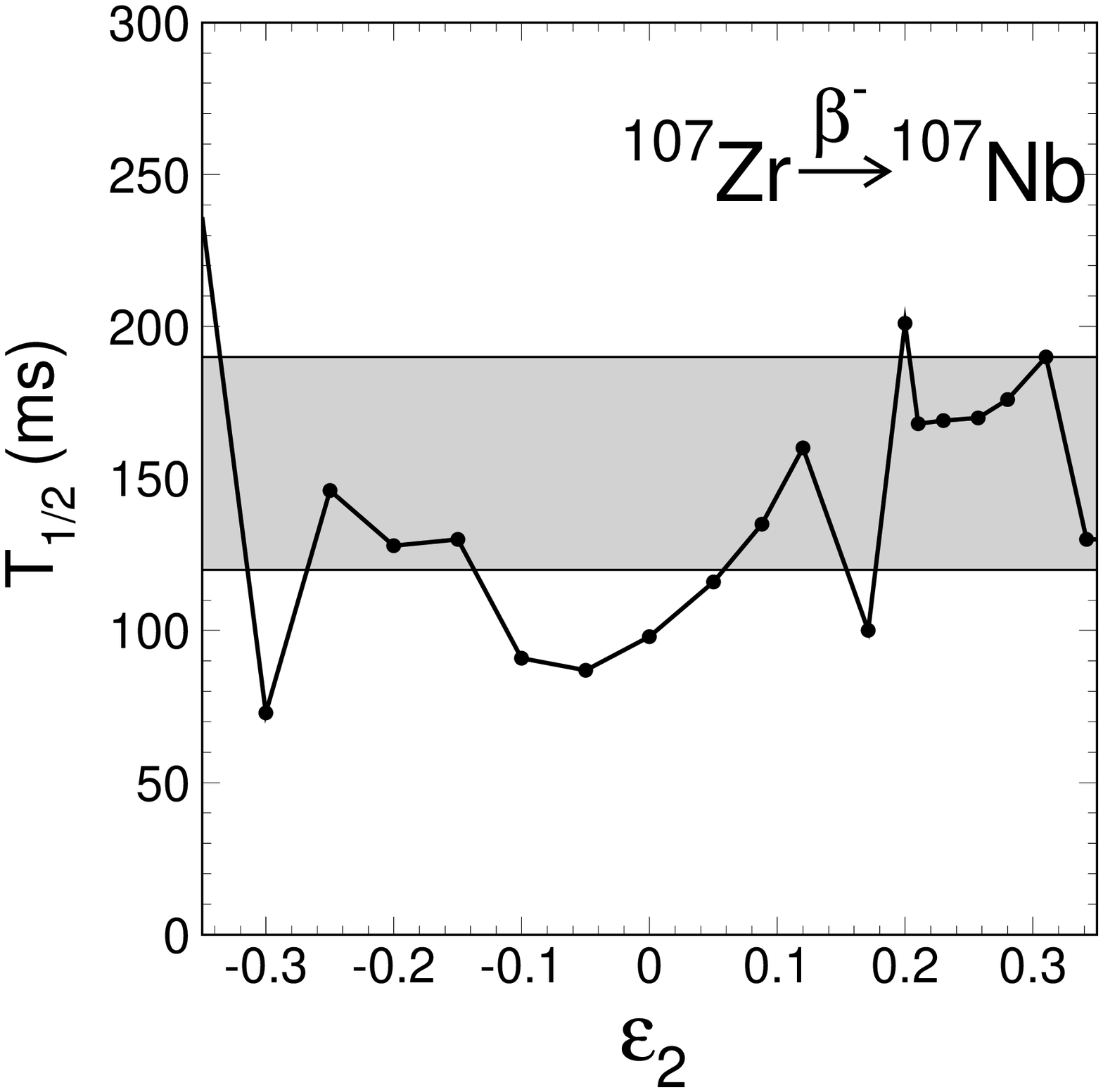}
\includegraphics[width=4.0cm]{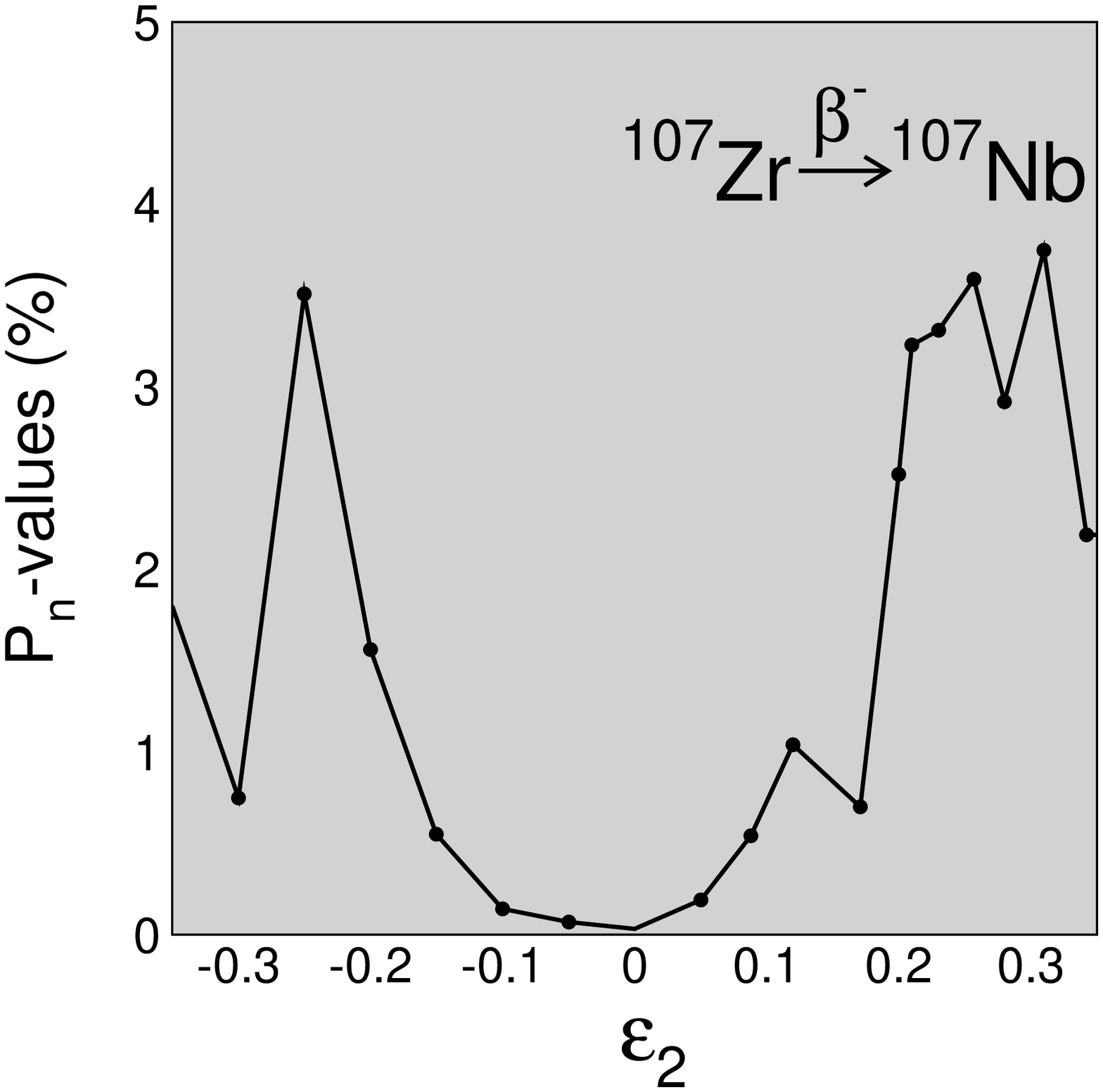}\\
\includegraphics[width=4.0cm]{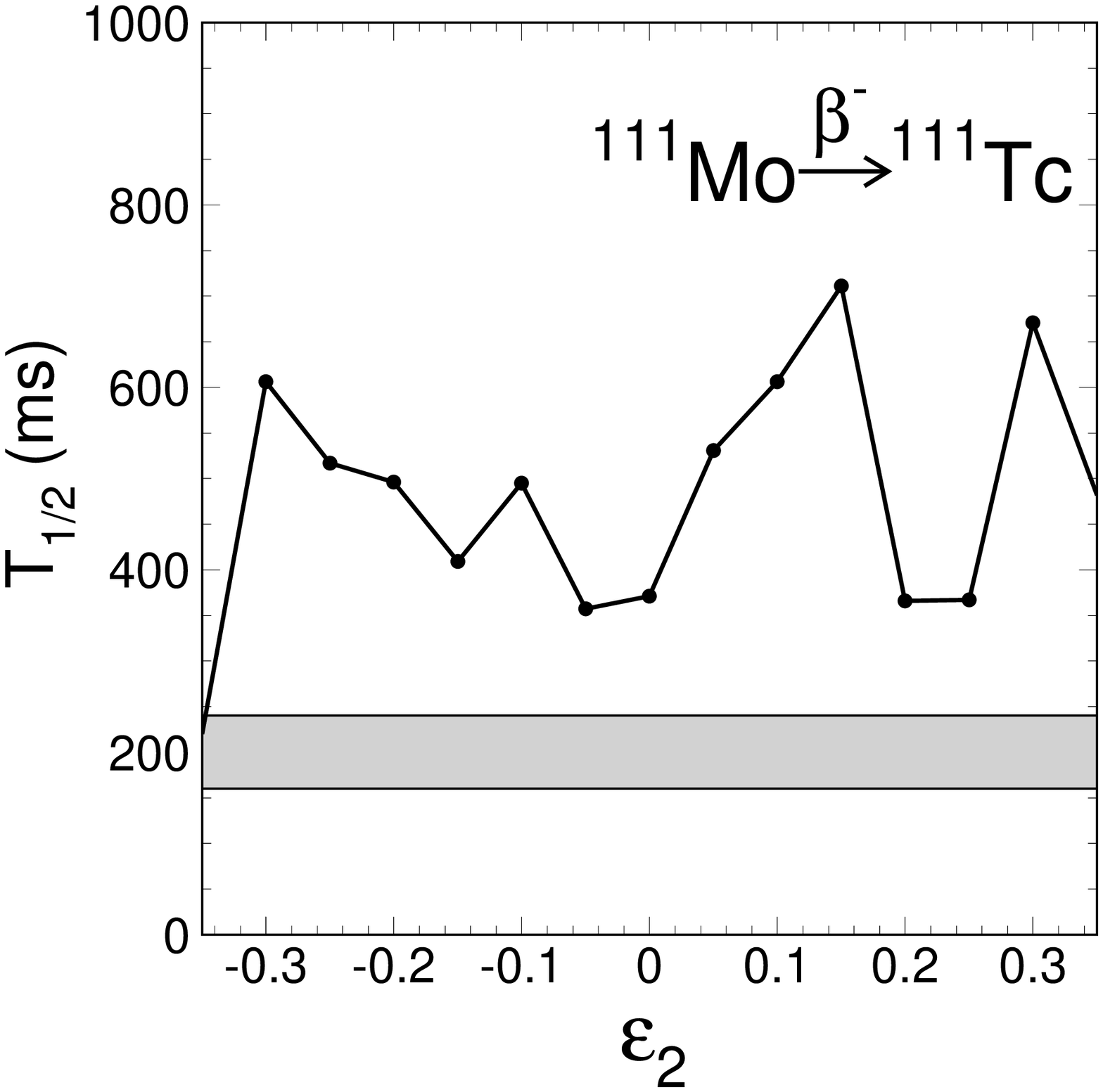}
\includegraphics[width=4.0cm]{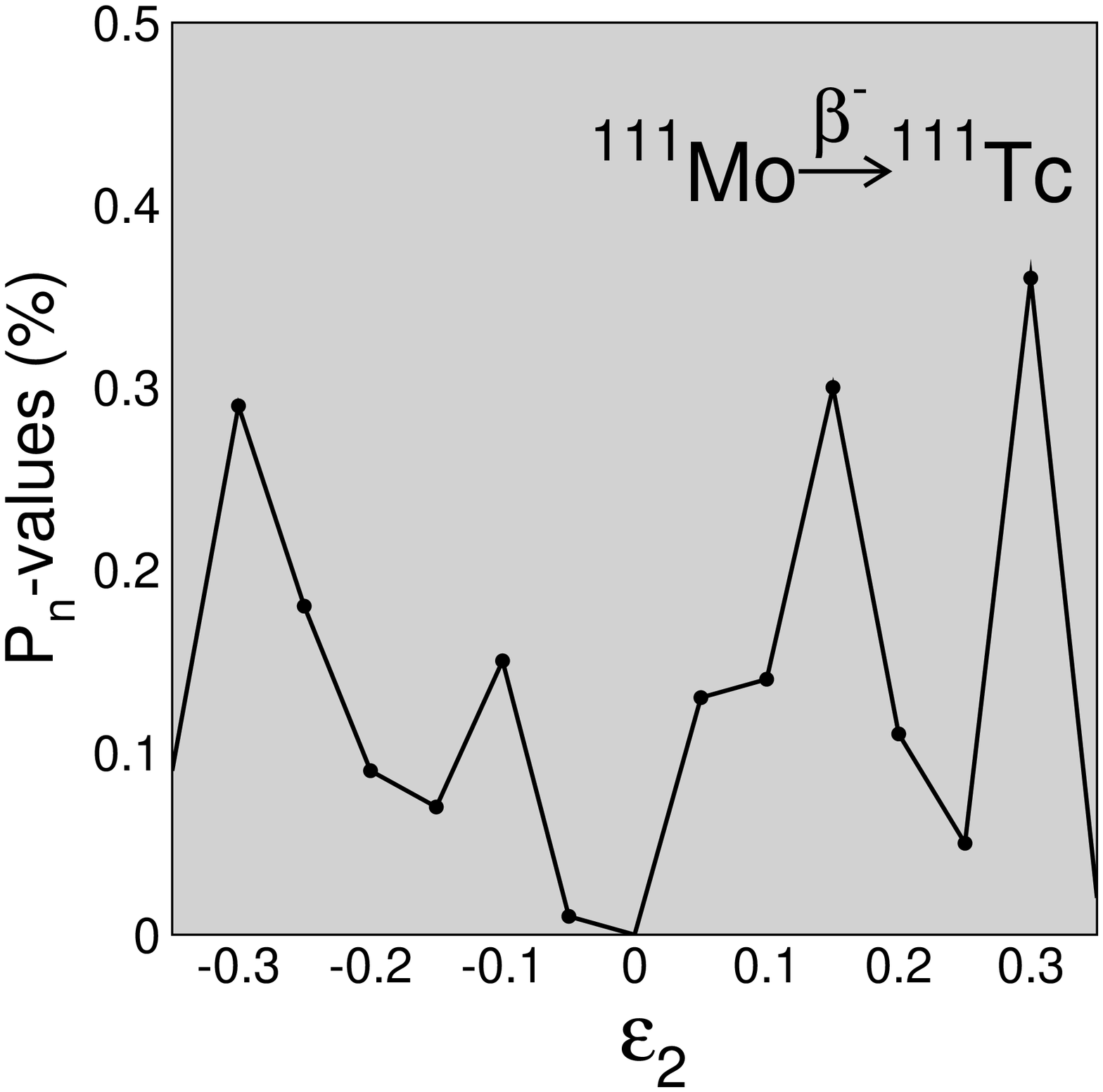}

\caption{QRPA-calculated $\beta$-decay half-lives and $P_{\rm n}$ values as a function of the quadrupole deformation of the daughter nuclei $\epsilon_{2}$ (solid line), compared with measured data within 1-$\sigma$ uncertainty (shaded area).}
\label{fig:QRPA-deformation}
\end{center}
\end{figure*}

In the case of $\beta$ decay of the Y isotopes, both models behave similarly, showing too short half-lives and too low $P_{\rm n}$ values. The improved treatment of deformation in QRPA06 had no major impact when compared with QRPA03, as triaxiality is not expected to develop for these nuclei. Indeed, spectroscopic studies of Zr isotopes between $N$$=$60 and $N$$=$64 showed that these nuclei are dominated by increasing prolate deformations with no indication of triaxial components~\cite{Smi96,Urb01,Tha03,Hua04}.
In an attempt to extend the analysis of nuclear shapes beyond $^{104}$Zr$_{64}$, we have re-calculated the $T_{1/2}$ and $P_{\rm n}$ of $^{104}$Y$_{65}$ and $^{105}$Y$_{66}$, assuming different pure prolate shapes for the corresponding mother-daughter systems. Results from this analysis are shown in Fig.~\ref{fig:QRPA-deformation}, where the measured $T_{1/2}$ and $P_{\rm n}$ are compared with calculations performed over a large range of quadrupole deformation ($-0.35 \leq \epsilon_{2} \leq 0.35$) of the daughter nuclei.

Three remarks from this analysis, in regards to the $\beta$ decay of $^{104}$Y$_{65}$ into $^{104}$Zr$_{64}$. Firstly, the calculated values of $T_{1/2}$ and $P_{\rm n}$ experience an abrupt transition from their maxima, for a spherical daughter $^{104}$Zr$_{64}$, to very low values at deformations around $\epsilon_{2}$$\simeq$$0.1$. Secondly, experimental $T_{1/2}$ and $P_{\rm n}$ are reproduced assuming a prolate deformation $\epsilon_{2}$$\sim$0.20. Thirdly, for larger deformations beyond $\epsilon_{2}$$\gtrsim$$0.25$ the $\beta$ decay becomes faster with decreasing probabilities for $\beta$-delayed neutron emissions. The good agreement of the calculations at $\epsilon_{2}$$\sim$$0.20$ is ruled by GT transitions into four-quasiparticle levels at energies around $S_{\rm n}$. Conversely, the too low predicted $T_{1/2}$ and $P_{\rm n}$ are governed by the fragmentation of $S_{\beta}$ into low-energy (i.e., well below $S_{\rm n}$) two-quasiparticle states involving the coupling of $\pi g_{9/2}$ levels with $\nu s_{1/2}$ (at $\epsilon_{2}$$\simeq$$0.1$) or with high-$\Omega$ $\nu d_{5/2}$ Nilsson orbitals (at $\epsilon_{2}$$\gtrsim$$0.25$).
Finally, the high $P_{\rm n}$ values around 100$\%$ for $\beta$ decay into a spherical $^{104}$Zr$_{64}$ arises from one single GT transition to the $\pi g_{9/2} \otimes \nu g_{7/2}$ level at 7.25~MeV, well above $S_{\rm n}$.

Similarly to $^{104}$Y$_{65}$, the $\beta$-decay half-life of $^{105}$Y$_{66}$ can only be correctly reproduced for deformations of the daughter $^{105}$Zr$_{65}$ given by $\epsilon_{2}$$\simeq$$0.2$, whereas the $P_{\rm n}$ upper limit serves only to rule out spherical deformations. At $\epsilon_{2}$$\simeq$$0.2$, the calculated $T_{1/2}$ is governed by transitions into prolate three-quasiparticle $\pi g_{9/2} \otimes \nu d_{5/2}$ states at energies around $S_{\rm n}$. Conversely, at larger and smaller deformations, the calculated $\beta$ decay is too fast due to the presence of different low-lying one-quasiparticle states that push $S_{\beta}$ down in energy, well below $S_{\rm n}$. Although the measured $T_{1/2}$ is also compatible with oblate deformations $\epsilon_{2}$$\lesssim$$-0.3$, we ruled out such scenario as there is no experimental evidence of oblate shapes in lighter Zr isotopes~\cite{Smi96,Urb01,Tha03,Hua04}.

In summary, the measured $T_{1/2}$ and $P_{\rm n}$ of $^{104}$Y$_{65}$ and $^{105}$Y$_{66}$ can only be correctly reproduced if one assumes $\epsilon_{2}$$\sim$0.2 for the corresponding daughter $^{104}$Zr$_{64}$ and $^{105}$Zr$_{65}$. On the basis of the larger elongations found for lighter Zr isotopes~\cite{Smi96,Urb01,Tha03}, the new values of $\epsilon_{2}$ for $^{104}$Zr and $^{105}$Zr$_{65}$ contradict somehow the rather expected maximum saturated deformations at the $N$$=$66 mid-shell. Interestingly enough, the maximum allowed quadrupole deformation of $^{104}$Zr$_{64}$ deduced from our data ($\epsilon_{2}$$\simeq$0.25) disagrees with the large $\beta_{2}\simeq$0.4 value obtained from analysis of the quadrupole moment $Q_{0}$ of the yrast band~\cite{Smi96}, and from measurements of $B(E2;2^{+}_{1}$$\rightarrow$$0^{+}_{1})$~\cite{Goo07}. Since $S_{\beta}$ is sensitive to the nuclear structure of the daughter nucleus---including, besides the yrast band, any other level from its ground state to energies just below $Q_{\beta}$---our result points to the possible presence of spherical or weakly-deformed low-lying bands coexisting with a highly-deformed yrast band of $^{104}$Zr$_{64}$. Spectroscopic studies of $^{100}$Zr$_{60}$ by Mach \emph{et al.}~\cite{Mac89} revealed such coexisting bands with $\beta_{2}$$\sim$0.4 and $\beta_{2}$$\sim$0.2. In addition, analysis of the $Q_{0}$ and $B(E2)$ systematics for Zr isotopes from $N$$=$50 to $N$$=$64 by Urban \emph{et al.}~\cite{Urb01} showed that coexisting spherical or weakly-deformed structures may be present beyond $^{100}$Zr$_{60}$, although these authors claim that such phenomenon may end at $N$$=$64. Since the QRPA formalism used in our analysis does not include deformation of excited levels, the $\epsilon_{2}$$\sim$$0.2$ presented here should be considered as an \textquotedblleft effective$\textquotedblright$ ground-state deformation resulting from the mixture of weakly and highly deformed bands in the daughter nuclei. Such a result suggests that shape coexistence may still be present at $^{104}$Zr$_{64}$ and $^{105}$Zr$_{65}$. This in turn may reflect the \textquotedblleft tailing effect$\textquotedblright$ of the predicted re-occurrence of the $Z$$=$40 sub-shell, together with a new sub-shell $N$$=$70 very far from stability~\cite{Dob94,Dob96,Pfe96}, or the development of a more exotic tetrahedral shape at $^{110}$Zr$_{70}$~\cite{Sch04}.

The calculated $T_{1/2}$ and $P_{\rm n}$ for $^{106}$Zr$_{66}$, $^{107}$Zr$_{67}$ and $^{111}$Mo$_{69}$ as a function of $\epsilon_{2}$ for the corresponding mother-daughter systems are shown in Fig.~\ref{fig:QRPA-deformation}. Here again, comparisons between the measured $T_{1/2}$ with calculations assuming pure quadrupole deformation allow to constrain the possible $\epsilon_{2}$ values of $^{106}$Nb$_{65}$ and $^{107}$Nb$_{66}$, whereas the calculated $P_{\rm n}$ values show not enough variation to distinguish between different deformations within uncertainties. The calculated $T_{1/2}$ for $^{111}$Mo$_{68}$ disagrees with the data for any pure quadrupole deformation of $^{111}$Tc$_{68}$. In this context, the FRLDM model predicts a triaxial component $\gamma$$=$15$^{\circ}$ for $^{106}$Nb$_{65}$, $^{107}$Nb$_{66}$ and $^{111}$Tc$_{68}$, which agrees with the values deduced by Luo \emph{et al.} for $^{105}$Nb$_{64}$ ($\gamma$$=$13$^{^{\circ}}$~\cite{Luo05}) and $^{111}$Tc$_{68}$ ($\gamma$$=$26$^{\circ}$~\cite{Luo06}).
Interestingly, the measured $T_{1/2}$ and $P_{\rm n}$ of $^{106}$Zr$_{66}$ and $^{107}$Zr$_{67}$ are in excellent agreement with the results from QRPA06 (which includes triaxiality), as shown in Fig.~\ref{fig:halflives-results}. Similarly, the calculated $T_{1/2}$ of $^{111}$Mo$_{69}$ is significantly improved when triaxiality is included, although no agreement with the measured value was yet found.

In summary, analysis of the new measured data for $^{106}$Zr$_{66}$, $^{107}$Zr$_{67}$ and $^{111}$Mo$_{69}$ using our QRPA06 calculations indicates triaxial deformations for the corresponding daughter nuclei $^{106}$Nb$_{65}$, $^{107}$Nb$_{66}$ and $^{111}$Tc$_{68}$.

\section{Summary}
We have reported on the measurements of $\beta$-decay properties of neutron-rich Y, Zr, Nb, Mo and Tc, which include new half-lives for $^{105}$Y, $^{106,107}$Zr, and $^{111}$Mo, along with new $P_{\rm n}$ values for $^{104}$Y and $^{109,110}$Mo and $P_{\rm n}$ upper limits for $^{103-107}$Zr and $^{108,111}$Mo. The new data could be attained due to the low $\beta$-decay background and $\beta$-delayed neutron background rates obtained with the BCS/NERO detection setup. The high selectivity of the A1900 in-flight separator at NSCL was also an instrumental achievement for the unambiguous identification of the exotic nuclei, thus demonstrating the optimum capabilities of this experiment setup to reach very exotic regions, near (and at) the r-process path.

The half-lives were analyzed using the MLH method and, in cases with enough statistics, least-squares fits of the decay curves. Agreement between both analysis brings confidence to the results. Analysis of the measured $T_{1/2}$ and $P_{\rm n}$ based on QRPA model calculations brings new insights to explore this interesting region in terms of deformations. The measured $T_{1/2}$ and $P_{\rm n}$ of $^{104,105}$Y$_{65,66}$ isotopes could only be reproduced for quadrupole deformation parameters $\epsilon_{2}$ of the corresponding daughter nuclei $^{104,105}$Zr$_{64,65}$ below the values reported in the literature for $^{104}$Zr$_{64}$ and lighter isotopes. Since the $\beta$-strength function $S_{\beta}$ governing the $\beta$-decay from $^{104}$Y$_{65}$ and $^{105}$Y$_{66}$ is sensitive to the level structure of the corresponding daughter nuclei, we believe that the low $\epsilon_{2}$ derived in the present work for $^{104}$Zr$_{64}$ and $^{105}$Zr$_{65}$ is a probable signature of coexisting weakly-deformed bands. Such an interpretation is supported by previous independent analysis of deformations based on measurements of yrast-band quadrupole moments $Q_{0}$ and $B(E2)$ for Zr isotopes between $N$$=$50 and $N$$=$64. The deformations reported in the present paper, however, show that weakly-deformed bands may still be present for $^{104,105}$Zr$_{64,65}$. The persistence of shape-coexistence for $^{104}$Zr$_{64}$ and $^{105}$Zr$_{65}$ may indicate the existence of a (near-) spherical doubly-magic $^{110}$Zr$_{70}$ nucleus; a result that is compatible with the quenching of the $N$$=$82 shell gap necessary to correct the unrealistic $A$$\simeq$110 r-process abundance trough predicted by r-process model calculations.

The QRPA calculations also show that triaxial shapes play a critical role in the $\beta$-decay to $^{106,107}$Nb$_{65,66}$ and $^{111}$Tc$_{68}$. The inclusion of this new deformation degree of freedom---on the basis of the new FRLDM of M\"oller \emph{et al.}--- significantly improves the calculated $T_{1/2}$ and $P_{\rm n}$ with respect to the new measured values. In addition, the FRLDM-predicted triaxial components are compatible with values reported in the literature for nuclei in this region.

Extension of $\beta$-decay and spectroscopic experimental studies to full r-process nuclei requires new high-intensity fragmentation-beam facilities like FRIB and FAIR. These measurements are necessary to understand the nuclear physics governing the r-process. New measurements of masses and $Q_{\beta}$ values of $N$$=$82 r-process isotones below Sn will clarify the role of shell quenching in the synthesis of heavy nuclei.

\begin{acknowledgments}
The authors wish to thank the NSCL operations staff for providing the primary beam, as well as T.~Ginter and the A1900 staff for the planning and development of the fragments analyzed in the present paper. Fruitful discussions with R.~Fox during the preparation of the experiment are acknowledged. The authors are also grateful to T.~Hebbeker for letting them use his method and code to calculate upper limits.

This work was supported in part by the Joint Institute for Nuclear Astrophysics (JINA) under NSF Grant PHY-02-16783 and the National Superconducting Cyclotron Laboratory (NSCL) under NSF Grant PHY-01-10253.
\end{acknowledgments}

\appendix
\section{Maximum Likelihood analysis of half-lives}~\label{sec:AppendixA}
The maximum likelihood method is the mathematical correct description even in
cases of poor statistics.  Let a set of independently
measured quantities $x_i$ originate from a probability density function
$f(x_i, \boldsymbol{\alpha})$, where $\boldsymbol{\alpha}$ is a set of unknown parameters. The maximum likelihood method consists of finding a
parameter set $\boldsymbol{\alpha}$ which maximizes the joint
probability density

\begin{equation*}
{\cal L}(\boldsymbol{\alpha}) = \prod_i f(x_i, \boldsymbol{\alpha}),
\end{equation*}

for all measured data points $x_i$. ${\cal L}$ is also called the
likelihood function. In most cases, it is easier to use $\ln {\cal L}$ instead
and to solve the likelihood equation

\begin{equation*}
\frac{\partial\ln {\cal L}}{\partial \boldsymbol{\alpha}} = 0.
\end{equation*}

Normalization factors, which depend on the set of parameters
$\boldsymbol{\alpha}$ have to be included in the maximization process.
All other multiplicative constants in the $f(x_i, \boldsymbol{\alpha})$
can be neglected, even if they depend on the measured quantities $x_i$.

The individual decay events of a decay sequence are not statistically independent, therefore
the likelihood function has to be defined. Additional corrections must
be used to compensate for the neglected late decay events, if the
correlation time window is small compared to the mean life time of the
investigated nuclei.  A method only using the first measured decay
event within the correlation window is reported in~\cite{Ber90}.
Such a method does not make use of all available information and
therefore might be disadvantageous in the case of poor statistics.
Based on the work in~\cite{Sch96}, the mathematical correct
probability density function for up to three decay events within the
correlation time window was developed.

We assumed that all the decay events after an implantation of an
identified nucleus within a position and time correlation interval
belong to the first three decay generation (mother, daughter and granddaughter decay). Additionally, coincidentally assigned background
events with a constant rate might occur. Up to three events within the
time correlation window are considered. For the sake of simplicity, we will exclude in the present discussion the
$\beta$-delayed neutron branchings.

Let $\lambda_1$, $\lambda_2$, and $\lambda_3$  be the decay constants
for the decay of a mother, daughter and granddaughter nucleus, respectively.
It is important to distinguish the probability for a decay within a
time $t$ characterized by a decay constant $\lambda_1$

\begin{equation*}
F_1(\lambda_1,t) = 1 - e^{-\lambda_1 t},
\end{equation*}

from the probability density function for a decay at exact time $t$
characterized by a decay constant $\lambda_1$

\begin{equation*}
f_1(\lambda_1,t) = \lambda_1   e^{-\lambda_1 t}.
\end{equation*}

For the detection of the second decay generation, we use the
probability for a decay within a time $t$ of a daughter nuclei with a
decay constant $\lambda_2$, which was populated by a mother decay with
decay constant $\lambda_1$:

\begin{equation*}
F_2(\lambda_1,\lambda_2,t) =  1 -
\frac{\lambda_1 \lambda_2}{\lambda_2 - \lambda_1}
\left[\frac{e^{-\lambda_1 t}}{\lambda_1}  -
\frac{e^{-\lambda_2 t}}{\lambda_2}\right],
\end{equation*}

and the probability density function for a decay of a daughter nuclei with a
decay constant $\lambda_2$ at time $t$, which was populated by a mother
decay with decay constant $\lambda_1$:

\begin{equation*}
f_2(\lambda_1,\lambda_2,t) =
\frac{\lambda_1 \lambda_2}{\lambda_2 - \lambda_1}
\Bigl[e^{-\lambda_1 t} - e^{-\lambda_2 t}\Bigr].
\end{equation*}

Similarly, the probability $F_3$, and the corresponding probability density function $f_3$, for a decay within a time $t$ of a granddaughter nuclei with a decay constant $\lambda_3$, which was populated by a mother and daughter decay characterized by decay constants $\lambda_1$ and $\lambda_2$ are given by:
\begin{widetext}
{\small
\[
\begin{split}
F_3(\lambda_1,\lambda_2,\lambda_3,t) =  1 -
\frac{\lambda_1 \lambda_2 \lambda_3}{(\lambda_2\!-\!\lambda_1)
(\lambda_3\!-\!\lambda_1) (\lambda_3\!-\!\lambda_2)}
\Biggl[\frac{(\lambda_3\!-\!\lambda_2)}{\lambda_1} e^{-\lambda_1 t} -
\frac{(\lambda_3\!-\!\lambda_1)}{\lambda_2} e^{-\lambda_2 t} + \\
+ \frac{(\lambda_2\!-\!\lambda_1)}{\lambda_3} e^{-\lambda_3 t}\Biggr],
\end{split}
\]
}
and
{\small
\[
\begin{split}
f_3(\lambda_1,\lambda_2,\lambda_3,t) =
\frac{\lambda_1 \lambda_2 \lambda_3}{(\lambda_2\!-\!\lambda_1)
(\lambda_3\!-\!\lambda_1) (\lambda_3\!-\!\lambda_2)}
\Bigl[(\lambda_3\!-\!\lambda_2) e^{-\lambda_1 t} -
(\lambda_3\!-\!\lambda_1) e^{-\lambda_2 t} + \\
+ (\lambda_2\!-\!\lambda_1) e^{-\lambda_3 t} \Bigr].
\end{split}
\]
}
\end{widetext}

Finally, for background events, the average rate and the expected number of
events within the correlation time is known.  The probability for the
observation of exact $r$ background events within a correlation time
$t_c$ and a background rate $b$ can be calculated using Poisson
statistics:
\begin{equation*}
B_r = \frac{(b t_c)^r e^{-b t_c}}{r!}.
\end{equation*}

Depending on the number of observed decay events within the correlation
time, one has to consider all possible scenarios leading to the
observation. In the following, we use a short notation to identify the
composition of the probability terms of the various scenarios.  $D_i$
stands for the probability that a decay of the i-th generation occurs,
$O_i$ that an occurring decay is observed.  $\epsilon_1$, $\epsilon_2$
and $\epsilon_3$ designate the detection efficiencies for the
respective decays, this is the probability for the observation of an
occurring decay. In addition, the notations $\bar{F}(\lambda,t) = 1 -
F(\lambda,t)$ and $\bar{\epsilon} = 1 - \epsilon$ are used.

The probability for the observation of no decay event within the
correlation time can be calculated as follows:

\begin{widetext}
\begin{align*}
P_0(\lambda_1) & = (\bar{D}_1 + D_1\bar{O}_1\bar{D}_2 +
D_1\bar{O}_1D_2\bar{O}_2\Bar{D}_3 + D_1\bar{O}_1D_2\bar{O}_2D_3\bar{O}_3)
  B_0\\[1ex]
P_0(\lambda_1) & =
\bigl[\bar{F}_1(\lambda_1,t_c) +
\bigl(\bar{F}_2(\lambda_1,\lambda_2,t_c) - \bar{F}_1(\lambda_1,t_c)\bigr)
  \bar{\epsilon}_1 +
\bigl(\bar{F}_3(\lambda_1,\lambda_2,\lambda_3,t_c) - \\
&\qquad -
\bar{F}_2(\lambda_1,\lambda_2,t_c)\bigr)
  \bar{\epsilon}_1   \bar{\epsilon}_2 + F_3(\lambda_1,\lambda_2,\lambda_3,t_c)
  \bar{\epsilon}_1   \bar{\epsilon}_2   \bar{\epsilon}_3 \bigl]
  B_0\\
& =
\bigl[1 - F_1(\lambda_1,t_c)   \epsilon_1 -
F_2(\lambda_1,\lambda_2,t_c)   \bar{\epsilon}_1   \epsilon_2 -
F_3(\lambda_1,\lambda_2,\lambda_3,t_c)
  \bar{\epsilon}_1   \bar{\epsilon}_2   \epsilon_3 \bigr]
  B_0.
\end{align*}
\end{widetext}

For the case of the observation of only one decay event within the
correlation time, four scenarios are possible:

1) The decay of the mother
was observed, daughter and granddaughter decay did either not occur or
these decays were not observed:
\[
P_{101} = P(d_1) =
D_1O_1   (\bar{D}_2 + D_2\bar{O}_2\bar{D}_3 + D_2\bar{O}_2D_3\bar{O}_3)
  B_0.
\]

2) The decay of mother and daughter did occur, but only the daughter decay
was observed, whereas the granddaughter decay did not occur or was not observed:
\[
P_{102} = P(d_2) =
D_1\bar{O}_1D_2O_2   (\bar{D}_3 + D_3\bar{O}_3)
  B_0.
\]

3) All three decays did occur, but only the granddaughter decay was observed:
\[
P_{103} = P(d_3) =
D_1\bar{O}_1D_2\bar{O}_2D_3O_3
  B_0.
\]
Until now, we assumed that there was no background event within the
correlation time.

4) The last scenario describes the observation of a
background event, all three decays did not occur or were not observed:

\begin{widetext}
\[
P_{104} = P(b) =
(\bar{D}_1 + D_1\bar{O}_1\bar{D}_2 + D_1\bar{O}_1D_2\bar{O}_2\bar{D}_3 + D_1\bar{O}_1D_2\bar{O}_2D_3\bar{O}_3)
  B_1.
\]
\end{widetext}

Calculation of the likelihood function requires the probability
density functions for the observations of a single decay event at time
$t_1$:

\begin{widetext}
\begin{align*}
p_{101}(\lambda_1) & = C_1
f_1(\lambda_1,t_1)   \epsilon_1
\bigl[\bar{F}_1(\lambda_2,t_c-t_1) + \bigl(\bar{F}_2(\lambda_2,\lambda_3,t_c-t_1) -
\bar{F}_1(\lambda_2,t_c-t_1)\bigr) \\
&\qquad    \bar{\epsilon}_2 + F_2(\lambda_2,\lambda_3,t_c-t_1)
  \bar{\epsilon}_2   \bar{\epsilon}_3
\bigr]   B_0\\
& = C_1
f_1(\lambda_1,t_1)   \epsilon_1
\bigl[1 - F_1(\lambda_2,t_c-t_1)   \epsilon_2 -
F_2(\lambda_2,\lambda_3,t_c-t_1)   \bar{\epsilon}_2   \epsilon_3 \bigr]
  B_0,\\[1ex]
p_{102}(\lambda_1) & = C_1
f_2(\lambda_1,\lambda_2,t_1)   \bar{\epsilon}_1   \epsilon_2
\bigl[\bar{F}_1(\lambda_3,t_c-t_1) + F_1(\lambda_3,t_c-t_1)   \bar{\epsilon}_3\bigr]
  B_0\\
& = C_1
f_2(\lambda_1,\lambda_2,t_1)   \bar{\epsilon}_1   \epsilon_2
\bigl[1 - F_1(\lambda_3,t_c-t_1)   \epsilon_3 \bigr]
  B_0,\\[1ex]
p_{103}(\lambda_1) & = C_1
f_3(\lambda_1,\lambda_2,\lambda_3,t_1)   \bar{\epsilon}_1
\bar{\epsilon}_2   \epsilon_3
  B_0,\\[1ex]
p_{104}(\lambda_1) & = C_1
\bigl[\bar{F}_1(\lambda_1,t_c) +
\bigl(\bar{F}_2(\lambda_1,\lambda_2,t_c) - \bar{F}_1(\lambda_1,t_c)\bigr)
\bar{\epsilon}_1 +
\bigl(\bar{F}_3(\lambda_1,\lambda_2,\lambda_3,t_c) - \\
& \qquad -
\bar{F}_2(\lambda_1,\lambda_2,t_c)\bigr)   \bar{\epsilon}_1   \bar{\epsilon}_2 +
F_3(\lambda_1,\lambda_2,\lambda_3,t_c)
  \bar{\epsilon}_1   \bar{\epsilon}_2   \bar{\epsilon}_3\bigr]
  B_1   t_c^{-1}\\
& = C_1
\bigl[1 - F_1(\lambda_1,t_c)   \epsilon_1 -
F_2(\lambda_1,\lambda_2,t_c)   \bar{\epsilon}_1   \epsilon_2 -
F_3(\lambda_1,\lambda_2,\lambda_3,t_c)
  \bar{\epsilon}_1   \bar{\epsilon}_2   \epsilon_3 \bigr] \\
& \quad   B_1   t_c^{-1}.
\end{align*}
\end{widetext}
The joint probability density function for observing one decay event at
time $t_1$ is the sum of the single probability densities:

\begin{equation*}
p_1(\lambda_1) = p_{101}(\lambda_1) + p_{102}(\lambda_1) +
p_{103}(\lambda_1) + p_{104}(\lambda_1),
\end{equation*}
where the normalization constant $C_1$ fulfills the equation:

\begin{equation*}
\label{eqn:mlh_norm}
\int_0^{t_c} p_1(\lambda_1)\, dt_1 = 1.
\end{equation*}

Ten different scenarios need to be considered when two decay events occur within $t_{c}$, and twenty different scenarios for three decay events. A detailed description of
all scenarios and the resulting normalized joint probability functions
$p_2(\lambda_1)$ and $p_3(\lambda_1)$ can be found in Refs.~\cite{Sch96,Sto01}.

The analysis program assigns decay events to preceding
implantations, only events with one ($n_i = 1$), two ($n_i = 2$) or
three $n_i = 3$) decay events within the correlation time are
considered.  Therefore, we initially maximize the likelihood function
for $N_{123}$ observed decay sequences:

\begin{widetext}
\begin{equation*}
{\cal L}_{123}(\lambda_1) =
\prod_{i=1}^{N_{123}} \Bigl[ \delta(n_i-1)   p_1(\lambda_1) +
\delta(n_i-2)   p_2(\lambda_1) +
\delta(n_i-3)   p_3(\lambda_1) \Bigr].
\end{equation*}
The solution $\hat{\lambda}_{1_0}$ of the maximization equation
\end{widetext}

\begin{equation*}
\frac{\partial {\cal L}_{123}(\lambda_1)}{\partial \lambda_1} = 0
\end{equation*}

has to be corrected for events with
no observed decay events within the correlation time window.
The most likely number of events $N_0$ with no observed decay events
within the correlation time depends on $P_0(\lambda_1)$ and therefore
on $\lambda_1$ itself:

\begin{equation*}
N_0 = \frac{P_0(\lambda_1)}{1-P_0(\lambda_1)}   N_{123}.
\end{equation*}

To find the solution $\hat{\lambda}_{1_{j+1}}$ of the maximization equation
of the joint likelihood function ${\cal L}$, an iterative numerical method is
used until $\hat{\lambda}_1$ converges:

\begin{equation*}
{\cal L}_{(j+1)}(\lambda_1)={\cal L}_{123}(\lambda_1)
P_0(\lambda_1)^{N_0(\hat{\lambda}_{1_j})},
\end{equation*}

\begin{equation*}
\left.\frac{\partial {\cal L}_{(j+1)}(\lambda_1)}{\partial \lambda_1}
\right| _{\lambda_1=\hat{\lambda}_{1_{(j+1)}}} = 0.
\end{equation*}

The correlation time, therefore, should be long compared to the mean life time of the
mother nuclei to avoid large correction factors due to this iterative
method. If the background rate is low enough, a correlation time equal
to ten half-lives should be used. If the time window is too long, the
assumption of a maximum number of three decays within the correlation
time is no longer valid and the maximum likelihood method might
fail.

The validity of this approximation as well as a check of the whole
procedure for the analysis of decay sequences was thoroughly discussed in
Ref.~\cite{Sto01}.

%

\end{document}